\documentclass[aps,prl,twocolumn,showpacs,superscriptaddress,
groupedaddress,10pt]{revtex4}
\usepackage{amsmath}
\usepackage{amssymb}
\usepackage{graphicx}
\usepackage{epsfig}
\usepackage{dcolumn}
\usepackage{bm}
\usepackage{bm}
\usepackage{blindtext}
\usepackage{natbib}
\usepackage{siunitx} 

\usepackage{multirow}
\usepackage{bbm}

\usepackage[usenames,dvipsnames]{xcolor}
\usepackage{amsthm}
\usepackage{booktabs}
\usepackage{hyperref}
\usepackage{tikz}
\usetikzlibrary{calc}
\usepackage[printwatermark]{xwatermark}
\newcommand{\tikzmark}[1]{\tikz[overlay,remember picture] \node (#1) {};}
\newcommand{\DrawBox}[3][]{%
    \tikz[overlay,remember picture]{
    \draw[black,#1]
      ($(#2)+(-0.1em,2.0ex)$) rectangle
      ($(#3)+(0.1em,-0.5ex)$);}
}
\usepackage{mathtools}

\def\be{\begin{equation}} \def\ee{\end{equation}}
\def\bea{\begin{eqnarray}} \def\eea{\end{eqnarray}}

\def\nn{\nonumber}

\newcommand{\ket}[1]{| #1 \rangle}
\newcommand{\bra}[1]{\langle #1 |}

\begin{document}
\title{
Phase diagram of the spin-1/2 Kitaev-Gamma chain and emergent  SU(2) symmetry
}

\author{Wang Yang}
\affiliation{Department of Physics and Astronomy and Stewart Blusson Quantum Matter Institute,
University of British Columbia, Vancouver, B.C., Canada, V6T 1Z1}

\author{Alberto Nocera}
\affiliation{Department of Physics and Astronomy and Stewart Blusson Quantum Matter Institute, 
University of British Columbia, Vancouver, B.C., Canada, V6T 1Z1}

\author{Tarun Tummuru}
\affiliation{Department of Physics and Astronomy and Stewart Blusson Quantum Matter Institute, 
University of British Columbia, Vancouver, B.C., Canada, V6T 1Z1}

\author{Hae-Young Kee}
\affiliation{Department of Physics, University of Toronto, Ontario M5S 1A7, Canada}

\author{Ian Affleck}
\affiliation{Department of Physics and Astronomy and Stewart Blusson Quantum Matter Institute, 
University of British Columbia, Vancouver, B.C., Canada, V6T 1Z1}

\begin{abstract}
We study the phase diagram of a one-dimensional version of the Kitaev spin-1/2 model with an extra ``$\Gamma$-term",
using analytical, density matrix renormalization group and exact diagonalization methods.
Two intriguing phases are found.
In the gapless phase, 
although the exact symmetry group of the system is discrete,
the low energy theory is described by an emergent SU(2)$_1$ Wess-Zumino-Witten (WZW)  model.
On the other hand, the spin-spin correlation functions exhibit SU(2) breaking  prefactors,
even though the exponents and the logarithmic corrections are consistent with the SU(2)$_1$ predictions.
A modified nonabelian bosonization formula is proposed to capture such exotic emergent ``partial" SU(2) symmetry.
In the ordered phase, 
there is numerical evidence for an $O_h\rightarrow D_4$ spontaneous symmetry breaking.

\end{abstract}
%\pacs{75.10.Pq, 05.30.Rt, 71.10.Hf, 75.10.Jm}
\maketitle

A quantum spin liquid is a phase of matter in which the constituent spins are highly entangled with each other without exhibiting any long range order \cite{Balents2010,Witczak-Krempa2014,Rau2016,Winter2017,Zhou2017,Savary2017}.
The Kitaev spin-1/2 model on the two-dimensional (2D) honeycomb lattice 
is an exactly solvable spin liquid model \cite{Kitaev2006},
which has received considerable theoretical and experimental interest in the past decade  \cite{Jackeli2009,Chaloupka2010,Singh2010,Price2012,Singh2012,Plumb2014,Kim2015,Winter2016,Baek2017,Leahy2017,Sears2017,Wolter2017,Zheng2017,Rousochatzakis2017,Kasahara2018,Catuneanu2018,Gohlke2018} due to its potential in realizing quantum computers \cite{Nayak2008}.
On the other hand,  
additional spin interactions allowed by the lattice symmetries inevitably exist in real materials.
The Heisenberg term was first considered as a supplement to the Kitaev model \cite{Chaloupka2010}.
Later it had been proposed that another off-diagonal term, dubbed the ``$\Gamma$-term", naturally arises in Kitaev materials  \cite{Rau2014},
which dominates over the Heisenberg term and may even be crucial to stabilize the ferromagnetic (FM) Kitaev phase \cite{Gordon2019}.

In one dimension (1D), strong quantum fluctuations make 1D spin liquids even more ubiquitous \cite{Haldane1983,Haldane1983a,Affleck1987,Affleck1988,Affleck1989}. 
In many circumstances, strongly interacting 1D systems  are more amenable to both analytical and numerical treatments.
Methods having been proven successful in 1D include conformal field theory (CFT) \cite{Belavin1984,Knizhnik1984,Affleck1987a,Affleck1988,Affleck1995}, bosonization \cite{Haldane1981,Haldane1981a,Witten1984}, Bethe ansatz \cite{Bethe1931,Yang1966,Yang1966a,Faddeev1979,Baxter1982}, and the density matrix renormalization group (DMRG) method \cite{White1992,White1993,Schollwock2011}.
Due to this reason, 1D physics is not only interesting in its own respect,
but also useful in providing hints on strongly correlated physics in higher dimensions
when an exact or controllable approach is lacking.

%-------------------------------------------- 
\begin{figure}[h]
\includegraphics[width=0.48\textwidth]{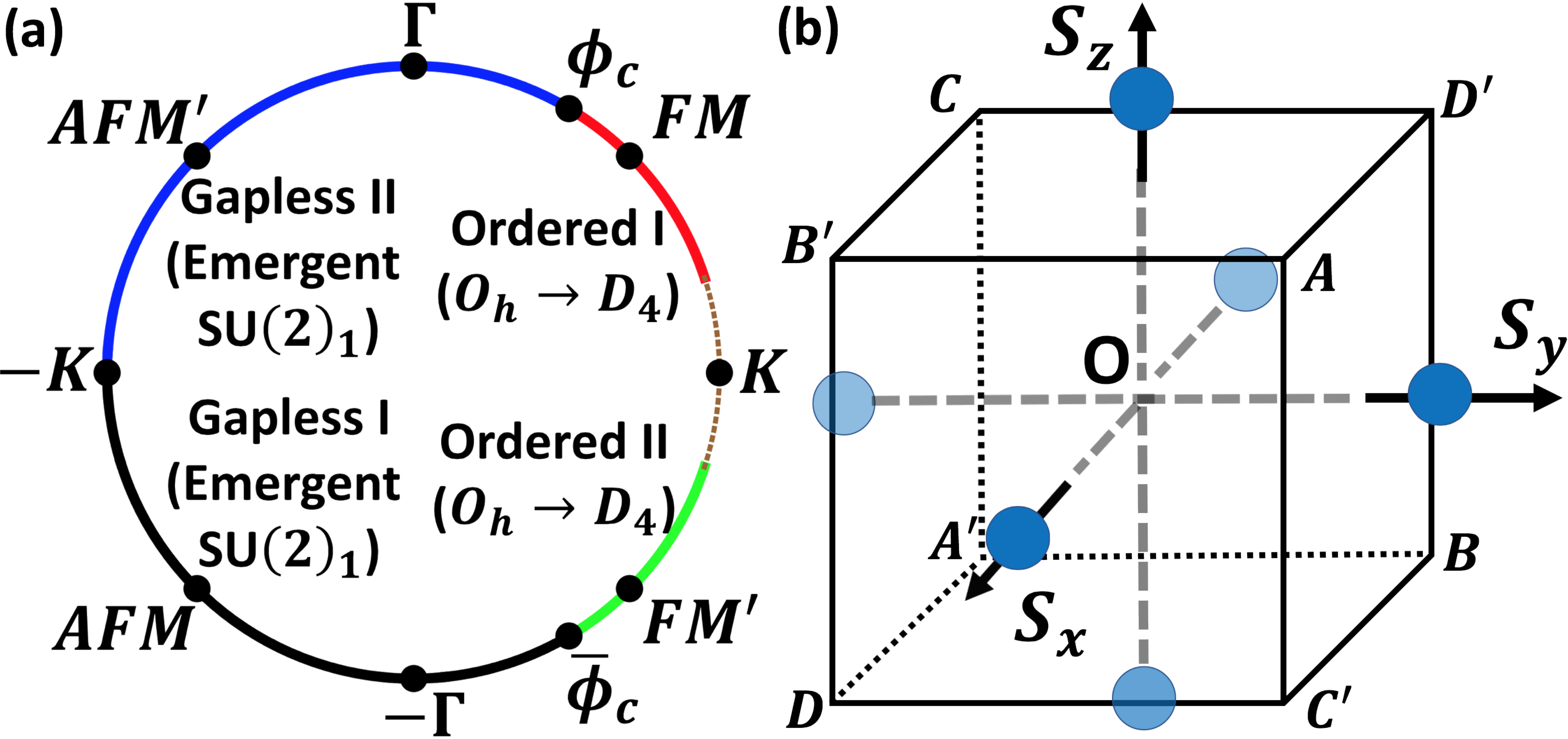}
\caption{
(a) Phase diagram, and (b) a cube in spin space.
In (a), the regions ``Gapless I, II" are related by a three-sublattice rotation, and so are ``Ordered I, II".
The region close to $K$ represented by the thin dashed line remains to be explored further.
%The hidden SU(2) symmetric points $FM$, $AFM^\prime$, $AFM$, $FM^\prime$ are located at $\phi=\pi/4,3\pi/4,5\pi/4,7\pi/4$, respectively.
%In (b), the solid blue circles represent the ``center of mass" spin directions in a unit cell for the six degenerate ground states in the ordered phase.
} \label{fig:phase}
\end{figure}
%--------------------------------------------

In this work, by combining the analytical, exact diagonalization (ED) and DMRG methods,
we study the phase diagram of a 1D version of the Kitaev model with an extra $\Gamma$-term, termed as the spin-1/2 ``Kitaev-Gamma chain".
The phase diagram is divided into a gapless phase and an ordered phase as shown in Fig. \ref{fig:phase} (a).
In the gapless phase, we find that the low energy theory is described by an emergent SU(2)$_1$ Wess-Zumino-Witten (WZW) model, 
although the exact symmetry group $G\cong O_h \ltimes \mathbb{Z}$  is discrete where $O_h$ is the full octahedral group. 
On the other hand, the spin-spin correlation functions exhibit SU(2) breaking  prefactors as revealed by DMRG numerics,
even though the exponents and the logarithmic corrections are consistent with SU(2)$_1$ predictions.
A modified nonabelian bosonization formula for the spin operators is proposed to incorporate such emergent ``partial" SU(2) symmetry.
Based on a renormalization group (RG) analysis, 
the SU(2) breaking coefficients  in the ``bridge" between the local spin operators and the low energy degrees of freedom is attributed to a multiplicative renormalization of the spin operators in the high energy region along the RG flow before the SU(2)$_1$ low energy theory applies.
In the ordered phase, 
ED and DMRG calculations show evidence of an $O_h \rightarrow D_4$ spontaneous symmetry breaking
where $D_{n}$ represents the dihedral group of order $2n$,
except in a small region close to the Kitaev point which remains to be explored further.
%In both phases of the spin-1/2 chain, the system is either gapless or has a long range order.
%Hence, our study provides an interesting example beyond the Lieb-Schultz-Mattis theorem \cite{Lieb1961} since the model has no continuous spin rotational symmetry.
We also discuss the relevance of our work to real materials and higher dimensional systems.

%-------------------------------------------------------
\begin{figure}
\includegraphics[width=0.48\textwidth]{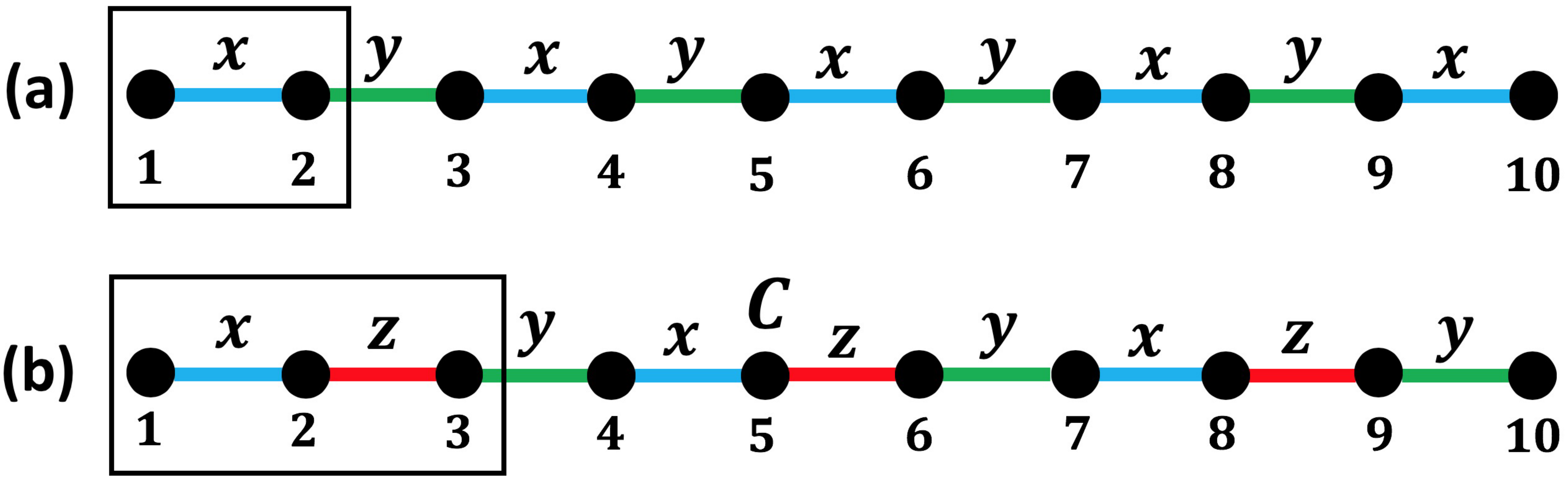}
\caption{Bond structures (a) before and (b) after the six-sublattice rotation.
The rectangular boxes denote unit cells.
}
\label{fig:bonds}
\end{figure}
%-------------------------------------------------------

{\it The Model}  The Hamiltonian of the spin-1/2 Kitaev-Gamma chain is defined as 
\begin{eqnarray}
H=\sum_{<ij>\in \gamma\,\text{bond}} \big[ K S_i^{\gamma}S_j^{\gamma}   +\Gamma (S_i^{\alpha}S_j^{\beta}+S_i^{\beta}S_j^{\alpha})\big],
\label{eq:Ham}
\end{eqnarray}
in which $i$ and $j$ are two nearest neighboring sites, 
$\gamma=x$ (or $y$) is the spin direction associated with the $\gamma$-bond,
$\alpha,\beta=y,z$ (or $z,x$) on the $\gamma$-bond are the other two spin directions different from $\gamma$,
and $K=\cos(\phi)$ and $\Gamma=\sin(\phi)$ are the Kitaev and Gamma couplings, respectively.
The bond pattern shown in Fig. \ref{fig:bonds} (a) is generated by selecting out one row of the 2D Kitaev model on the honeycomb lattice \cite{Kitaev2006}.
%A global spin rotation $R(\hat{z},\pi)$ around $z$-axis by $\pi$ maps $(K,\Gamma)$ to $(K,-\Gamma)$, and as a result, $\phi$ is equivalent to $2\pi-\phi$.
Under a six-sublattice rotation \cite{Chaloupka2015,Stavropoulos2018}, the model can be mapped into the following form (see Supplementary Materials (SM) \cite{suppl}, which includes Refs. \cite{Dresselhaus2008,Calabrese2009,Coxeter1965,Eggert1992,Affleck1989a,Cardy1996}),
\begin{eqnarray}
H^\prime=\sum_{<ij>\in \gamma}\big[ -KS_i^\gamma S_j^\gamma-\Gamma (S_i^\alpha S_j^\alpha+S_i^\beta S_j^\beta) \big],
\label{eq:rotated}
\end{eqnarray}
in which the bonds $\gamma=x,z,y$  depicted in Fig. \ref{fig:bonds} (b) have a three-site periodicity. 
Clearly, $\phi=\pi/4$ and $5\pi/4$ are SU(2) symmetric in the rotated frame with FM and antiferromagnetic (AFM) couplings, respectively.
%, corresponding to the $FM$ and $AFM$ points in Fig. \ref{fig:phase} (a).
%Due to the equivalence established by $R(\hat{z},\pi)$, $\phi=7\pi/4$ and $3\pi/4$ also have hidden SU(2) symmetry.

From here on, we will stick to the rotated Hamiltonian in Eq. (\ref{eq:rotated}). 
%\cite{rotback}.
The phase diagram is shown in Fig. \ref{fig:phase} (a), in which
$\phi_c$  separating the gapless and ordered phases takes
the value $\approx 0.335 \pi$ 
determined by numerics with evidence to be a first order phase transition (see SM \cite{suppl}).
Another three-sublattice rotation $\mathcal{T}_3$ maps $(K,\Gamma)$ to $(K,-\Gamma)$ (see SM \cite{suppl}),
and as a result, $\phi$ is equivalent to $2\pi-\phi$.
%Therefore, $\phi=7\pi/4$ and $3\pi/4$ also have hidden SU(2) symmetry, as shown by the $FM^\prime$ and $AFM^\prime$ in Fig. \ref{fig:phase} (a).
There is numerical evidence for the exactly solvable \cite{Brzezicki2007} and self-dual $K$ and $-K$ points in Fig. \ref{fig:phase} (a) to be 
a continuous and a first order phase transition point, respectively, as discussed in SM \cite{suppl}.
Due to the equivalence established by $\mathcal{T}_3$, we will focus on the range $\phi\in[\pi,2\pi]$ in the subsequent discussions.
The Hamiltonian will be taken as $H^\prime$ and $\mathcal{T}_3H^\prime \mathcal{T}_3^{-1}$ in the ranges $[\pi,\bar{\phi}_c]$
% (the ``Gapless I" phase in Fig. \ref{fig:phase})
and $[\bar{\phi}_c,2\pi]$ 
%(``Ordered II" in Fig. \ref{fig:phase})
, respectively,
where $\bar{\phi}_c=2\pi-\phi_c$.

Next we discuss the symmetries of the model.
The Hamiltonian in Eq. (\ref{eq:rotated}) has time reversal symmetry $T$.
The bond structure is transformed back by applying either the screw operation $R_aT_a$ or the combined operation $R_I I$,
in which $T_a$ is translation by one site, $I$ is the spatial inversion around the site $C$ shown in Fig. \ref{fig:bonds} (b),
$R_a:(S_i^x,S_i^y,S_i^z)\rightarrow (S_i^z,S_i^x,S_i^y)$ and $R_I:(S_i^x,S_i^y,S_i^z)\rightarrow (-S_i^z,-S_i^y,-S_i^x)$ 
are rotations in spin space.
The Hamiltonian is also invariant under spin rotations $R(\hat{\alpha},\pi)$ ($\alpha=x,y,z$)
where $R(\hat{n},\theta)$ represents a rotation around the $\hat{n}$-axis by an angle $\theta$ in spin space.
%Furthermore, the bond structure is transformed back by applying either  $R_aT_a$ or $R_I I$,
%in which $T_a$ is translation by one site, $I$ is the inversion to the plane $L$ shown in Fig. \ref{fig:bonds} (b),
%With this notation, $R_a$ and $R_I$ can also be expressed as
%$R_a=R(\hat{n}_a,-2\pi/3)$ and $R_I=R(\hat{n}_I,\pi)$, where  $\hat{n}_a=\frac{1}{\sqrt{3}}(1,1,1)^T$ and $\hat{n}_I=\frac{1}{\sqrt{2}}(1,0,-1)^T$.
In summary, the symmetry group $G$ is generated by the above mentioned operations, {\it i.e}.,
$G=\mathopen{<}  T,R_aT_a,R_I I, R(\hat{x},\pi),R(\hat{y},\pi),R(\hat{z},\pi) \mathclose{>}$.

The group structure of $G$ can be worked out.
A translation by three sites $T_{3a}=(R_aT_a)^3$ generates an abelian normal subgroup $\mathopen{<} T_{3a} \mathclose{>}$ of G,
hence it's legitimate to consider the quotient group $G/\mathopen{<} T_{3a}\mathclose{>}$.
%If the spatial transformations $T_a$ and $I$ are suppressed,
The spin operations $R_a,R_I,R(\hat{\alpha},\pi)$ ($\alpha=x,y,z$) are all symmetries of a cube in the spin space
shown in  Fig. \ref{fig:phase} (b).
%(see SM \cite{suppl}).
%as depicted in Fig. \ref{fig:FM} (a).
Furthermore, $T:\vec{S}_i\rightarrow -\vec{S}_i$ can be viewed as an improper element with determinant $-1$ 
which also leaves the cube invariant.
Indeed, these operations within the spin space generate the full octahedral group $O_h$.
On the other hand, it's a pleasant result that $G/\mathopen{<} T_{3a}\mathclose{>}$ is isomorphic to $O_h$ even if the spatial operations $T_a$ and $I$ are also included (see SM \cite{suppl}).
%which is analyzed in detail in Appendix \ref{app:group}.
Thus $G\cong O_h \ltimes \mathbb{Z}$ where $\mathbb{Z}\cong \mathopen{<} T_{3a}\mathclose{>}$. 
%We also note that the Hilbert space of the spin-$1/2$ chain is a projective representation of $O_h$.

{\it The gapless phase} 
We first briefly review the low energy theory at the SU(2) symmetric point $\phi_{AF}=5\pi/4$.
The low energy physics is described by the SU(2)$_1$ WZW model \cite{Affleck1985,Affleck1988,Affleck1989}.
The nonabelian bosonization formula \cite{Affleck1988}
$\frac{1}{a} S_n^\alpha = J_L^{\alpha}(x) +J_R^{\alpha}(x)+\text{const.} (-)^{x/a} \frac{1}{\sqrt{a}}i \text{tr} (g(x)\sigma^{\alpha})$
relates the local spin operators to the low energy degrees of freedom in the SU(2)$_1$ WZW model,
in which $a$ is the lattice constant, $x=na$ is the coordinate in the continuum limit,
$\vec{J}_L$ and $\vec{J}_R$ are the left and right WZW currents, respectively, 
%the SU(2) matrix 
$g$ is an SU(2) matrix  which is also the WZW primary field,
and $\sigma^{\alpha}$ ($\alpha=x,y,z$) are the three Pauli matrices.
Since the scaling dimensions of $g$ and the currents $\vec{J}_L, \vec{J}_R$ are $1/2$ and $1$, respectively, 
the zero temperature equal-time spin-spin correlation functions in the long distance limit  can be derived as 
$\langle S^{\alpha}_i S^{\beta}_j  \rangle = \delta_{\alpha \beta} \big[ -\frac{1}{4\pi^2}\frac{1}{r^2} +(-)^r\frac{1}{(2\pi)^{3/2}}\frac{\ln^{1/2}(r/r_0)}{r} \big]$,
in which $r=|i-j|\gg 1$, $\langle \cdots \rangle$ is an average over the ground state,
 $r_0$ is some microscopic length scale, 
and the logarithmic correction arises from the marginally irrelevant term $-\lambda\vec{J_L}\cdot \vec{J_R}$ \cite{Affleck1989a,Affleck1998}
where $\lambda>0$ is the coupling constant.
For a periodic system of size $L$, 
the correlation functions can be obtained by replacing $r$ 
with $\frac{L}{\pi} \sin(\frac{\pi r}{L})$ \cite{Barzykin1999}.

Now we analyze the low energy field theory away from $\phi_{AF}$.
In the vicinity of $\phi_{AF}$, the SU(2) breaking term $(\Gamma-K)\sum_{<ij>\in \gamma} S_i^\gamma S_j^\gamma$ 
can be treated as a perturbation to the SU(2)$_1$ WZW model.
The dimension $1/2$ operators $\epsilon,N^\alpha$ and the dimension $3/2$ operators $J^\alpha_L\epsilon,J^\alpha_R\epsilon, J_L^{\alpha} N^{\beta},J_R^\alpha N^\beta$
flip the sign under $T_{3a}$ since $T_a:g\rightarrow -g$ \cite{Affleck1988}, where $\epsilon(x)=\text{tr}g(x)$ and $\vec{N}(x)=i\text{tr}(g(x)\vec{\sigma})$ are the dimer and Neel order fields, respectively \cite{Affleck1988}.
The dimension $1$ operators $J^\alpha_L,J^\alpha_R$ acquire a sign change under $R(\hat{\beta},\pi)$ ($\beta\neq \alpha$).
Hence they are all forbidden in the low energy theory.
Among the dimension $2$ operators, only the SU(2) invariant combinations $\vec{J}_L\cdot \vec{J}_L+\vec{J}_R\cdot \vec{J}_R$ and $\vec{J}_L\cdot \vec{J}_R$
are allowed by the $O_h$ symmetry.
Higher dimensional operators are irrelevant at the CFT fixed point.
Thus we conclude that there is an emergent SU(2)$_1$ symmetry in a neighborhood of $\phi_{AF}$. %\cite{conformaltower}.
Indeed, the finite size spectrum exhibits a conformal tower structure consistent with an emergent SU(2)$_1$ symmetry (see SM \cite{suppl}).

On the other hand, when $\phi\neq\phi_{AF}$, 
we propose the following modified nonabelian bosonization formula
with SU(2) breaking coefficients:
%\begin{widetext}
\begin{eqnarray}
&\frac{1}{a}S_n^{\alpha}=\big[c_0+c_4\cos(\frac{2\pi}{3a}x+\frac{2\pi}{3}(\alpha-1)) \big] \big(J_L^\alpha(x)+J_R^\alpha(x) \big)\nn\\
&+\big[c_6(-)^{\frac{x}{a}}+c_2\cos(\frac{\pi}{3a}x+\frac{2\pi}{3}(\alpha-1))\big]\frac{1}{\sqrt{a}}i\text{tr}(g(x)\sigma^\alpha),
\label{eq:modified_bosonize}
\end{eqnarray}
%\end{widetext}
in which $\alpha=1,2,3$ corresponding to $\alpha=x,y,z$, and $x=na$.
As can be seen from Eq. (\ref{eq:modified_bosonize}),
 $\langle S^{\alpha}_1 S^{\alpha}_{r+1}  \rangle$ now contains momentum $\pm\pi/3$ and $\pm2\pi/3$ oscillating components
in addition to the uniform part $u_{\alpha\alpha} (r)$ and the staggered part $(-)^r s_{\alpha\alpha}(r)$ where both $u_{\alpha\alpha}$ and $s_{\alpha\alpha}$ are smooth.
Alternatively,
\begin{eqnarray}
\langle S^{\alpha}_i S^{\beta}_j  \rangle = \delta_{\alpha \beta} \big[ -D_{[i]}^\alpha D_{[j]}^\alpha\frac{1}{r^2} +(-)^r C_{[i]}^\alpha C_{[j]}^\alpha\frac{\ln^{1/2}(r/r_0)}{r} \big],
\label{eq:modified_correlation}
\end{eqnarray}
in which $1\leq[i]\leq 3$, $i\equiv [i] \mod 3$, $r=|i-j|$,
$D_1^z=D_2^y=D_3^x=D_1=c_0+c_4$, $D_1^x=D_1^y=D_2^x=D_2^z=D_3^y=D_3^z=D_2=c_0-c_4/2$, 
and similar equalities hold for the $C$'s with $C_1=c_6+c_2$ and $C_2=c_6-c_2/2$.
Particularly, the relations between the $D$'s (also $C$'s) are dictated by symmetries (see SM \cite{suppl}).

%-------------------------------------
\begin{figure}
\includegraphics[width=0.49\textwidth]{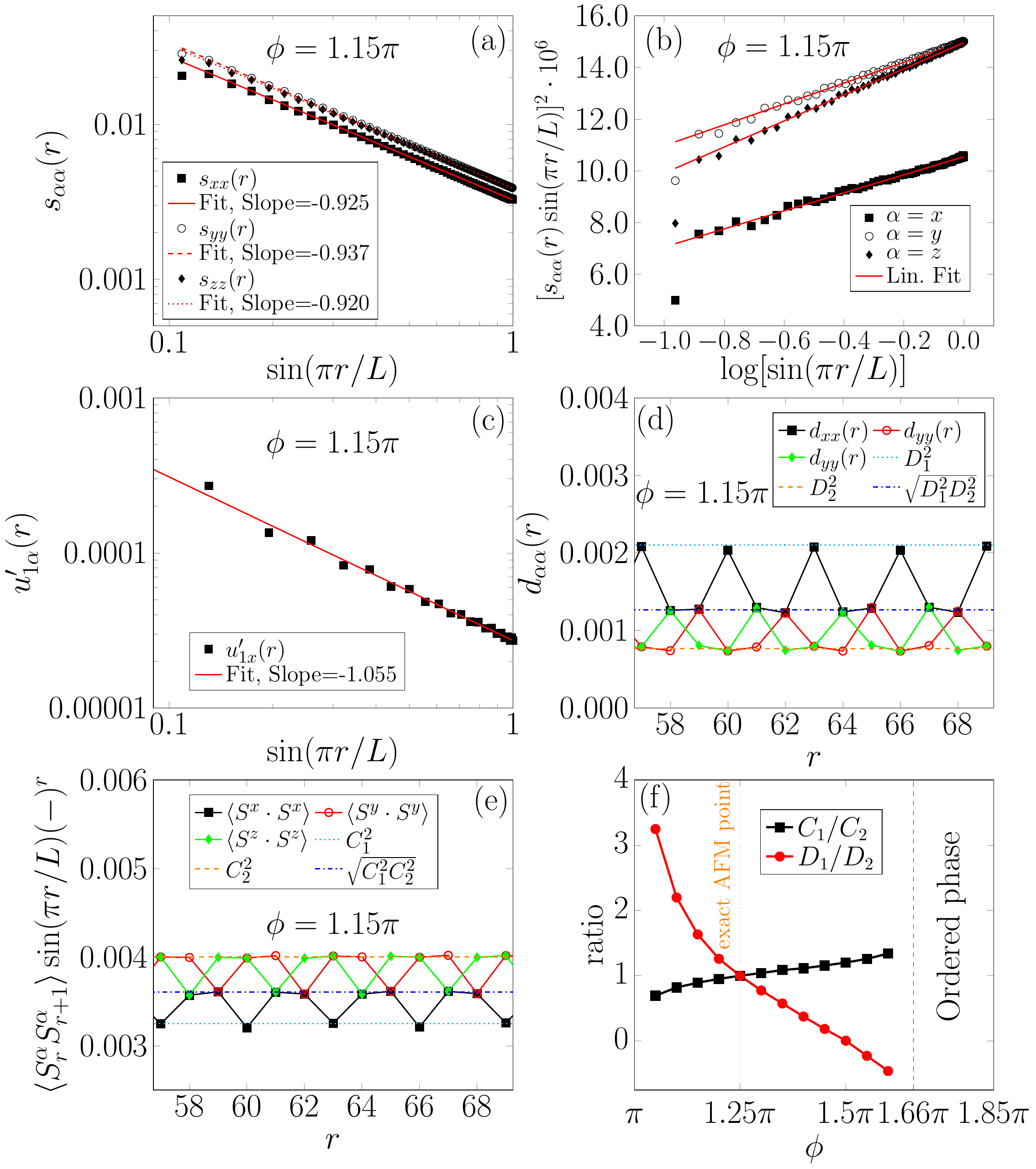}
\caption{
(a) $s_{\alpha\alpha}(r)$  vs. $\tilde{r}_L$ ($=\sin(\pi r/L)$) 
on a log-log scale, (b) $(\tilde{r}_L s_{\alpha\alpha}(r))^2$ vs. $\log \tilde{r}_L$,
(c) $u^\prime_{1x}(r)$ vs. $\tilde{r}_L$ on a log-log scale, (d) $d_{\alpha\alpha}(r)$ vs. $r$, 
(e) $(-)^r \tilde{r}_L \langle S^{\alpha}_{1} S^{\alpha}_{r+1}  \rangle$  vs. $r$, 
and (f) $C_1/C_2$ and $D_1/D_2$ as varying $\phi$,
in which $s_{\alpha\alpha}(r)$ is the staggered component of the correlation function $\langle S^{\alpha}_{1} S^{\alpha}_{r+1}  \rangle$;
 $u^\prime_{1x}(r)$ is the uniform component of $\tilde{r}_L\langle S^{x}_{1} S^{x}_{r+1}  \rangle$ where $r$ is chosen every three sites;
$d_{\alpha\alpha}(r)$ ($\alpha=x,y,z$) is reconstructed from $\tilde{r}_L u^\prime_{i\alpha}(r)$ by combining $i=1,2,3$ together; 
$C_i,D_i$ ($i=1,2,$)  are the SU(2) breaking coefficients in the modified nonabelian bosonization formula.
In all of (a,b,c,d,e), $\phi=1.15\pi$.
%DMRG numerics are performed on a chain of $L=144$ sites.
} 
\label{fig:CnD}
\end{figure}
%-------------------------------------

Next we proceed to numerics.
Throughout the gapless phase, DMRG numerics are performed on a chain of $L=144$ sites with a periodic boundary condition.
The DMRG results for the staggered parts $s_{\alpha\alpha}(r)$ at $\phi=1.15\pi$
are displayed in Fig. \ref{fig:CnD} (a,b).
%and the results for the SU(2) symmetric point are displayed in Fig. \ref{fig:staggered} (a,b) for comparison.
$s_{\alpha\alpha}$ ($\alpha=x,y,z$) are extracted from the numerical data using a nine-point formula 
derived in SM \cite{suppl}.
In Fig. \ref{fig:CnD} (a), the slopes determined from a linear fit of $\log s_{\alpha\alpha}(r)$ vs. $\log(\sin(\pi r/L))$
are all close to $-1$ within 5-10\%,
compatible with Eq. (\ref{eq:modified_correlation}).
To study the logarithmic corrections,  $(\sin(\pi r/L) s_{\alpha\alpha}(r))^2$ vs. $\log\big(\sin(\pi r/L)\big)$ are plotted in Fig. \ref{fig:CnD} (b) all exhibiting good linear relations,
which confirm the logarithmic factor with a power of $1/2$.
%In particular, the splitting of the lines in Fig. \ref{fig:CnD} (b) indicates an SU(2) breaking in the prefactors of $(-)^r\ln^{1/2}(r/r_0)/r$.

We also study numerically in detail the SU(2) breaking factors in Eq. (\ref{eq:modified_correlation})  at $\phi=1.15\pi$.
In Fig. \ref{fig:CnD} (e), $(-)^r\sin(\pi r/L)\langle S^{\alpha}_1 S^{\alpha}_{r+1}  \rangle$ are observed to
approach constant values at large $r$
%(but still within a small range of $r$ such that the logarithmic differences can be neglected),
which are proportional to $C_{1}^\alpha C_{[r+1]}^\alpha$ according to Eq. (\ref{eq:modified_correlation}) \cite{Logarithmic}.
To obtain $D_1D_2,(D_1)^2,(D_2)^2$,
we first pick out the data for every three sites,
then $u^{\prime}_{i\alpha}(i+3n-1)$ defined as
the uniform component  of $\{ \sin(\pi (i+3n-1)/L) \langle S^{\alpha}_1 S^{\alpha}_{i+3n}  \rangle\}_{n\in \mathbb{Z}}$  as a function of $n$ for fixed $i$ and $\alpha$
is extracted from a three-point formula \cite{suppl},
where $\alpha=x,y,z$, and $i=1,2,3$.
%Fig. \ref{fig:CnD} (a) shows $u^\prime_{1x}(n)$ vs. $\sin(\pi (i+3n-1)/L)$ on a log-log scale.
As is clear from Fig. \ref{fig:CnD} (c), a good linear fit for $\log u^\prime_{1x}(r)$ vs. $\log\big( \sin(\pi r)/L)\big)$
is obtained with a slope close to $-1$ within $\sim 5\%$.
Hence the exponent of $u^\prime_{1x}(r)$ is determined to be around $2$.
Fig. \ref{fig:CnD} (d) displays the function $d_{\alpha\alpha}(r)$ which is reconstructed from $\sin(\pi r/L) u^{\prime}_{i\alpha}(r)$ for a fixed $\alpha$ but combining $i=1,2,3$ together,
and $D_1^{\alpha}D_{i}^{\alpha}$ can be determined from the asymptotic value of $\sin(\pi r/L) u^{\prime}_{i\alpha}(r)$ at large $n$, 
in which $r=i+3n-1$. 
We have verified that the independently extracted values of $C_1C_2=a_1,(C_1)^2=a_2,(C_2)^2=a_3$ and $D_1D_2=b_1,(D_1)^2=b_2,(D_2)^2=b_3$ 
indeed satisfy the relationships $a_1=\sqrt{a_2a_3}$ and $|b_1|=\sqrt{b_2b_3}$,
which provide direct numerical evidence for Eq. (\ref{eq:modified_correlation}).
In addition, the ratios $C_1/C_2$ and $D_1/D_2$ are studied as a function of the angle $\phi$
as shown in Fig. \ref{fig:CnD} (f),
which provide evidence for emergent SU(2)$_1$ in the entire AFM phase.

Finally, we present an RG argument to understand the origin of the SU(2) breaking coefficients.
The following fermion Hamiltonian is considered
\begin{eqnarray}
H_F&=-t\sum_{<ij>,\alpha} (c_{i\alpha}^\dagger c_{j\alpha}+\text{h.c.})-\mu\sum_{i\alpha}c_{i\alpha}^\dagger c_{i\alpha}\nn\\
&+U\sum_i n_{\uparrow}n_{i\downarrow}+\Delta \sum_{<ij>\in \gamma} S_{i}^\gamma S_{j}^\gamma
\label{eq:fermion}
\end{eqnarray}
in which $t$ is the hopping term, $\mu$ the chemical potential tuned to half-filling,
 $U>0$ is a repulsive Hubbard coupling, $\Delta=\Gamma-K$, and $S_{i}^\gamma=c^\dagger_{i\alpha}\frac{1}{2} \sigma^\gamma_{\alpha\beta} c_{i\beta}$.
%Hence,
In the limit $|\Delta|\ll U\ll t$, $H_F$ contains the same low energy physics in the spin sector as 
that of an SU(2) symmetric AFM Heisenberg chain  perturbed by $\Delta \sum_{<ij>\in \gamma} S_{i}^\gamma S_{j}^\gamma$ \cite{LargeU}.
%the second line of Eq. (\ref{eq:fermion}).
%Hence, taking derivatives to the scaling fields $h_{k}^{\alpha}(n)$ of the free energy computed from $H_F$ 
%is able to give the long-distance behaviors of the spin-spin correlation functions of Eq. (\ref{eq:Ham}) when $\phi$ is close to $\phi_{AF}$.
To study the renormalization of spin operators, a term $-\sum_{kn\alpha} h_{k}^\alpha(n) S_{k+3n}^\alpha$ can be added to $H_F$,
in which $k=1,2,3$ are the site indices within a unit cell (u.c.),
$n$ is summed over the unit cells,
and $h_{k}^\alpha(n)$ are the scaling fields coupled to $S_{i+3n}^\alpha$ 
which can be separated to a uniform part $h_{u,k}^\alpha(n)$
and a staggered part $(-)^{k+3n} h_{s,k}^\alpha(n)$.

We study the flow of $h_{\eta, k}^\alpha(n)$ ($\eta=u,s$, $k=1,2,3$) by lowering the cutoff from $\Lambda_0=\pi/a$ to $\Lambda_1\sim \Lambda_0/b_1$ close to the free fermion fixed point.
 The RG scale $b_1$ can be taken as $\sim 3$ where the three sites within a unit cell can no longer be clearly distinguished. 
%Note that a linearization of the fermion dispersion is not valid at $\Lambda_1$, hence the SU(2)$_1$ low energy description has not emerged yet.
Neglecting the flow of the marginal operator $\Delta$ and keeping terms up to first order in $\Delta$, 
the flow equations of $h_{\eta,k}^x(b,n)$ at scale $\Lambda_0/b$ are  
\begin{eqnarray}
\frac{d h_{\eta,i}^x}{d\ln b} &=& (1-\nu_\eta \Delta)h_{\eta,i}^x - \lambda_\eta \Delta h_{\eta,j}^x- \nu_\eta \Delta h_{\eta,3}^x,\nn\\
\frac{d h_{\eta,3}^x}{d\ln b} &=& h_{\eta,3}^x,
\label{eq:flow0}
\end{eqnarray}
in which $i,j=1,2$, $i\neq j$,  $\lambda_u=0.14/t$, $\nu_u=-0.07/t$, $\lambda_s= -0.04/t$, $\nu_s=0.06/t$ (see SM \cite{suppl}).
In particular, the absence of $\Delta$ in $dh_{\eta,3}^x/d\ln b$ is due to the absence of $S^x_{3+3n}$ in the $\Delta$-term of $H_F$.
By solving Eq. (\ref{eq:flow0}), %(see SM \cite{suppl}), 
the coupling to the scaling fields at scale $\Lambda_1$ 
can be obtained by coupling all the three $h_{\eta, k}^x(b_1,n)$ to a same smeared spin operator $S_{\eta,n}^x$, 
\begin{eqnarray}
\sum_{n\in \text{u.c.}} b_1 \big[ A_\eta (h_{\eta,1}^x(n)+h_{\eta,2}^x(n))+ B_\eta h_{\eta,3}^x(n)\big] (-)^{\epsilon_\eta n} S_{\eta,n}^x,
\end{eqnarray}
in which $\epsilon_\eta=0 ,1$ for $\eta= u, s$, $A_\eta=(1-(\nu_\eta+\lambda_\eta)\Delta \ln b_1)$,  $B_\eta=(1-2\nu_\eta \Delta_\eta \ln b_1)$, 
and $h_{\eta,k}^x(n)$ are the bare fields at the scale $\Lambda_0$.
Hence, a difference in the coefficients develops below the scale $\Lambda_1$.
The ratio $D_1/D_2$ ($C_1/C_2$) is equal to $B_\eta/A_\eta=1+(\lambda_\eta-\nu_\eta)\Delta$ with $\eta=u$ ($\eta=s$),
which is linear in $\phi$ for $|\phi-\phi_{AF}|\ll 1$ of a negative (positive) slope,
 consistent with the numerical results shown in Fig. \ref{fig:CnD} (f).
Similar analysis can be applied to the $y$- and $z$-directions.

{\it The ordered phase}
We find numerical evidence of an $O_h \rightarrow D_4$ spontaneous symmetry breaking   for $\bar{\phi}_c\leq\phi\lessapprox 1.88\pi$,
except the $FM^\prime$ point at $\phi=7\pi/4$ where it is SU(2)$\rightarrow$ U(1).
Since $T_{3a}$ is unbroken, 
%in the FM phase, 
the spin orientations within a unit cell %and the quotient group $G/\mathopen{<} T_{3a}\mathclose{>}$ 
will be considered.
The spin polarizations in one of the symmetry breaking ground states are
\begin{eqnarray}
\langle \vec{S}_1 \rangle=S^\prime \hat{z}, \langle \vec{S}_2 \rangle=S^{\prime\prime} \hat{z},\langle \vec{S}_3\rangle=S^{\prime\prime} \hat{z},  
\label{eq:spin_orient2}
\end{eqnarray}
in which $\langle \vec{S}_i\rangle$ represents the expectation value of $\vec{S}_i$ in the corresponding state,
 $S^\prime$ and $S^{\prime\prime}$ are the magnitudes of the spin orders, 
and the little group of Eq. (\ref{eq:spin_orient2}) is $\mathopen{<} R_aT_a R(\hat{z},\pi)R_I I R(\hat{z},\pi), T(R_aT_a)^{-1}R_I I R_aT_a \mathclose{>}\cong D_4$ modulo $T_{3a}$ (see SM \cite{suppl}).  
%which is isomorphic to the symmetry group of a square (see SM \cite{suppl}).
%where the two generators of $D_8$ act in the spin space as $R(\hat{z},\pi/2)$ and the reflection to the plane $ACA^\prime C^\prime$ shown in Fig. \ref{fig:phase} (b), respectively.
Hence, the symmetry breaking of Eq. (\ref{eq:spin_orient2}) is $O_h\rightarrow D_4$, 
and the ``center of mass" spin directions in the six (since $|O_h/D_4|=6$) distinct symmetry breaking ground states are along $\pm \hat{x},\pm\hat{y},\pm \hat{z}$ shown as the six solid blue circles in Fig. \ref{fig:phase} (b).

Next we discuss numerics.
%The phase for $\bar{\phi}_c\leq\phi\lessapprox 1.88\pi$ is identified. 
ED calculations show evidence of a six-fold ground state degeneracy at zero field,
and one-fold, two-fold, three-fold degeneracies under small uniform fields $h_z$ along $\hat{z}$, $h_{n_I}$ along $\hat{n}_I$, and $h_{n_a}$ along $\hat{n}_a$, respectively. 
The fields $h$'s are chosen to satisfy $\Delta E\ll hL \ll E_g$, in which $L$ is the system size, $E_g$ is the excitation gap, and $\Delta E$ is the finite size splitting of the ground state sextet at zero field.  
In addition, DMRG numerics show evidence of vanishing cross correlation functions $\langle S^{\alpha}_1 S^{\beta}_{r+1}  \rangle$ ($\alpha \neq \beta$) at any field,
and nonzero diagonal correlations $\langle S^{\alpha}_1 S^{\alpha}_{r+1}  \rangle$ for $\alpha=z$ with $h_z$, $\alpha=x,z$ with $h_{n_I}$ and $\alpha=x,y,z$ with $h_{n_a}$.
%The numerical data is included in SM \cite{suppl}.
These results are all consistent with $O_h\rightarrow D_4$.
However, ED  show evidence of a four-fold ground state degeneracy at zero field when $\phi\gtrapprox 1.88\pi$ represented by the thin dashed line in Fig. \ref{fig:phase} (a), 
which is incompatible with $O_h\rightarrow D_4$.
Whether this is a finite size artifact or represents a different ordered phase remains to be explored. 
%and the ground state energy as varying $\phi$ shows no signature of phase transition within $\phi\in[0,\phi_c]$.
%and the numerical data are presented in SM \cite{suppl}.
%Furthermore, we have performed a classical analysis and found an $O_h\rightarrow D_6$ symmetry breaking as discussed in SM (\cite{suppl}).
%In fact, numerics provides evidence for $O_h\rightarrow D_6$ of spins higher than $1/2$ (see SM \cite{suppl}).
%More numerical investigations close to $\phi=0$ and how quantum fluctuations invalidate the classical analysis in the spin-1/2 case will be left for a future study.

Finally, we discuss the relevance of our study to real materials and higher dimensions.
Since Kitaev and Gamma interactions are dominant in $\alpha$-RuCl$_3$, 
1D Kitaev-Gamma chain of Ruthenium stripes can be tailored using a- or b-axis oriented superlattices made of the Mott insulator RuCl$_3$.
Engineering of such 1D systems %from superlattices 
out of 2D layered materials has 
been successful in fabricating Iridium chain systems \cite{Gruenewald2017}.
Furthermore, our results provide a starting point for an extrapolation to higher dimensions.
The parameters relevant to real materials in 2D are FM for Kitaev and AFM for Gamma couplings \cite{Kim2016,Janssen2017,Sears2019,Gordon2019}.
Thus the evolution of  the 1D emergent SU(2)$_1$ phase by increasing couplings between the chains is worth future studies which may offer further insights into possible spin liquid phases in $\alpha$-RuCl$_3$ systems. 

In summary, we have studied the phase diagram of the spin-1/2 Kitaev-Gamma chain.
In the gapless phase, the low energy physics is described by an emergent SU(2)$_1$ WZW model, with SU(2) breaking coefficients in the modified nonabelian bosonization formula.
%but the prefactors in the correlation functions exhibit SU(2) breaking.
%A modified nonabelian bosonization formula is proposed to capture the SU(2) breaking effect.
In the ordered phase, DMRG and ED numerics provide evidence for an $O_h\rightarrow D_4$ symmetry breaking
except in a small region close to the Kitaev point.

%%%%%%%%%%%%%%%%%%%%%%
%{\it Acknowledgments}
\begin{acknowledgments}
We would like to thank  A. Catuneanu for
early contributions to this project.
W. Y. and I. A. acknowledge support
from NSERC Discovery Grant 04033-2016.
H.-Y. K. acknowledges support from NSERC
Discovery Grant 06089-2016, the Centre for Quantum
Materials at the University of Toronto and
the Canadian Institute for Advanced Research.
A. N. acknowledges computational resources and services provided by 
 Compute Canada and
Advanced Research Computing at the University of British Columbia.
A. N. is supported by the Canada First
Research Excellence Fund.
\end{acknowledgments}

%%%%%%%%%%%%%%%%%%%%%%
\let\oldaddcontentsline\addcontentsline% Store \addcontentsline
\renewcommand{\addcontentsline}[3]{}% Make \addcontentsline a no-op

%%%%%%%%%%%%%%%%%%%%%%

\let\addcontentsline\oldaddcontentsline% Restore \addcontentsline

%%%%%%%%%%%%%%%%%%%%%%%%%%%%%%%%%%%%%%%%%%%%%%%%%%%%%%%%%%%%%%
%%%%%%%%%%%%%%%%%%%%%%%%%%%%%%%%%%%%%%%%%%%%%%%%%%%%%%%%%%%%%%

\begin{widetext}
\clearpage

\centerline{ {\Large \bf Supplementary Materials} }

%\appendix
%\title{Supplementary Materials: 
%Phase diagram of the spin-1/2 Kitaev-Gamma chain
%and emergent SU(2) symmetry
%}

%\maketitle
\tableofcontents

%%%%%%%%%%%%%%%%%%%
\section{The sublattice rotations}

\subsection{The six-sublattice rotation}
\label{app:six-sublattice}

Denoting $H_{ij}$ to be the term in the Hamiltonian corresponding to the bond $<ij>$, 
the unrotated Hamiltonian of the Kitaev-Gamma chain is
\bea
H_{12} &=&  K S^x_1 S^x_2 + \Gamma (S^y_1 S^z_2+S^z_1 S^y_2),\nn \\
H_{23} &=& K S^y_2 S^y_3 + \Gamma (S^z_2 S^x_3+S^x_2 S^z_3),\nn \\
H_{34} &=& H_{12} ,(3\rightarrow 1, 4\rightarrow 2),\,\text{\it etc}.
\label{eq:H}
\eea
The six-sublattice rotation is defined as
\bea
\text{Sublattice $1$}: & (x,y,z) & \rightarrow (x^{\prime},y^{\prime},z^{\prime}),\nn\\ 
\text{Sublattice $2$}: & (x,y,z) & \rightarrow (-x^{\prime},-z^{\prime},-y^{\prime}),\nn\\
\text{Sublattice $3$}: & (x,y,z) & \rightarrow (y^{\prime},z^{\prime},x^{\prime}),\nn\\
\text{Sublattice $4$}: & (x,y,z) & \rightarrow (-y^{\prime},-x^{\prime},-z^{\prime}),\nn\\
\text{Sublattice $5$}: & (x,y,z) & \rightarrow (z^{\prime},x^{\prime},y^{\prime}),\nn\\
\text{Sublattice $6$}: & (x,y,z) & \rightarrow (-z^{\prime},-y^{\prime},-x^{\prime}),
\label{eq:6rotation}
\eea
in which we have dropped the spin symbol $S$ for simplicity.
The Hamiltonian $H^{\prime}$ in the rotated basis becomes
\bea
H^{\prime}_{12} &=&  -K S^{\prime x}_1 S^{\prime x}_2-\Gamma (S^{\prime y}_1 S^{\prime y}_2+S^{\prime z}_1 S^{\prime z}_2), \nn \\
H^{\prime}_{23} &=& -K S^{z \prime}_2 S^{\prime z}_3-\Gamma (S^{\prime x}_2 S^{\prime x}_3+S^{\prime y}_2 S^{\prime y}_3), \nn \\
H^{\prime}_{34} &=& -K S^{\prime y}_3 S^{\prime y}_4-\Gamma (S^{\prime z}_3 S^{\prime z}_4+S^{\prime x}_3 S^{\prime x}_4) \nn \\
H^{\prime}_{45} &=& H^{\prime K}_{12} ,(4\rightarrow 1, 5\rightarrow 2), \,\text{\it etc}.
\label{eq:Hprime}
\eea

%-------------------------------------
\begin{figure}[h]
\includegraphics[width=12cm]{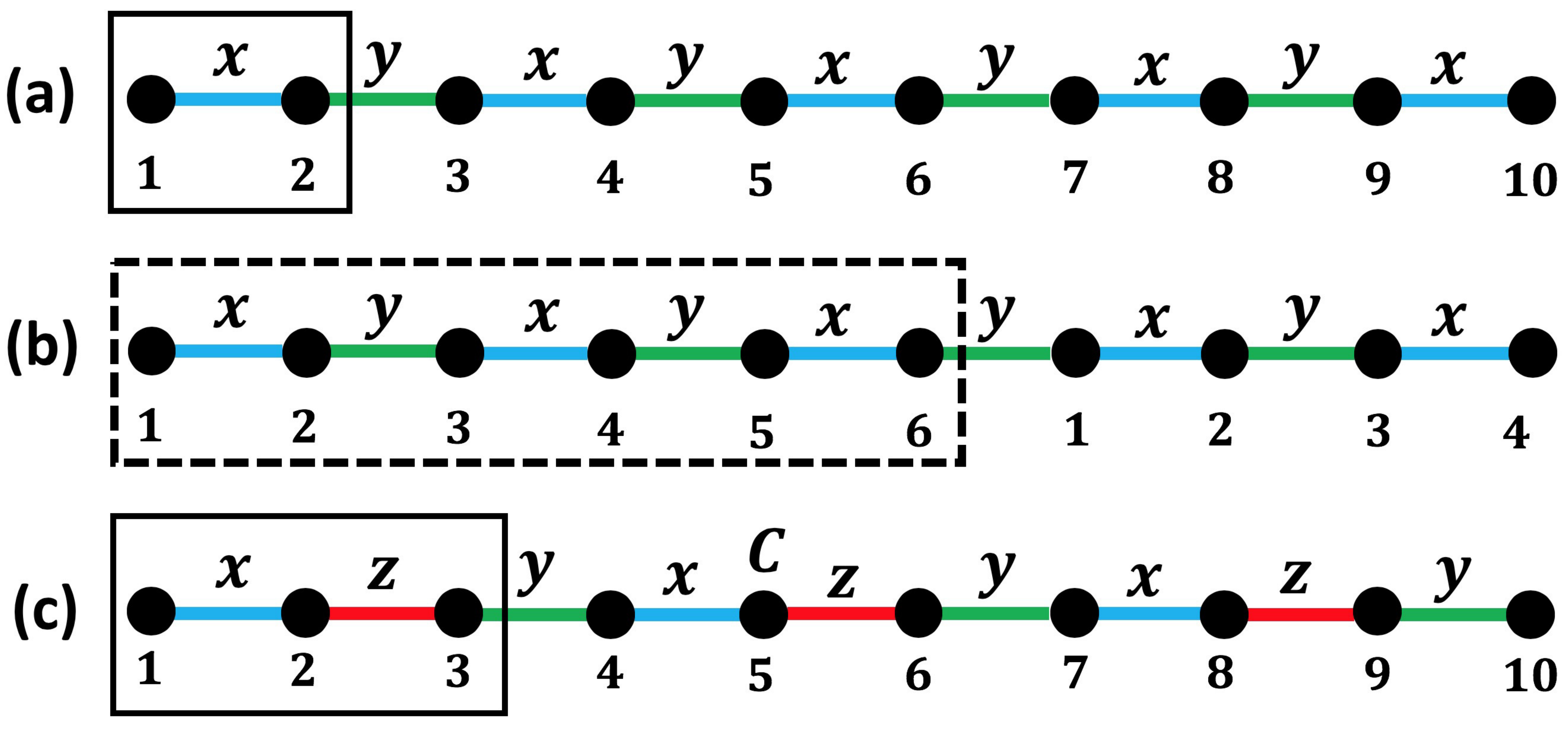}
\caption{(a,b) Bonds before and (c) after the six-sublattice rotation.
In (a,c), the rectangular boxes represent unit cells of the lattice before and after six-sublattice rotation, respectivley.
In (b), the box represents a  ``unit" cell of the six-sublattice rotation, 
and the number at each site denotes the sublattice index in Eq. (\ref{eq:6rotation}).
} 
\label{fig:bonds_suppl}
\end{figure}
%-------------------------------------

%%%%%%%%%%%%%%%%%%%
\subsection{The three-sublattice rotation and $(K,\Gamma)\cong (K,-\Gamma)$}
\label{app:three-sublattice}

In the unrotated frame, the three-sublattice rotation is defined as
\bea
\text{Sublattice $1$}: & (x,y,z) & \rightarrow (x^{\prime\prime},y^{\prime\prime},z^{\prime\prime}),\nn\\ 
\text{Sublattice $2$}: & (x,y,z) & \rightarrow (x^{\prime\prime},-y^{\prime\prime},-z^{\prime\prime}),\nn\\ 
\text{Sublattice $3$}: & (x,y,z) & \rightarrow (x^{\prime\prime},-y^{\prime\prime},-z^{\prime\prime}),\nn\\ 
\eea
and the Hamiltonian $H$ is transformed into
\bea
H^{\prime\prime}_{12} &=& K S_1^{\prime\prime x} S_2^{\prime\prime x}-\Gamma (S_1^{\prime\prime y} S_2^{\prime\prime z}+S_1^{\prime\prime z} S_2^{\prime\prime y}),\nn\\
H^{\prime\prime}_{23} &=& K S_2^{\prime\prime y} S_3^{\prime\prime y}-\Gamma (S_2^{\prime\prime z} S_3^{\prime\prime x}+S_2^{\prime\prime x} S_3^{\prime\prime z}),\nn\\
H^{\prime\prime}_{34} &=& H^{\prime\prime}_{12} ,(3\rightarrow 1, 4\rightarrow 2), \,\text{\it etc}.
\eea
Alternatively, by performing the following three-sublattice rotation
\bea
\text{Sublattice $1$}: & (x,y,z) & \rightarrow (x^{\prime\prime\prime},y^{\prime\prime\prime},z^{\prime\prime\prime}),\nn\\ 
\text{Sublattice $2$}: & (x,y,z) & \rightarrow (x^{\prime\prime\prime},-y^{\prime\prime\prime},-z^{\prime\prime\prime}),\nn\\ 
\text{Sublattice $3$}: & (x,y,z) & \rightarrow (-x^{\prime\prime\prime},y^{\prime\prime\prime},-z^{\prime\prime\prime}),\nn\\ 
\eea
to the rotated Hamiltonian $H^{\prime}$, the transformed Hamiltonian $H^{\prime\prime\prime}$ is
\bea
H^{\prime\prime\prime}_{12} &=&  -K S^{\prime\prime\prime x}_1 S^{\prime\prime\prime x}_2+\Gamma (S^{\prime\prime\prime y}_1 S^{\prime\prime\prime y}_2+S^{\prime\prime\prime z}_1 S^{\prime\prime\prime z}_2), \nn \\
H^{\prime\prime\prime}_{23} &=& -K S^{z \prime\prime\prime}_2 S^{\prime\prime\prime z}_3+\Gamma (S^{\prime\prime\prime x}_2 S^{\prime\prime\prime x}_3+S^{\prime\prime\prime y}_2 S^{\prime\prime\prime y}_3), \nn \\
H^{\prime\prime\prime}_{34} &=& -K S^{\prime\prime\prime y}_3 S^{\prime\prime\prime y}_4+\Gamma (S^{\prime\prime\prime z}_3 S^{\prime\prime\prime z}_4+S^{\prime\prime\prime x}_3 S^{\prime\prime\prime x}_4) \nn \\
H^{\prime\prime\prime}_{45} &=& H^{\prime\prime\prime K}_{12} ,(4\rightarrow 1, 5\rightarrow 2), \,\text{\it etc}.
\eea
Indeed, in either case we verify that $(K,\Gamma)\cong (K,-\Gamma)$.

%%%%%%%%%%%%%%%%%%%
\subsection{Noncollinear FM-like quasi-long-range order in the unrotated frame}

All physical properties of $H$ in Eq. (\ref{eq:H}) can be obtained from those of $H^\prime$ in Eq. (\ref{eq:Hprime})
by performing the inverse of the six-sublattice rotation in Eq. (\ref{eq:6rotation}).
In this section, we discuss the spin-spin correlation functions in the unrotated frame in the AFM phase
based on the results within the rotated frame presented in the maintext. 
Throughout this section, $S_i^\alpha$ and $S^{\prime\alpha}_i$ represent
a spin operator in the unrotated and rotated frames, respectively.

In the rotated frame, all the cross correlation functions vanish, {i.e.,} 
\bea
\bra{\Omega^\prime} S^{\prime \alpha}_i S^{\prime \beta}_j \ket{\Omega^\prime}=0, \alpha\neq\beta,
\label{eq:cross_vanish}
\eea
in which $\ket{\Omega^\prime}$ is the ground state of $H^\prime$.
This is a consequence of the invariance of $H^\prime$ under $R(\hat{\alpha^\prime},\pi)$ ($\alpha=x,y,z$) 
if the ground state does not break these symmetries (which is true in the AFM phase).
For example, suppose $R(\hat{x}^\prime,\pi)\ket{\Omega^\prime}=\ket{\Omega^\prime}$.
Then 
\bea
\bra{\Omega^\prime}S^{\prime x}_i S^{\prime y}_j \ket{\Omega^\prime} &=&
\bra{\Omega^\prime}R^{-1}(\hat{x}^\prime,\pi) S^{\prime x}_i S^{\prime y}_j R(\hat{x}^\prime,\pi)\ket{\Omega^\prime}\nn\\
&=&\bra{\Omega^\prime}R^{-1}(\hat{x}^\prime,\pi) S^{\prime x}_iR(\hat{x}^\prime,\pi)\cdot R^{-1}(\hat{x}^\prime,\pi) S^{\prime y}_j R(\hat{x}^\prime,\pi)\ket{\Omega^\prime}\nn\\
&=&-\bra{\Omega^\prime} S^{\prime x}_i S^{\prime y}_j \ket{\Omega^\prime},
\label{eq:xy_vanish}
\eea
in which $R^{-1}(\hat{x}^\prime,\pi) S^{\prime x}_iR(\hat{x}^\prime,\pi)=S_i^{\prime x}$ and $R^{-1}(\hat{x}^\prime,\pi) S^{\prime y}_iR(\hat{x}^\prime,\pi)=-S^{\prime y}_i$ are used.
From Eq. (\ref{eq:xy_vanish}), it is clear that $\bra{\Omega^\prime} S^{\prime x}_i S^{\prime y}_j \ket{\Omega^\prime}=0$.

%--------------------------------------------------------------------------------------------------------------------------------------
\begin{table}
\begin{center}
\begin{tabular}{ |c | c | c | c | c | }
\hline
$D_4$ & $E$ & $C_2(z)$ & $C_2(y)$ & $C_2(x)$ \\
\hline
$A$ & $1$ & $1$ & $1$ & $1$ \\
\hline
$B_1$ & $1$ & $1$ & $-1$ & $-1$ \\
\hline
$B_2$ & $1$ & $-1$ & $1$ & $-1$ \\
\hline
$B_3$ & $1$ & $-1$ & $-1$ & $1$ \\
\hline
\end{tabular}
\end{center}
\caption{Character table of $D_2$ \cite{Dresselhaus2008}.
\label{table:D4}
}
\end{table}
%--------------------------------------------------------------------------------------------------------------------------------------

There is a group theoretical way to understand Eq. (\ref{eq:cross_vanish}).
Consider the group $F$ generated by $R(\hat{\alpha}^\prime,\pi)$ ($\alpha=x,y,z$).
$F$ is isomorphic to $D_2$ which is the dihedral group of order $4$.
To see this, note that the generator-relation representation of $D_{2}$ is 
\bea
D_2=\mathopen{<}a,b|a^2=b^2=(ab)^2=e\mathclose{>},
\label{eq:D4}
\eea
in which $a,b$ are the two generators of $D_2$, and $e$ is the identity element.
In the group $F$, we identify $a=R(\hat{x}^\prime,\pi)$ and $b=R(\hat{y}^\prime,\pi)$,
and as a result, the product is $ab=R(\hat{z}^\prime,\pi)$.
Hence, all the relations in Eq. (\ref{eq:D4}) are satisfied,
which shows that $F$ is a subgroup of $D_2$.
On the other hand, there are at least four elements in $F$, {i.e.}, $1$ and $R(\hat{\alpha}^\prime,\pi)$ ($\alpha=x,y,z$).
Thus we conclude $F\cong D_2$.
We also note that $D_2$ has a direct product decomposition $D_2\cong \mathbb{Z}_2\times \mathbb{Z}_2$,
and the corresponding decomposition of $F$ is $F=\{1,R(\hat{x}^\prime,\pi)\}\times \{1,R(\hat{y}^\prime,\pi)\}$.
With this preparation, we are able to interpret Eq. (\ref{eq:D4}) in the language of group theory.
The group $D_2$ is abelian, hence only has one-dimensional irreducible representations. 
The character table of $D_2$ \cite{Dresselhaus2008} is shown in Table \ref{table:D4}, containing four irreducible representations $A,B_1,B_2,B_3$.
Notice that $S_i^{\prime z},S_i^{\prime y},S_i^{\prime x}$ are in the $B_1,B_2,B_3$ representations, respectively.
Since the inner product between different irreducible representations vanishes 
\footnote{Let $V_1$ and $V_2$ be two $\mathbb{R}$-linear irreducible representations of the group $F$. 
Let $\phi:V_1\times V_2 \rightarrow \mathbb{R}$ be a bilinear form  satisfying $\phi(f v_1,v_2)=\phi( v_1,f^{-1}v_2)$ for any $v_1\in V_1,v_2\in V_2,f\in F$. 
Then $\phi$ induces a map $\phi^*: V_1\rightarrow V_2^*$, in which $V_2^*$ is the dual space of $V_2$, {\it i.e.,} 
$V_2^*$ consists of all $\mathbb{R}$-linear functions on $V_2$.
Note that $V_2^*$ is naturally a representation of $F$.
The property of $\phi$ ensures that $\phi^*$ commutes with $F$. 
Then by Schur's lemma, $\phi^*$ is either $0$ or an isomorphism.
Thus if $V_1$ and $V_2^*$ are two different irreducible representations, $\phi^*$ has to be zero, hence $\phi=0$.
Since $\mathbb{C}$-linear spaces are also $\mathbb{R}$-linear 
and the inner product on a complex Hilbert space is bilinear in $\mathbb{R}$
by restricting the coefficients in $\mathbb{C}$ to $\mathbb{R}$,
the above conclusion holds also for complex irreducible representations.
},
we conclude Eq. (\ref{eq:cross_vanish}).

In the unrotated frame, Eq. (\ref{eq:cross_vanish}) imposes constraints on the correlation functions by virtue of the six-sublattice rotation in Eq. (\ref{eq:6rotation}).
However, unlike Eq. (\ref{eq:cross_vanish}),
in the present case some of the cross correlations will be nonvanishing,
while some of the diagonal correlations ({\it i.e.,} $\langle S_i^\alpha S_i^\alpha\rangle$) are zero.
Consider $\bra{\Omega}S_1^xS^\alpha_{j+6n}\ket{\Omega}$ as an example, in which $\ket{\Omega}$ is the ground state of $H$ in the unrotated frame,
$1\leq j\leq 6$, and $n\in \mathbb{Z}$.
Denote $U_6$ as the six-sublattice rotation, so that $U_6HU^{-1}_6=H^\prime$, and $\ket{\Omega^\prime}=U_6 \ket{\Omega}$.
Then 
\bea
\bra{\Omega}S_1^xS^\alpha_{j+3n}\ket{\Omega}&=&\bra{\Omega^\prime}U_6 S_1^xS^\alpha_{j+3n}U_6^{-1}\ket{\Omega^\prime}\nn\\
&=&\bra{\Omega^\prime}U_6 S_1^xU_6^{-1}\cdot U_6S^\alpha_{j+3n}U_6^{-1}\ket{\Omega^\prime}.
\eea
Using Eqs. (\ref{eq:6rotation},\ref{eq:cross_vanish}), we see that only the following correlations are nonzero,
\bea
\bra{\Omega}S_1^xS^x_{1+3n}\ket{\Omega}&=&\bra{\Omega^\prime}S^{\prime x}_1 S^{\prime x}_{1+3n}\ket{\Omega^\prime},\nn\\
\bra{\Omega}S_1^xS^x_{2+3n}\ket{\Omega}&=&-\bra{\Omega^\prime}S^{\prime x}_1 S^{\prime x}_{2+3n}\ket{\Omega^\prime},\nn\\
\bra{\Omega}S_1^xS^z_{3+3n}\ket{\Omega}&=&\bra{\Omega^\prime}S^{\prime x}_1 S^{\prime x}_{3+3n}\ket{\Omega^\prime},\nn\\
\bra{\Omega}S_1^xS^y_{4+3n}\ket{\Omega}&=&-\bra{\Omega^\prime}S^{\prime x}_1 S^{\prime x}_{4+3n}\ket{\Omega^\prime},\nn\\
\bra{\Omega}S_1^xS^y_{5+3n}\ket{\Omega}&=&\bra{\Omega^\prime}S^{\prime x}_1 S^{\prime x}_{5+3n}\ket{\Omega^\prime},\nn\\
\bra{\Omega}S_1^xS^z_{6+3n}\ket{\Omega}&=&-\bra{\Omega^\prime}S^{\prime x}_1 S^{\prime x}_{6+3n}\ket{\Omega^\prime},
\label{eq:xx_unrotate}
\eea
and all other correlations vanish.
In the long distance limit $n\gg 1$, all the six correlations in Eq. (\ref{eq:xx_unrotate}) are positive due to the 
staggered nature of the correlation functions in the rotated frame.
Due to the $\ln^{1/2}(r)/r$ power law decay behavior,
Eq. (\ref{eq:xx_unrotate}) exhibits a noncollinear FM-like quasi-long-range order.
Similar analysis can be performed on $\bra{\Omega}S_1^yS^\alpha_{j+6n}\ket{\Omega}$ and $\bra{\Omega}S_1^zS^\alpha_{j+6n}\ket{\Omega}$.

Before closing this section, we further note that all the above analysis in the unrotated frame can alternatively 
be performed using the $D_4$ group in the unrotated frame,
which is given by $\{1,U_6^{-1}R(\hat{x}^\prime,\pi)U_6,U_6^{-1}R(\hat{y}^\prime,\pi)U_6,U_6^{-1}R(\hat{z}^\prime,\pi)U_6\}$.
It is straightforward to work out the actions of $U_6^{-1}R(\hat{\alpha}^\prime,\pi)U_6$:
\bea
 \begin{array}{cccccc}
U_6^{-1}R(\hat{x}^\prime,\pi)U_6:&\text{Sublattice 1.} & R(\hat{x},\pi): & (x,y,z) & \rightarrow & (x,-y,-z) \\
&\text{Sublattice 2.} & R(\hat{x},\pi): & (x,y,z) & \rightarrow & (x,-y,-z) \\
&\text{Sublattice 3.} & R(\hat{z},\pi): & (x,y,z) & \rightarrow & (-x,-y,z) \\
&\text{Sublattice 4.} & R(\hat{y},\pi): & (x,y,z) & \rightarrow & (-x,y,-z) \\
&\text{Sublattice 5.} & R(\hat{y},\pi): & (x,y,z) & \rightarrow & (-x,y,-z) \\
&\text{Sublattice 3.} & R(\hat{z},\pi): & (x,y,z) & \rightarrow & (-x,-y,z), 
\end{array}
\label{eq:Rx_unrot}
\eea
\bea
 \begin{array}{cccccc}
U_6^{-1}R(\hat{y}^\prime,\pi)U_6:&\text{Sublattice 1.} & R(\hat{y},\pi): & (x,y,z) & \rightarrow & (-x,y,-z) \\
&\text{Sublattice 2.} & R(\hat{z},\pi): & (x,y,z) & \rightarrow & (-x,-y,z) \\
&\text{Sublattice 3.} & R(\hat{x},\pi): & (x,y,z) & \rightarrow & (x,-y,-z) \\
&\text{Sublattice 4.} & R(\hat{x},\pi): & (x,y,z) & \rightarrow & (x,-y,-z) \\
&\text{Sublattice 5.} & R(\hat{z},\pi): & (x,y,z) & \rightarrow & (-x,-y,z) \\
&\text{Sublattice 3.} & R(\hat{y},\pi): & (x,y,z) & \rightarrow & (-x,y,-z), 
\end{array}
\label{eq:Ry_unrot}
\eea
\bea
 \begin{array}{cccccc}
U_6^{-1}R(\hat{\alpha}^\prime,\pi)U_6:&\text{Sublattice 1.} & R(\hat{z},\pi): & (x,y,z) & \rightarrow & (-x,-y,z) \\
&\text{Sublattice 2.} & R(\hat{y},\pi): & (x,y,z) & \rightarrow & (-x,y,-z) \\
&\text{Sublattice 3.} & R(\hat{y},\pi): & (x,y,z) & \rightarrow & (-x,y,-z) \\
&\text{Sublattice 4.} & R(\hat{z},\pi): & (x,y,z) & \rightarrow & (-x,-y,z) \\
&\text{Sublattice 5.} & R(\hat{x},\pi): & (x,y,z) & \rightarrow & (x,-y,-z) \\
&\text{Sublattice 3.} & R(\hat{x},\pi): & (x,y,z) & \rightarrow & (x,-y,-z),
\end{array}
\label{eq:Rz_unrot}
\eea
in which ``sublattice i" means all the lattice sites $i+6n$ where $n\in \mathbb{Z}$,
and $S_{i+3n}^\alpha$ is denoted as $\alpha$ under ``sublattice i" for short.
Notice that the transformations acquire rather complicated forms in the unrotated frame.
However, the group structure is still $D_4$, 
and equations like Eq. (\ref{eq:xx_unrotate}) are direct consequence of 
the symmetry operations in Eqs. (\ref{eq:Rx_unrot},\ref{eq:Ry_unrot},\ref{eq:Rz_unrot}).

%%%%%%%%%%%%%%%%%%%
%\section{Phase transitions}

%\subsection{Ground state energy as varying $\phi$}

%-------------------------------------
%\begin{figure}[h]
%\includegraphics[width=14cm]{E0.eps}
%\caption{Ground state energy per site (red circles) and first order derivative $\partial{E_0}/\partial \phi$ (blue circles)
%as varying $\phi$.
%ED calculations are performed for a system of $N=24$ with a periodic boundary condition.
%} 
%\label{fig:E0}
%\end{figure}
%-------------------------------------

%Fig. \ref{fig:E0} shows the ground state energy per site $E_0$ and the first order derivative $\partial E_0/\partial \phi$ as a function of $\phi$ calculated using ED on a system of $N=24$ sites with a periodic boundary condition.
%As can be seen from Fig. \ref{fig:E0}, $\partial E_0/\partial \phi$ is discontinuous at $\phi\sim 0.35\pi,\pi,1.65\pi$.
%$\phi=\pi$ is a Kitaev point with negative coupling,
%which separates the phase of emergent explicit SU(2)$_1$ ($\phi\in [\pi,2\pi-\phi_c]$)
%and the phase of emergent implicit SU(2)$_1$  ($\phi\in [\phi_c,\pi]$).
%Here by ``implicit", we mean the low energy theory is unitarily equivalent to SU(2)$_1$ under the three-sublattice rotation.
%$\phi=\phi_c\sim 0.35 \pi$ is the phase transition point separating the emergent implicit SU(2)$_1$ gapless phase
%and the FM phase, and $\phi=\bar{\phi}_c\sim 1.65 \pi$ is the image of $\phi_c$ under the reflection around $\phi=\pi$.
%More precise numerical value of $\phi_c$ is determined in the next section.

%%%%%%%%%%%%%%%%%%%
\section{Phase transitions}

%%%%%%%%%%%%%%%%%%%
\subsection{Ground state energy as a function of $\phi$}

The ground-state energy $E_0$ is a thermodynamic quantity that serves as an indicator of zero temperature phase transitions. 
The order of the transition is given by the first discontinuous derivative of $E_0$. 
In Fig. \ref{fig:gs_ener} (a), the ground state energy density $E_0/L$ is shown as a function of $\phi$ in the full parameter range calculated on a system of $L=36$ sites.
Fig. \ref{fig:gs_ener} (b) plots the first order derivative $\partial E_0/\partial \phi$ for three selective system sizes $L=12,24,36$.
There are clearly sudden jumps at $\phi=\phi_c\sim0.33\pi$ and $\phi=\pi$,
indicating first order phase transitions. 
In addition, the size of the jump scales linearly with $L$,
therefore, the discontinuity of the energy density $E_0/L$ has a definite value in the thermodynamic limit.
Fig. \ref{fig:gs_ener} (c) and (d) plot the second order derivative $\partial^2 E_0/\partial \phi^2$ in the vicinities of $\phi=\pi$ and $\phi=\phi_c$, respectively.
The divergences at these two angles provide further confirmations for the occurence of first order phase transitions. 

%As is clear from Fig. \ref{fig:gs_ener}, there are signatures of discontinuous  first order derivative $\partial\epsilon_0/\partial\phi$ (and hence diverging $\partial^2\epsilon_0/\partial\phi^2$) at $\phi=\phi_c,\pi,\bar{\phi}_c$.
%On the other hand, there is no evidence of discontinuity up to $\partial^2\epsilon_0/\partial\phi^2$ at $\phi=0$.
%Thus, we conclude that there are numerical evidence for $\phi_c$, $-K$, and $\bar{\phi}_c$ to be first order transition points, while the transition at $K$ is possibly continuous.

%--------------------------------------------------------------------
\begin{figure}[h!]
	\centering
	\includegraphics[width=16cm]{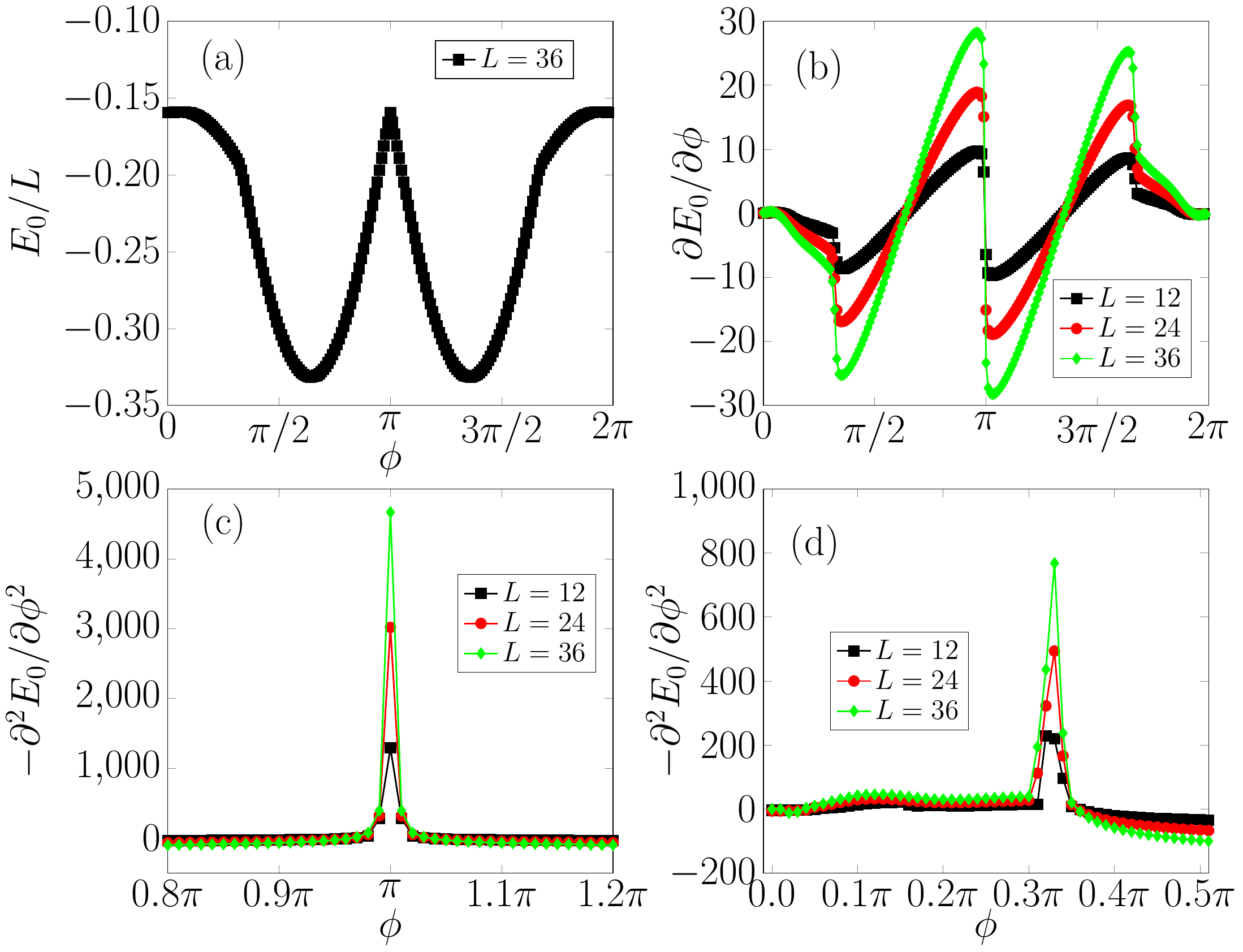}
	\caption{
	(a) $E_0/L$ and (b) $\partial E_0/\partial \phi$
	 as functions of  $\phi$ in the full parameter range; $\partial^2 E_0/\partial \phi^2$ in the vicinities of (c) $\phi=\pi$  and (d)  $\phi=\phi_c$.
	In (a), the system size is taken as $L=36$.
	In (b,c,d), the plots include data for three different system sizes $L=12,24,36$.  
	}
	\label{fig:gs_ener}
\end{figure}
%--------------------------------------------------------------------

%%%%%%%%%%%%%%%%%%%
\subsection{Numerical determination of $\phi_c$}
\label{app:phic}

%--------------------------------------------------------
\begin{figure*}[h]
\includegraphics[width=\textwidth]{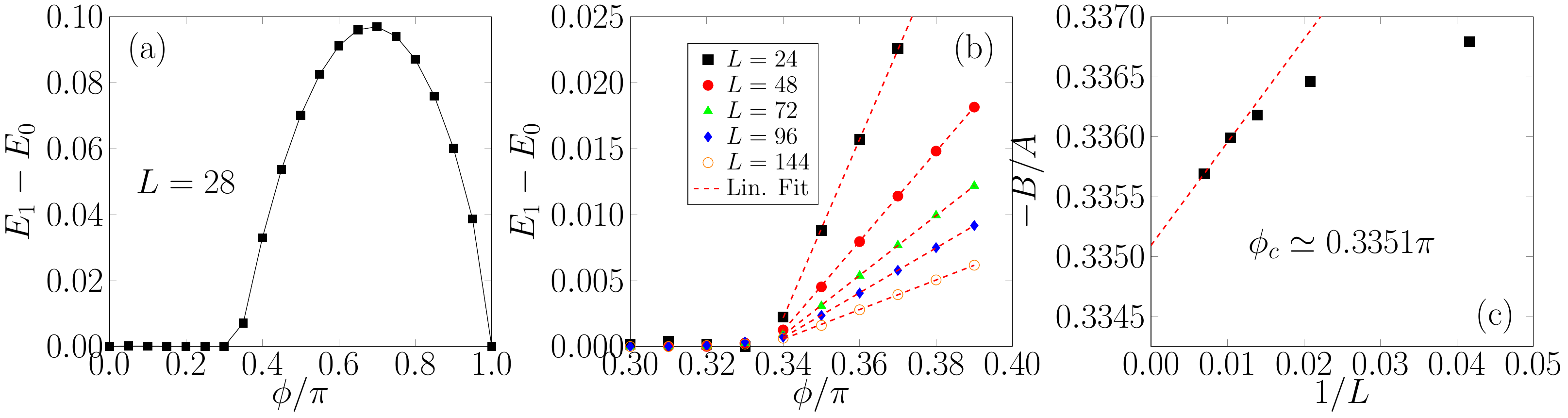}
\caption{
(a) $\Delta=E_1-E_0$ as a function of $\phi$ where $E_0$ is the energy of the ground state and $E_1$ the energy of the first excited state,
(b) $\Delta$ vs. $\phi$ in the range $\phi\in [0.3\pi,0.4\pi]$ close to the phase transition point for several different system sizes $L$, 
and (c) the extrapolated value of $\phi_c$ as a function of $1/L$.
In (a), $\Delta$ is calculated using DMRG on a system of $L=28$ sites with open boundary conditions ($m=600$ DMRG states were kept and up to 10 finite size sweeps were performed to reach convergence). In (b), $\Delta$ is computed on a chain with periodic boundary conditions. In this case, to reach numerical convergence, up to $m=1000$ DMRG states were kept and tens of finite size sweeps were performed with a final truncation error of $10^{-6}$.
In (c), $\phi_c=-B/A$ is determined by extrapolating the fitted red linear dashed line in (b) to $\Delta=0$ at the corresponding $L$.
The eventual value of $\phi_c\approx 0.3351\pi$ in (c) is obtained by a linear extrapolation of the finite size $\phi_c$ to $L\rightarrow \infty$.
} \label{fig:4suppl}
\end{figure*}
%--------------------------------------------------------

We determine $\phi_c$ as the phase transition point separating the gapless and ordered phases with the following procedure.
Fig. \ref{fig:4suppl} (a) shows the gap $\Delta=E_1-E_0$ in the finite size spectrum  for a system size of $L=28$ sites calculated using DMRG  with open boundary conditions (OBC),
where $E_0$ is the energy of the ground state and $E_1$ the energy of the first excited state. When chains with OBC are considered, $m=600$ DMRG states and up to 10 finite size sweeps were performed, with a final truncation error smaller than $10^{-9}$.
The rounded dome structure appearing for $\phi> 0.33\pi$ corresponds to the finite size gap in the gapless AFM phase. 
To determine the value of $\phi_c$ more accurately, we zoomed in the interval $0.3<\phi<0.4$ (Fig. \ref{fig:4suppl} (b)). 
Here, several system sizes from $L=24$ up to $L=144$ are considered with periodic boundary conditions (PBC).
The convergence of DMRG results was checked using up to $m=1000$ DMRG states and performing several tens of finite size sweeps, with a final truncation error smaller than $10^{-6}$.
We emphasize that the use of chains with periodic boundaries is not necessary for the 
determination of the value of $\phi_c$. In fact, Fig. \ref{fig:4suppl} (a) and Fig. \ref{fig:4suppl} (b) 
provide evidence that the numerical results do not depend on the choice of boundary conditions.

In the gapless phase the finite size gap shows a perfect linear behavior as $\Delta=A\phi+B$. 
Fig. \ref{fig:4suppl} (c) shows the extrapolated values of $\phi_c=-B/A$ as a function of the inverse system size $1/L$. 
Finally the value $\phi_c\simeq0.3351\pi$ is obtained by extrapolating to the thermodynamic limit as shown by the red dashed line in Fig. \ref{fig:4suppl} (c). 

%%%%%%%%%%%%%%%%%%%
\subsection{The central charge in the gapless region}

%---------------------------------------------------------
\begin{figure}[h!]
	\centering
	\includegraphics[width=10cm]{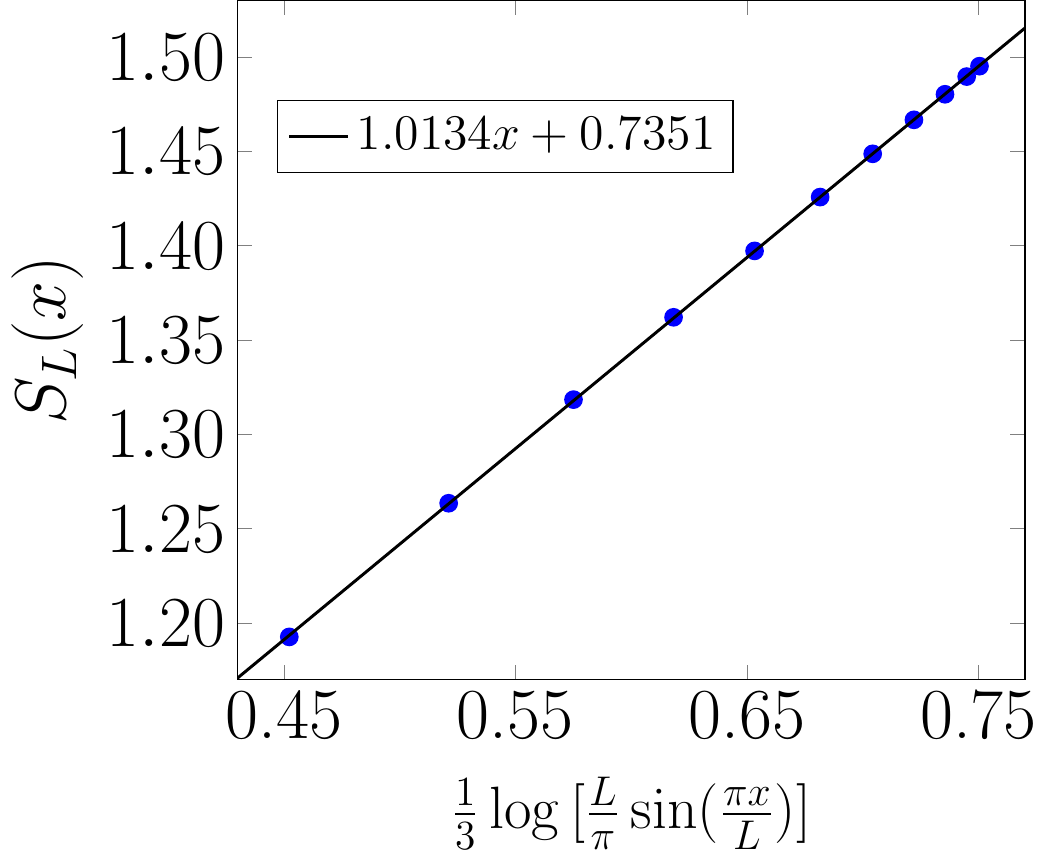}
	\caption{Entanglement entropy $S_L(x)$ of a subregion $x$, with rest of the system, in a periodic chain of length $L=30$. This data has been obtained for the choice $\phi = 0.85 \pi$. By fitting against the conformal distance on the horizontal axis, we obtain $c=1$, which is consistent with the analytical analysis in the maintext.}
	\label{fig:cnt_chg}
\end{figure}
%---------------------------------------------------------

To extract the central charge of the critical phase, we study periodic systems of length $L$ and compute the entanglement entropy $S_L(x)$ of a subregion $x$. The entanglement, for a CFT with central charge $c$, is expected to scale as \cite{Calabrese2009}
\begin{align}
    S_L(x) = \frac{c}{3} \ln \left[ \frac{L}{\pi} \sin\left( \frac{\pi x}{L} \right) \right] + \cdots.
\end{align}
A typical numerical fit for central charge, which we verified for multiple points in the gapless phase, is shown in Fig.\,\ref{fig:cnt_chg}. We find the $c=1$ to very good accuracy, thereby further corroborating the phase diagram.

%%%%%%%%%%%%%%%%%%%
\section{The symmetry operations}

\subsection{Proof of $G/<T_{3a}>\cong O_h$}
\label{app:group}

%-------------------------------------
\begin{figure}[h]
\includegraphics[width=7.5cm]{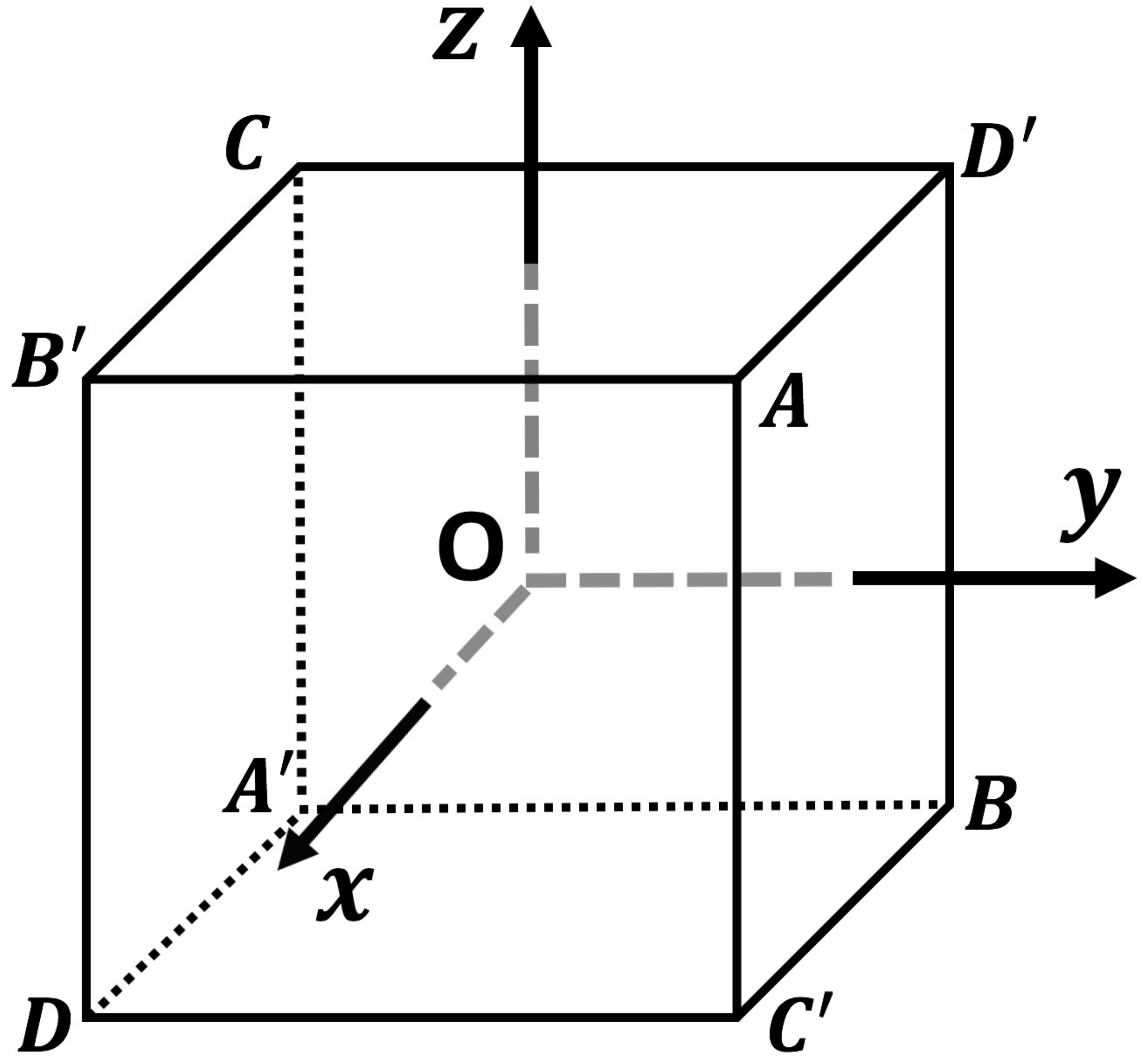}
\caption{A cube.
} 
\label{fig:cube}
\end{figure}
%-------------------------------------

%--------------------------------------------------------------------------------------------------------------------------------------
\begin{table}
\begin{center}
\begin{tabular}{| c | c | c | c | c | }
\hline
 $E$ & $1$ & $(x,y,z)$ & $\mathbbm{1}$ & e  \\
 \hline
 \multirow{3}{*}{$3C_2$}& $2$ & $(x,-y,-z)$ & $R(OX,\pi)$ & $r(ts)^2r$\\
 \cline{2-5}
& $3$ & $(-x,y,-z)$ & $R(OY,\pi)$ & $sr(st)^2rs$ \\
 \cline{2-5}
& $4$ & $(-x,-y,z)$ & $R(OZ,\pi)$ & $(st)^2$ \\
 \hline
 \multirow{6}{*}{$6C_4$}  & $5$ & $(x,z,-y)$ & $R(OX,\frac{\pi}{2})$ & $trsr$ \\
 \cline{2-5}
 & $6$ & $(x,-z,y)$ & $R(OX^{\prime},\frac{\pi}{2})$ & $rsrt$ \\
 \cline{2-5}
 & $7$ & $(-z,y,x)$ & $R(OY,\frac{\pi}{2})$ & $rsts$ \\
 \cline{2-5}
 & $8$ & $(z,y,-x)$ & $R(OY^{\prime},\frac{\pi}{2})$ & $stsr$ \\
 \cline{2-5}
 & $9$ & $(y,-x,z)$ & $R(OZ,\frac{\pi}{2})$  & $st$\\
 \cline{2-5}
 & $10$ & $(-y,x,z)$ & $R(OZ^{\prime},\frac{\pi}{2})$ & $ts$ \\
 \hline
 \multirow{6}{*}{$6C_{2}^{'}$} & $11$ & $(y,x,-z)$ & $R([AC],\pi)$ & $rsrtsr$ \\
 \cline{2-5}
 & $12$ & $(-y,-x,-z)$ & $R([BD],\pi)$ & $r(ts)^2rst$\\
 \cline{2-5}
 & $13$ &$(z,-y,x)$ & $R([AB],\pi)$  & $rt$\\
 \cline{2-5}
 & $14$ & $(-z,-y,-x)$ & $R([CD],\pi)$ & $(st)^2rsts$\\
 \cline{2-5}
 & $15$ & $(-x,z,y)$ & $R([AD],\pi)$ & $strs$\\
 \cline{2-5}
 & $16$ & $(-x,-z,-y)$ & $R([BC],\pi)$ & $tsrtst$\\
 \hline
 \multirow{8}{*}{$8C_3$}&  $17$ & $(y,z,x)$ & $R(OA,\frac{2\pi}{3})$ & $rs$ \\
 \cline{2-5}
 & $18$ & $(z,x,y)$ & $R(OA^{\prime},\frac{2\pi}{3})$ & $sr$\\
 \cline{2-5}
 & $19$ & $(-y,-z,x)$ & $R(OB,\frac{2\pi}{3})$ & $trst$ \\
 \cline{2-5}
 & $20$ & $(z,-x,-y)$ &  $R(OB^{\prime},\frac{2\pi}{3})$ & $tsrt$ \\
 \cline{2-5}
 & $21$ & $(y,-z,-x)$ &  $R(OC,\frac{2\pi}{3})$ & $stsrst$\\
 \cline{2-5}
 & $22$ & $(-z,x,-y)$ &  $R(OC^{\prime},\frac{2\pi}{3})$ & $tsrsts$\\
 \cline{2-5}
 & $23$ & $(-y,z,-x)$  &  $R(OD,\frac{2\pi}{3})$ & $(st)^2rs$ \\
 \cline{2-5}
 & $24$ & $(-z,-x,y)$ &  $R(OD^{\prime},\frac{2\pi}{3})$ & $sr(ts)^2$ \\
 \hline
\end{tabular}
\caption{List of $24$ the group elements of the point group $O$.
In accordance with the notations in Fig. \ref{fig:cube}, $OM$ represents the vector pointing from the center of the cube (i.e. the point $O$) to the vertex or the direction $M$, where $M$ is one of $A,\,A^{\prime},\,B,\,B^{\prime},\,C,\,C^{\prime},\,D,\,D^{\prime}$ when it is a vertex of the cube, and is one of $X,\,Y,\,Z,\,X^{\prime},\,Y^{\prime},\,Z^{\prime}$ when it represents a direction.
$X,\,Y,\,Z$ represent the positive directions of the three axes $x,\,y,\,z$, and $X^{\prime},\,Y^{\prime}\,Z^{\prime}$ represent the negative directions of the three axes.
The symbol $[MN]$ represents the line passing through the point that bisects the edge $MN^{\prime}$ and the point that bisects $M^{\prime}N$,
where $M,\,N,\,M^{\prime},\,N^{\prime}$ are all vertices of the cube.
The caption and the first four columns  of the table are taken from W. Yang, T. Xiang, and C. Wu, Phys. Rev. B {\bf 96}, 144514 (2017).
\label{table:O}
}
\end{center}
\end{table}
%--------------------------------------------------------------------------------------------------------------------------------------

The full octahedral group $O_h$ is the symmetry group of a cube as shown in Fig. \ref{fig:cube},
and is the largest among the five cubic point groups in three dimensional space.
$O_h$ contains $48$ group elements.
In $O_h$, there are $24$ rotations which can be classified into five conjugacy classes $\{E, 3C_2, 6C_4,6C_2^\prime,8C_3\}$ where $E$ represents the identity element. 
The actions of these $24$ rotations on the x-, y- and z-axes and their geometrical meanings as symmetry operations of a cube 
are summarized in Table \ref{table:O}.
The other $24$ elements of $O_h$ are improper transformations with determinant $-1$ which can be obtained by multiplying the $24$ rotations with the spatial inversion operation $i$.
Correspondingly, the improper elements can also be classified into five conjugacy classes, {\it i.e.}, $\{i,3\sigma_h,6S_4,6\sigma_d,8S_6\}$.

There is a generator-relation representation for the $O_h$ group \cite{Coxeter1965}:
\bea
O_h=\mathopen{<} r,s,t| r^2=s^2=t^2=(rs)^3=(st)^4=(rt)^2=e \mathclose{>},
\label{eq:Generator_Oh}
\eea
in which $e$ is the identity element, and the geometrical meanings of the generators $r,s,t$ as symmetry operations of a cube are three reflections.
We are going to construct $r,s,t$ out of $T,R_aT_a,R_I I, R(\hat{x},\pi),R(\hat{y},\pi),R(\hat{z},\pi)$.
Then we will show that on the one hand they indeed satisfy the above relations modulo $T_{3a}$,
and on the other hand, the group generated by the constructed $r,s,t$ contains at least $48$ elements.
Since $|O_h|=48$,
this proves that $G/\mathopen{<}T_{3a}\mathclose{>}$ is isomorphic to $O_h$.

Before proceeding on, we fix some notations.
Let $\mathcal{R}$ be a rotation in spin space defined as $(\mathcal{R}(S^x),\mathcal{R}(S^y),\mathcal{R}(S^z))=(S^x,S^y,S^z)R$,
in which  $R$ is a $3\times 3$ orthogonal matrix corresponding to $\mathcal{R}$.
Let $\mathcal{R}^\prime$ be another rotation with $R^\prime$ the corresponding matrix.
Then the composition $\mathcal{R}\mathcal{R}^\prime$ is given by
\bea
\mathcal{R}\mathcal{R}^\prime: (S^x,S^y,S^z)\rightarrow (S^x,S^y,S^z) RR^\prime.
\label{eq:composition}
\eea
For later convenience, recall that $R_a=R(\hat{n}_a,-2\pi/3)$ and $R_I=R(\hat{n}_I,\pi)$ satisfy
\bea
R_a:&(S_i^x,S_i^y,S_i^z)&\rightarrow (S_i^z,S_i^x,S_i^y),\nn\\
R_I:&(S_i^x,S_i^y,S_i^z)&\rightarrow (-S_i^z,-S_i^y,-S_i^x),
\label{eq:RaRI}
\eea
in which $\hat{n}_a=\frac{1}{\sqrt{3}}(1,1,1)^T$ is parallel to the line of $OA$ in Fig. \ref{fig:cube}, 
and $\hat{n}_I=\frac{1}{\sqrt{2}}(1,0,-1)^T$ is parallel to the line passing through the point that bisects the edge $CD^{\prime}$ and the point that bisects $C^{\prime}D$ in Fig. \ref{fig:cube}.
Hereafter within this section,
the site index $i$ will be dropped in subsequent discussions for simplifications of notations.

The constructions of $r,s,t$ are as follows,
\bea
\begin{array}{ c  c  c  c  }
\text{Generator}& \text{Expression} & \text{Spin space} & \text{Geometrical} \\
r & T\cdot R_I I & (x,y,z)\rightarrow (z,y,x) & \text{Reflection to $ABA^{\prime}B^{\prime}$-plane} \\
s & T\cdot (R_aT_a)^{-1}\cdot R_I I\cdot R_aT_a & (x,y,z)\rightarrow (y,x,z) & \text{Reflection to $ACA^{\prime}C^{\prime}$-plane} \\
t & T \cdot R(\hat{y},\pi) & (x,y,z)\rightarrow (x,-y,z) & \text{Reflection to $xz$-plane} \\
\end{array},
\label{eq:generators}
\eea
in which the second, the third and the fourth columns give the expressions of $r,s,t$ in terms of the symmetry operations $T,R_aT_a,R_I I, R(\hat{x},\pi),R(\hat{y},\pi),R(\hat{z},\pi)$ of the model, 
the actions in the spin space where $S^\alpha$ is denoted as $\alpha$ for short,
and the geometrical meanings as symmetries of a cube in Fig. \ref{fig:cube}, respectively.
Now we verify the relations $r^2=s^2=t^2=e$.
Firstly,
\bea
r^2=T^2\cdot I^2\cdot (R_I)^2=1, 
\eea
since $T^2=1,I^2=1$ and $(R_I)^2=[R(\hat{n}_I,\pi)]^2=R(\hat{n}_I,2\pi)=1$.
Secondly,
\bea
s^2 & =& T^2\cdot (T_a^{-1} I T_a)^2 \cdot (R_a^{-1} R_I R_a)^2\nn\\
&=& T_{-a}IT_a \cdot  T_{-a} I T_a \cdot [R(R_a^{-1}\hat{n_I},\pi)]^2\nn\\
&=& T_{-a} I^2 T_a\cdot R(R_a^{-1}\hat{n_I},2\pi)\nn\\
&=&1,
\eea
in which $T_a^{-1}=T_{-a}$, and $R_0 R(\hat{n},\theta) R_0^{-1}=R(R_0\hat{n},\theta)$ is used.
Finally for $t$, we obtain
\bea
t^2=T^2 \cdot [R(\hat{y},\pi)]^2=1.
\eea

Using the expressions of $r,s,t$, it is straightforward to work out the expressions of $rs,st,rt$, as
\bea
\begin{array}{ c  c  c  c  }
\text{Operation}& \text{Expression} & \text{Spin space} & \text{Geometrical} \\
rs & (R_aT_a)^{-1} & (x,y,z)\rightarrow (y,z,x) & R(OA,\frac{2\pi}{3}) \\
st &  R_aT_a\cdot R(\hat{z},\pi)\cdot R_I I \cdot R(\hat{z},\pi) & (x,y,z)\rightarrow (y,-x,z) & R(\hat{z},\frac{\pi}{2})\\
rt & R(\hat{z},\pi) \cdot R_I I\cdot R(\hat{z},\pi) & (x,y,z)\rightarrow (z,-y,x) & R([AB],\pi)
\end{array},
\label{eq:relation_ops}
\eea
in which $[AB]$ represents the line passing through the point that bisects the edge $AB^{\prime}$ and the point that bisects $A^{\prime}B$ in Fig. \ref{fig:cube}.
Next we verify the relations $(rs)^3=(st)^4=(rt)^2=e$.
Firstly,
\bea
(rs)^3&=&(R_a)^{-3}\cdot (T_a)^{-3}\nn\\
&=& [R(\hat{n}_a,-2\pi/3) ]^{-3} \cdot T_{-3a}\nn\\
&=& R(\hat{n}_a,2\pi) \cdot T_{-3a}\nn\\
&=& T_{-3a},
\eea
in which $R_a=R(\hat{n}_a,-2\pi/3)$ is used,
and clearly $(rs)^3=e$ modulo $T_{3a}$.
Secondly,
\bea
(st)^4&=& (T_a I)^4 \cdot [R_a R(\hat{z},\pi)R_I R(\hat{z},\pi)]^4\nn\\
&=& T_a(I T_aI) T_a (I T_aI) \cdot [R(\hat{z},\pi/2)]^4\nn\\
&=& T_aT_{-a}T_aT_{-a}\cdot R(\hat{z},2\pi)\nn\\
&=&1,
\eea
in which $IT_aI=T_{-a}$ and $R_a R(\hat{z},\pi)R_I R(\hat{z},\pi)=R(\hat{z},\pi/2)$ are used.
Finally,
\bea
(rt)^2&=&I^2\cdot \big[R(\hat{z},\pi)R_IR(\hat{z},\pi)\big]^2\nn\\
&=&[R(R(\hat{z},\pi)\hat{n}_I,\pi)]^2\nn\\
&=&1.
\eea
This proves that all the relations in Eq. (\ref{eq:Generator_Oh}) are satisfied.
Hence $G/\mathopen{<}T_{3a}\mathclose{>}$ is isomorphic to a subgroup of $O_h$.

We note that the time reversal operation acquires a rather complicated form in terms of the generators.
In fact, we have
\bea
\begin{array}{ c  c  c  c  }
\text{Operation}& \text{Expression} & \text{Spin space} & \text{Geometrical} \\
sr(st)^2r(st) & T & (x,y,z)\rightarrow (-x,-y,-z) & \text{Inversion of the cube} \\
\end{array},
\label{eq:time_reversal}
\eea
To verify the expression of $T$, using Eq. (\ref{eq:generators},\ref{eq:relation_ops}), one obtains
\bea
T&=&(rs)^{-1}(st)^2 r (st)\nn\\
&=& (T_{-a})^{-1}  (T_aI)^2 I  (T_aI)  \cdot R_a [R(\hat{z},\pi/2)]^2 (TR_I)  R(\hat{z},\pi/2)\nn\\
&=& (T_{a} T_a)I (T_a (I I)  T_a)I\cdot R_aR(\hat{z},\pi)R_IR(\hat{z},\pi/2)\cdot T.
\label{eq:timereverse}
\eea
The spatial part of Eq. (\ref{eq:timereverse}) is $T_{2a} I T_{2a}I=T_{2a}T_{-2a}=1$.
Using Eq. (\ref{eq:RaRI}), $R(\hat{z},\pi):(x,y,z)\rightarrow (-x,-y,z)$, $R(\hat{z},\pi/2):(x,y,z)\rightarrow (y,-x,z)$,
and the composition rule Eq. (\ref{eq:composition}),
it is a straightforward calculation to verify that $R_aR(\hat{z},\pi)R_IR(\hat{z},\pi/2): (x,y,z)\rightarrow (x,y,z)$.
Thus $sr(st)^2r(st)$ is equal to $T$.

Next we show that the quotient group $G/\mathopen{<}T_{3a}\mathclose{>}$ contains at least $48$ elements.
One can verify that 
by restricting the actions to the spin space,
the $24$ operations in the last column of Table \ref{table:O}
are exactly given by the third column of Table \ref{table:O} where $\alpha$ is $S^{\alpha}$ for short ($\alpha=x,y,z$).
This exhausts the $24$ proper elements of the $O_h$ group as a symmetry group of a cube in the spin space.
Furthermore, by multiplying the $24$ operations in the last column of Table \ref{table:O} with $T=(rs)^{-1}(st)^2 r (st)$,
and again restricting to the spin space, 
we obtain the other $24$ improper elements of the $O_h$ group acting in the spin space.
Then let's recover the spatial components of these $48$ operations generated by $r,s,t$,
and view them as elements in $G/\mathopen{<}T_{3a}\mathclose{>}$.
Since these $48$ operations already act differently in the spin space from each other,
they must be distinct elements in $G/\mathopen{<}T_{3a}\mathclose{>}$.
This shows that $G/\mathopen{<}T_{3a}\mathclose{>}$ has at least $48$ group elements.
Combining with the previously established fact that $G/\mathopen{<}T_{3a}\mathclose{>}$ is isomorphic to a subgroup of $O_h$,
we conclude that $G/\mathopen{<}T_{3a}\mathclose{>}$ is actually isomorphic to $O_h$.
We also note that the Hilbert space of the spin-$1/2$ Kitaev-Gamma chain is a projective representation of $O_h$,
since a rotation by $2\pi$ is $-1$ for half-odd-integer spins. 

We make a further comment on the group structure.
Note that $O_h=O\times \{1,i\}$ where ``i" is the inversion.
The cubic point group $O$ has $24$ elements and is isomorphic to $S_4$, the permutation group of four elements.
$S_4$ has a generator-relation representation as follows \cite{Coxeter1965},
\bea
S_4=<a,b,c|a^2=b^3=c^4=abc=e>,
\eea
or alternatively, it can also be generated by two generators,
\bea
O=<R,S|R^3=S^4=(RS)^2=E>.
\eea
We note that in our case, we can take
\bea
a=rt,b=rs,c=st,
\eea
and 
\bea
R=rs,S=st.
\eea

%%%%%%%%%%%%%%%%%%%
\subsection{Symmetry relations in the coefficients $C_{[i]}^\alpha$'s and in $D_{[i]}^\alpha$'s}
\label{app:sym}
Suppose the spin operators $S_i^\alpha$ at low energies can be written in terms of $J_L^\alpha,J_R^\alpha,g$ as,
\bea
\frac{1}{a}S_i^\alpha=D^\alpha_{[i]} J_L^\alpha+D^{\prime\alpha}_{[i]} J_R^\alpha+C^\alpha_{[i]} \frac{1}{\sqrt{a}} (-)^{x/a} N^\alpha,
\label{eq:relation}
\eea
in which $N^\alpha=i\text{tr} (g \sigma^\alpha)$.
We'll analyze what constraints the $O_h$ symmetry will put on the coefficients.

First, consider the time reversal symmetry $T$.
The transformations of $S_i^\alpha$, $J_L^\alpha,J_R^\alpha$ and $N^\alpha$ under $T$ are
\bea
T:&S_i^\alpha &\rightarrow -S_i^\alpha\nn\\
&J_L^\alpha(x)&\rightarrow -J_R^\alpha(x)\nn\\
&J_R^\alpha(x)&\rightarrow -J_L^\alpha(x)\nn\\
&N^\alpha(x)&\rightarrow -N^\alpha(x).
\eea
Performing time reversal transformation on both sides of Eq. (\ref{eq:relation}), we obtain,
\bea
-\frac{1}{a}S_i^\alpha=-D^\alpha_{[i]} J_R^\alpha-D^{\prime\alpha}_{[i]} J_L^\alpha-C^\alpha_{[i]} \frac{1}{\sqrt{a}} (-)^{x/a} N^\alpha.
\label{eq:Ttransform}
\eea
On the other hand, from  Eq. (\ref{eq:relation}),
\bea
-\frac{1}{a}S_i^\alpha=-D^\alpha_{[i]} J_L^\alpha-D^{\prime\alpha}_{[i]} J_R^\alpha-C^\alpha_{[i]} \frac{1}{\sqrt{a}} (-)^{x/a} N^\alpha.
\label{eq:Ttransform2}
\eea
Comparing Eq. (\ref{eq:Ttransform}) and Eq. (\ref{eq:Ttransform}), it is clear that 
\bea
D^\alpha_{[i]}=D^{\prime\alpha}_{[i]}.
\eea

Next consider the symmetry operation $R_aT_a$.
The transformations of $S_i^\alpha$, $J_L^\alpha,J_R^\alpha$ and $N^\alpha=\text{tr} g (\sigma^\alpha)$ under $R_aT_a$ are
\bea
R_aT_a:& S_i^\alpha &\rightarrow (R_a)^{\alpha}_{\,\,\beta} S_{i+1}^\beta\nn\\
& J_L^\alpha(x) &\rightarrow (R_a)^{\alpha}_{\,\,\beta} J^\beta_L(x)\nn\\
& J_R^\alpha(x) &\rightarrow (R_a)^{\alpha}_{\,\,\beta} J^\beta_R(x)\nn\\
& N^\alpha(x) &\rightarrow -(R_a)^{\alpha}_{\,\,\beta} N^\beta(x),
\eea
in which $(R_a)^{\alpha}_{\,\,\beta}$ is the matrix element of the vector representation of the rotation $R_a$,
and the minus sign in the transformation of $N^\alpha(x)$ is because $T_a:g\rightarrow -g$.
Applying $R_aT_a$ to $S_1^x$, under the same logic as the time reversal case, we obtain
\bea
D^z_{2} (J_L^z+ J_R^z)+C^z_{2} \frac{1}{\sqrt{a}} (-)^{x/a+1} N^z=D^x_{1} (J_L^z+ J_R^z)+C^x_{1} \frac{1}{\sqrt{a}} (-)^{x/a} (-)N^z,
\eea
from which
\bea
D_1^x=D_2^z,\,\,C_1^x=C_2^z.
\eea
Similar analysis on other spin operators gives
\bea
C_1^x=C_2^z=C_3^y, & D_1^x=D_2^z=D_3^y\nn\\
C_1^y=C_2^x=C_3^z, & D_1^y=D_2^x=D_3^z\nn\\
C_1^z=C_2^y=C_3^x, & D_1^z=D_2^y=D_3^x.\nn\\
\eea

There is another symmetry $R_I I$ in the $O_h$ group.
The transformations of $S_i^\alpha$, $J_L^\alpha,J_R^\alpha$ and $N^\alpha=\text{tr} g (\sigma^\alpha)$ under $R_I I$ are
\bea
R_I I:& S_i^\alpha &\rightarrow (R_I)^{\alpha}_{\,\,\beta} S_{-i}^\beta\nn\\
& J_L^\alpha(x) &\rightarrow (R_I)^{\alpha}_{\,\,\beta} J^\beta_R(-x)\nn\\
& J_R^\alpha(x)& \rightarrow (R_I)^{\alpha}_{\,\,\beta} J^\beta_L(-x)\nn\\
& N^\alpha(x) &\rightarrow (R_I)^{\alpha}_{\,\,\beta} N^\beta(-x).
\eea
Applying $R_I I$ to $S_1^x$, we obtain
\bea
-D^z_{3} (J_L^z(-x)+ J_R^z(-x))-C^z_{3} \frac{1}{\sqrt{a}} (-)^{-x/a} N^z(-x)=-D^x_{1} (J_L^z(-x)+ J_R^z(-x))-C^x_{1} \frac{1}{\sqrt{a}} (-)^{x/a} N^z(-x),
\eea
which gives
\bea
D_1^x=D_1^z,\,\,C_1^x=C_3^z.
\eea

In summary, by using the $O_h$ symmetry, we are able to show the following relations
\begin{flalign}
&D_1^z=D_2^y=D_3^x (=D_1),\nn\\
&D_1^x=D_2^z=D_3^y=D_1^y=D_2^x=D_3^z (=D_2),
\end{flalign}
and
\begin{flalign}
&C_1^z=C_2^y=C_3^x (=C_1),\nn\\
&C_1^x=C_2^z=C_3^y=C_1^y=C_2^x=C_3^z (=C_2).
\end{flalign}

Note that the difference in $D_1,\,D_2$ and in $C_1,\,C_2$ will introduce a $4k_f$ and a $2k_f$ oscillating component in the
nonabelian bosonization formula, respectively, where $k_f=\pi/6a$.
We now separate the components with different momenta in $D^x_{[i]}$ by performing a Fourier transformation.
Other directions and the $C_{[j]}^\alpha$'s can be treated in a similar manner.
Let 
\bea
D^x_{[j]}=A\cos(\frac{2\pi}{3}j+\psi)+B.
\eea
Then 
\bea
A\cos(\frac{2\pi}{3}+\psi)+B&=&D_2\nn\\
A\cos(\frac{4\pi}{3}+\psi)+B&=&D_2\nn\\
A\cos(\psi)+B&=&D_1,
\eea
which solves
\bea
A&=&\frac{2(D_1-D_2)}{3},\nn\\
B&=&\frac{2(D_2+D_1)}{3},\nn\\
\psi&=&0.\nn\\
\eea
In a compact form, we have
\bea
D^\alpha_{[j]}&=&\frac{2(D_1-D_2)}{3} \cos(\frac{2\pi}{3}j+\frac{2\pi}{3}(\alpha-1))+\frac{2D_2+D_1}{3},\nn\\
C^\alpha_{[j]}&=&\frac{2(C_1-C_2)}{3} \cos(\frac{2\pi}{3}j+\frac{2\pi}{3}(\alpha-1))+\frac{2C_2+C_1}{3},\nn\\
\eea
in which $\alpha=1,2,3$ corresponding to $x,y,z$.

Now we are able to write the nonabelian bosonization formula with different momenta separated, {\it i.e.}
\bea
\frac{1}{a} S_i^\alpha &= \big[ c_0+c_4\cos(\frac{2\pi}{3a}x+\frac{2\pi}{3}(\alpha-1)) \big] (J_L^\alpha(x)+J_R^\alpha(x))  \big[c_6(-)^{\frac{x}{a}}+c_2 \cos(\frac{\pi}{3a}x+\frac{2\pi}{3}(\alpha-1)) \big]\frac{1}{\sqrt{a}} \text{tr} (g(x)\sigma^\alpha),\nn\\
\eea
in which
\bea
c_0=\frac{2D_2+D_1}{3}, c_2=\frac{2(C_2-C_1)}{3}, c_4=\frac{2(D_1-D_2)}{3}, c_6=\frac{2C_2+C_1}{3}.
\eea

%%%%%%%%%%%%%%%%%%%
\section{The nine- and three-point formulas}

\subsection{The nine-point formula}
\label{sec:ninepoint}

%This time we only keep the $s,p,q$ components.
At large distances, the uniform and the $2\pi/3$-oscillating components decay much faster than the staggered and the $\pi/3$-oscillating components.
Due to this reason, we will derive a nine-point formula 
to extract the staggered and the $\pi/3$-oscillating components of the spin-spin correlation functions, 
assuming no uniform and the $2\pi/3$-oscillating components.

Let $f(j)$ ($j\in \mathbb{Z}$) be
\bea
f(j)=(-)^j s(j)+\cos(\frac{\pi}{3} j) p(j)+\sin(\frac{\pi}{3} j) q(j).
\eea  
Write $f(i)$ in terms of $i-j$, we have
\bea
f(i) =
(-)^{i-j} s^{\prime}(i-j)+\cos(\frac{\pi}{3} (i-j)) p^{\prime}(i-j)+\sin(\frac{\pi}{3} (i-j)) q^{\prime}(i-j),
\eea
in which 
\bea
s^{\prime}(i-j)&=&(-)^{j} s(i),\nn\\
p^{\prime}(i-j)&=&\cos(\frac{\pi}{3}j)p(i) +\sin(\frac{\pi}{3}j) q(i),\nn\\
q^{\prime}(i-j)&=&-\sin(\frac{\pi}{3}j)p(i) +\cos(\frac{\pi}{3}j) q(i).
\eea

Expanding $s^{\prime},p^{\prime},q^{\prime}$ as
\bea
s^{\prime}(k)&=&s_2 k^2+s_1 k+s_0\nn\\
p^{\prime}(k)&=&p_2 k^2+p_1 k+p_0\nn\\
q^{\prime}(k)&=&q_2 k^2+q_1 k+q_0,
\eea
 the constants $s_0,p_0,q_0$ can be determined as 
 \begin{flalign}
 s_0&=-\frac{1}{27} f (-4 + j)  + \frac{2}{27} f (-2 + j) -\frac{8}{27} f (-1 + j)+ \frac{1}{3} f (j)  -\frac{8}{27} f (1 + j) + \frac{2}{27} f (2 + j) -\frac{1}{27} f (4 + j) ,\nn\\
%\end{flalign}
 %\begin{flalign}
 p_0&=\frac{1}{27} f (-4 + j) - \frac{2}{27} f (-2 + j) + \frac{8}{27} f (-1 + j) + \frac{2}{3} f (j)  +\frac{8}{27} f (1 + j) - \frac{2}{27} f (2 + j)  +\frac{1}{27} f (4 + j), \nn\\
%\end{flalign}
% \begin{flalign}
 q_0&=-\frac{1}{9\sqrt{3}} f (-4 + j)  - \frac{2}{9\sqrt{3}} f (-2 + j) - \frac{8}{9\sqrt{3}} f (-1 + j) +\frac{8}{9\sqrt{3}} f (1 + j) + \frac{2}{9\sqrt{3}} f (2 + j)  +\frac{1}{9\sqrt{3}} f (4 + j). 
\end{flalign}

Then $s(j),p(j),q(j)$ can be expressed as
\bea
s(j)&=&(-)^js_0\nn\\
p(j)&=&\cos(\frac{\pi}{3}j)p_0-\sin(\frac{\pi}{3}j)q_0\nn\\
q(j)&=&\sin(\frac{\pi}{3}j)p_0+\cos(\frac{\pi}{3}j)q_0.
\eea

%%%%%%%%%%%%%%%%%%%
\subsection{The three-point formula}

Let $f(j)$ ($j\in \mathbb{Z}$)
\bea
f(j)=u(j)+(-)^j s(j).
\eea
The three-point formula can be used to extract the uniform part $u(j)$ and the stagger part $s(j)$, as
\bea
u(j)&=& \frac{1}{4}f(j-1)+\frac{1}{2}f(j)+\frac{1}{4}f(j+1),\nn\\
s(j)&=& (-)^j\big[-\frac{1}{4}f(j-1)+\frac{1}{2}f(j)-\frac{1}{4}f(j+1)\big].
\eea

%%%%%%%%%%%%%%%%%%%
\section{SU(2)$_1$ conformal tower in the finite size spectrum}

%-------------------------------------
\begin{figure}[h]
\includegraphics[width=14cm]{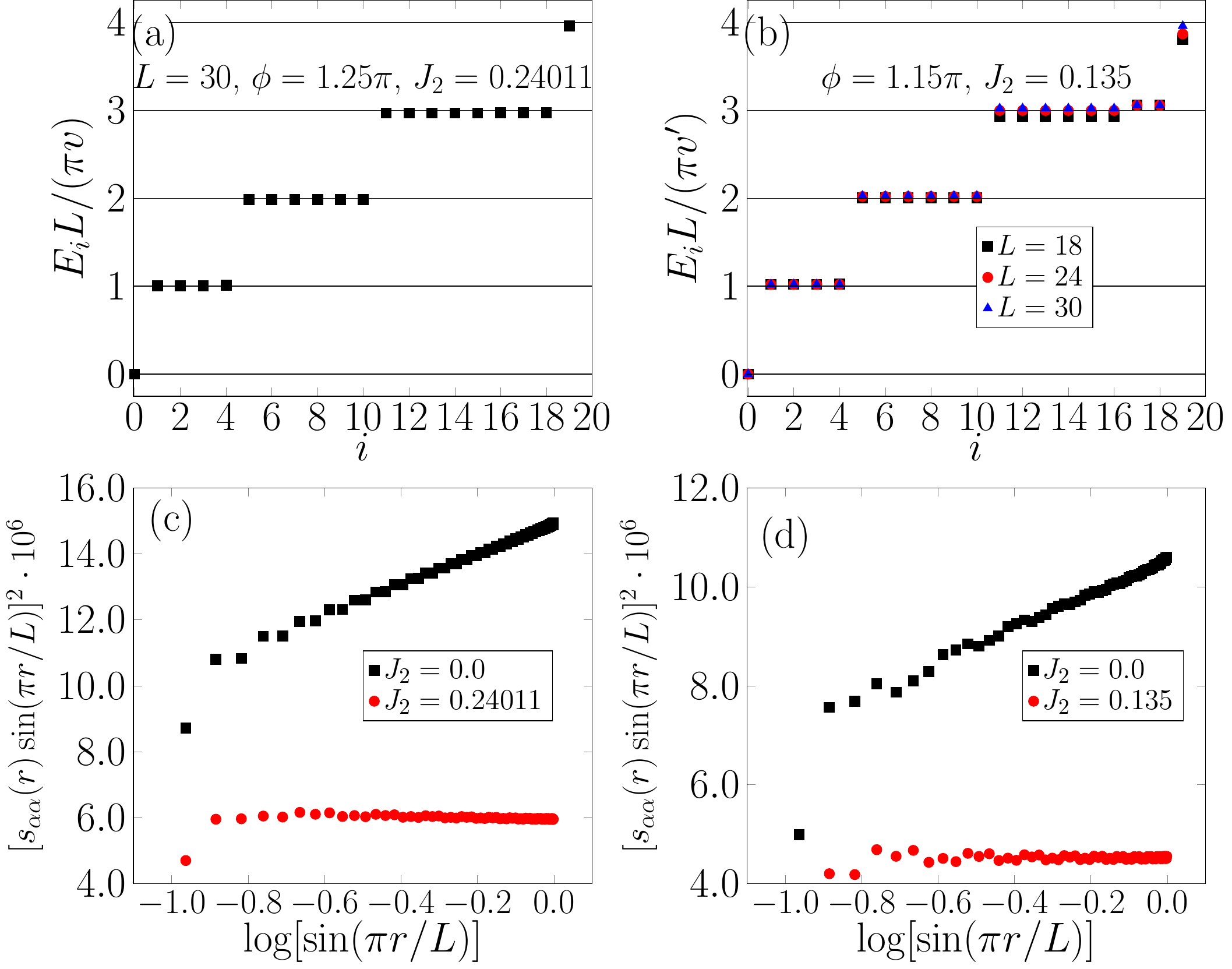}
\caption{Energies of the first $20$ eigenstates with appropriately chosen $J_{2c}$ for (a) $\phi= 1.25\pi$, and (b) $\phi= 1.15\pi$;
and $\big( s_{\alpha\alpha}(r) \sin(\pi r/L)\big)^2$ ($\alpha=x$) vs. $\log \sin(\pi r/L)$ with and without $J_2$ shown by black and red dots, respectively,
for (c) $\phi= 1.25\pi$ and (d) $\phi= 1.15\pi$.
In (a,b), the spectra are calculated using ED with a periodic boundary condition for both values of $\phi$.
The system size is taken as $L=30$ for $\phi=1.25\pi$ in (a), and $L=18,24,30$ for $\phi=1.15\pi$ in (b).
In (c,d), the correlation functions are computed using DMRG on a system of  $L=144$ sites with periodic boundary conditions.
$s_{\alpha\alpha}$ is then extracted from a nine-point formula in the same way as discussed  in the main text.
} 
\label{fig:3suppl}
\end{figure}
%-------------------------------------

In this section, we study numerically the finite size spectrum of an AFM Kitaev-Gamma chain,
and verify that the spectrum exhibits a conformal tower structure consistent with the emergent SU(2)$_1$ symmetry.

We first briefly review the SU(2) symmetric AFM Heisenberg point, {\it i.e.}, $\phi=5\pi/4$, following the treatment in Ref. \cite{Eggert1992}.
Due to the existence of the marginally irrelevant term $-u\vec{J}_L\cdot \vec{J}_R$ which breaks the chiral SU(2) symmetry,
the SU(2)$_1$ symmetry emerges only logarithmically along the RG flow.
In particular, there is a finite size correction to the energy spectrum only suppressed by $1/\ln L$ \cite{Affleck1989s}.
At small system size, the effects of such logarithmic correction are notable which obscures the emergent SU(2)$_1$ structure.
However, as shown in Refs. \cite{Affleck1989s,Eggert1992},
there is a clever trick to get around such problem.
One adds to the nearest neighbor Heisenberg Hamiltonian a next nearest neighbor $J_2$ term, so that the Hamiltonian now becomes
\bea
H_{AFM}^\prime = J \sum_n \vec{S}_n \cdot \vec{S}_{n+1} +J_2 \sum_n \vec{S}_n \cdot \vec{S}_{n+2}.
\eea
At certain value $J_{2c}$, the bare marginal operator is killed, {\it i.e.}, $u^\prime=0$ within $-u^\prime \vec{J}_L\cdot \vec{J}_R$.
In fact, $J_{2c}$ is the phase transition point between the gapless spin liquid phase and an ordered dimerized phase  \cite{Eggert1992}. 
According to Ref. \cite{Eggert1992}, when $J_2$ is tuned to $J_{2c}=0.2401J$, 
the finite size spectra are arranged to a nearly perfect conformal tower structure fully consistent with the SU(2)$_1$ predictions without any logarithmic corrections.
Using exact diagonalization, we have reproduced the results in Ref. \cite{Eggert1992} as shown in Fig. \ref{fig:3suppl} (a),
with the eigenenergies computed on a finite system with $L=30$ sites under periodic boundary conditions.
The energies are rescaled in unit of $\pi v/L$ where $v=1.1745 J$.
As can be seen from Fig. \ref{fig:3suppl} (a), the eigenenergies are grouped into equally spacing plateaus at $n \pi v/L$ with $n\in \mathbb{Z}$.
For several lowest $n$'s, the degeneracies are: $4$ for $n=1$, $6$ for $n=2$, $8$ for $n=3$,
all consistent with the emergent SU(2)$_1$ symmetry.

To confirm the absence of the marginal operator $-u\vec{J}_L\cdot \vec{J}_R$ at $J_{2c}$,
we further calculate the spin-spin correlation function $\langle S^{\alpha}_1 S^{\alpha}_{r+1} \rangle$ ($\alpha=x$)
using DMRG on a system of a size $L=144$ with a periodic boundary condition. 
We stress that, although it is well known that DMRG simulations 
are more challenging in the presence of periodic boundary conditions, these were only used to demonstrate 
evidence for the logarithmic corrections predicted by the bosonization expression in the main text. 
Our analysis indeed shows that system sizes of the order of $\sim$150 sites are sufficient for this purpose.
%In Fig.~\ref{fig:4}, DMRG results obtained on chains with open boundary conditions show a perfect $1/r$ behavior for the correlation functions in the emergent SU(2)$_1$ phase. 
To reach numerical convergence in the presence of PBC, up to $m=1000$ DMRG states were kept and tens of finite size sweeps were performed with a final truncation error of $10^{-6}$.
Since the momentum $\pi$ and $\pm \pi/3$ components decay as $1/r$ at long distances which dominates over the momentum $0$ and $\pm 2\pi/3$ components which decay as $1/r^2$,
we will assume that  $\langle S^{\alpha}_1 S^{\alpha}_{r+1} \rangle$ only contains the $\pi$ and $\pm\pi/3$ oscillating components.
Then the nine-point formula can be applied and the staggered part $s_{\alpha\alpha}(r)$ can be extracted
which should behave as $\sim 1/\sin(\pi r/L)$ with no logarithmic factor. 
Indeed, as shown by the red points in Fig. \ref{fig:3suppl} (c),
$\big( s_{\alpha\alpha}(r)\sin(\pi r/L)\big)^2$ vs. $\log \big(\sin(\pi r/L)\big)$  is nearly a flat line consistent with an absence of the logarithmic factor.
On the other hand, the black dots show the results of $\big( s_{\alpha\alpha}(r)\sin(\pi r/L)\big)^2$ vs. $\log\big(\sin(\pi r/L)\big)$
when $J_2=0$.
The linear relation of the black dots indicates a behavior of $s_{\alpha\alpha}(r)$ as $\sim \ln^{1/2}\big(\sin(\pi r/L)\big)/\sin(\pi r/L)$.
Hence, this provides evidence for the role of $J_{2c}$ in killing the marginal operator.

Next we apply the same methods to a representative point $\phi=1.15\pi$ away from the SU(2) symmetric point.
Again by adding a $J_2$ term, the Hamiltonian now is 
\bea
H_{K\Gamma}^\prime = -J \sum_{<ij>\in \gamma}\big[ \cos(\phi) S_i^\gamma S_j^\gamma+\sin(\phi) (S_i^\alpha S_j^\alpha+S_i^\beta S_j^\beta) \big] +J_2 \sum_n \vec{S}_n \cdot \vec{S}_{n+2}.
\eea
In Fig. \ref{fig:3suppl} (d), $\big( s_{\alpha\alpha}(r)\sin(\pi r/L)\big)^2$ vs. $\log\big(\sin(\pi r/L)\big)$ at $J_2=0$ are plotted with black dots 
showing a linear relation with a nonzero slope.
On the other hand, the slope is zero when $J_2=0.135J$ as can be seen from the red dots.
Thus, this time the critical $J^\prime_{2c}$ is $0.135J$ which is able to remove the marginal operator $-u\vec{J}_L \cdot \vec{J}_R$ in the low energy theory.
In Fig. \ref{fig:3suppl} (b), the energies of the first $20$ states are plotted for $L=18,24,30$ in units of $\pi v^\prime/L$.
Here $v^\prime=0.6479J$ is determined by an extrapolation of $E_1(L)-E_0(L)$ as a function of $1/L$ to $1/L\rightarrow 0$,
in which $E_0(L)$ and $E_1(L)$ are the energies of the ground state and the first excited state, respectively.
As can be seen from Fig. \ref{fig:3suppl} (b),
the SU(2)$_1$ conformal tower structure in the finite size spectrum is improved by increasing $L$.
And in fact, a good agreement with the SU(2) symmetric case in Fig. \ref{fig:3suppl} (a) is already obtained when $L=30$.
This provides strong evidence for the emergent SU(2)$_1$ symmetry at low energies even away from $\phi=1.25 \pi$.

%%%%%%%%%%%%%%%%%%%
\section{RG flows of the scaling fields}

\subsection{Derivation of the RG flow equations}
\label{app:flow}

Conceptually, the RG flow of the theory described by the following Hamiltonian ({\it i.e.}, $H_F$ in the maintext)
\bea
H_F=-t\sum_{<ij>,\alpha} (c_{i\alpha}^\dagger c_{j\alpha}+\text{h.c.})-\mu\sum_{i\alpha}c_{i\alpha}^\dagger c_{i\alpha}+U\sum_i n_{\uparrow}n_{i\downarrow}+\Delta \sum_{<ij>\in \gamma} S_{i}^\gamma S_{j}^\gamma-\sum_{kn\alpha} h_{k}^\alpha(n) S_{k+3n}^\alpha
\label{eq:fermion}
\eea
can be separated into three steps,
\bea
\Lambda_0 \rightarrow \Lambda_1 \rightarrow \Lambda_2 \rightarrow E(\gg \frac{1}{L}),
\eea
%first from $\Lambda_0$ to $\Lambda_1$ and then from $\Lambda_1$ to $E$,
where $\Lambda_0=\pi/a$ is the bare cutoff, 
$\Lambda_1$ is the energy scale at which the three sites within a unit cell get smeared, 
$\Lambda_2$ is the energy scale at which a linearization of the free fermion dispersion applies,
and $1/E$ is the length scale of the correlation functions. 
Below the energy scale $\Lambda_2$, the fermion becomes a Dirac fermion and can be written alternatively in terms of a charge boson and an SU(2) WZW boson using nonabelian bosonization.
However, now we study the flow in the energy region $\Lambda_0\rightarrow \Lambda_1$.

We first give a heuristic argument to the RG flow equations based on the operator product expansion (OPE) in real space.
The origin of the multiplicative renormalizations of the scaling fields is clear in this approach.
However, the OPE approach is not rigorous since it only applies to the continuum limit, 
and now the flow is within the high energy region.
Later we will derive the RG flow in a more rigorous manner in the framework of Wilsonian momentum shell RG.

First recall how the RG flow can be obtained from the OPE between operators \cite{Cardy1996}.
Let $\sum_i g_i \hat{O}_i$ be in the Hamiltonian,
in which  $\hat{O}_i$ is an operator with scaling dimension $x_i$.
If the OPEs between $\hat{O}_i$'s are given by
\bea
\hat{O}_i(x) \hat{O}_j(y) \sim \frac{c_{ijk}}{|x-y|^{x_i+x_j-x_k}} \hat{O}_k(\frac{x+y}{2}),
\eea
then the flow for the coupling $g_k$ up to one-loop level is
\bea
\frac{dg_k}{d\ln b} =(d-x_i)g_k-\frac{1}{2}S_d c_{ijk} g_ig_j,
\eea
in which $x,y$ are spacetime coordinates, $d$ is the spacetime dimension, and $S_d=(2\pi)^{d/2}/\Gamma(d/2)$ is the solid angle of the $(d-1)$-dimensional unit sphere.
In our case, take $h_{u,1}^x$ as an example.
Heuristically, if we take the discrete unit cell index $n$ as a continuous variable, and combine $n$ with $\tau$ into $x=(\tau,n)$,
then we obtain the following OPE ($j=1,2,3$),
\bea
S_1^x(x) S_2^x(x) \cdot S_j^x(y) &\sim& \langle S_2^x(x)  S_j^x(y) \rangle S_1^x(x).
\label{eq:SpinOPE}
\eea
Hence $h_{u,j}^x$ ($j=1,2,3$) all contribute to the renormalization of $h_{u,1}^x$.
The RG flow equation is then
\bea
\frac{d h_{u,1}^x}{d\ln b} &=& (1-\lambda_{21} \Delta)h_{u,1}^x - \lambda_{22} \Delta h_{u,2}^x- \lambda_{23} \Delta h_{u,3}^x,
\eea
in which $\lambda_{2j}$ is determined by the contraction $\langle S_2^x(x)  S_j^x(y) \rangle$.
Furthermore, $\lambda_{21}=\lambda_{23}$ due to the inversion symmetry of the free fermion band structure.
In addition, from a simple argument we expect that $\lambda_{22}>\lambda_{23}=\lambda_{21}$.
This is because at short distances the $j=2$ contraction in Eq. (\ref{eq:SpinOPE}) contains less oscillation than the $j=1,3$ cases,
since at $x=y$, the $j=2$ contraction is on-site while the $j=1,3$ terms are off-site.
Thus we are able to obtain the form of the RG equations presented in the main text in a simple manner,
and even the relation $\lambda>0>\nu$ is expected.
For the staggered part $h_{s,i}^\alpha$, roughly speaking, one needs to change $\Delta$ to $-\Delta$ since the spin operators on two adjacent sites differ by a sign.
Thus the slope of $C_2/C_1$ and $D_2/D_1$ around $\phi=\phi_{AF}$ should be opposite in sign.

In what follows, the flows of $h_{u,i}^\alpha$ and  $h_{s,i}^\alpha$
will be considered in a momentum shell RG approach.
The signs and the magnitudes of the coefficients $\lambda_j$'s will be determined.
We will neglect the flows of $U$ and $\Delta$ since they are marginal near the free fermion fixed point 
and the RG stopping scale $b_1\sim 3$ is not very large.
We also ignore the contribution from $U$ to the flows of the scaling fields, 
since this contribution is SU(2) symmetric.

%-------------------------------------------- 
\begin{figure}[h]
\includegraphics[width=6cm]{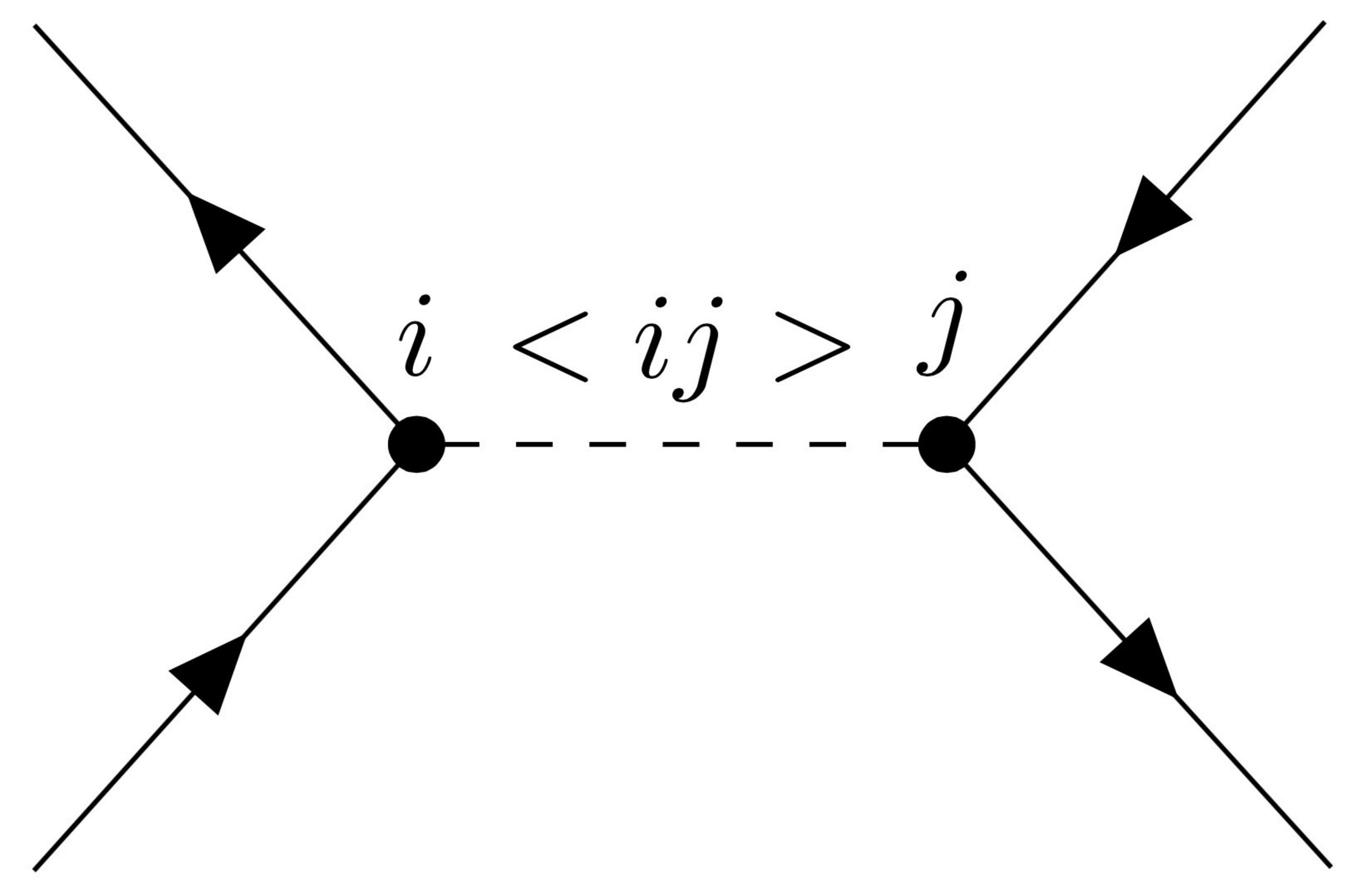}
\caption{Interaction vertex for the SU(2) breaking $\Delta$-term.
} \label{fig:delta}
\end{figure}
%--------------------------------------------
%-------------------------------------------- 
\begin{figure}[h]
\includegraphics[width=5cm]{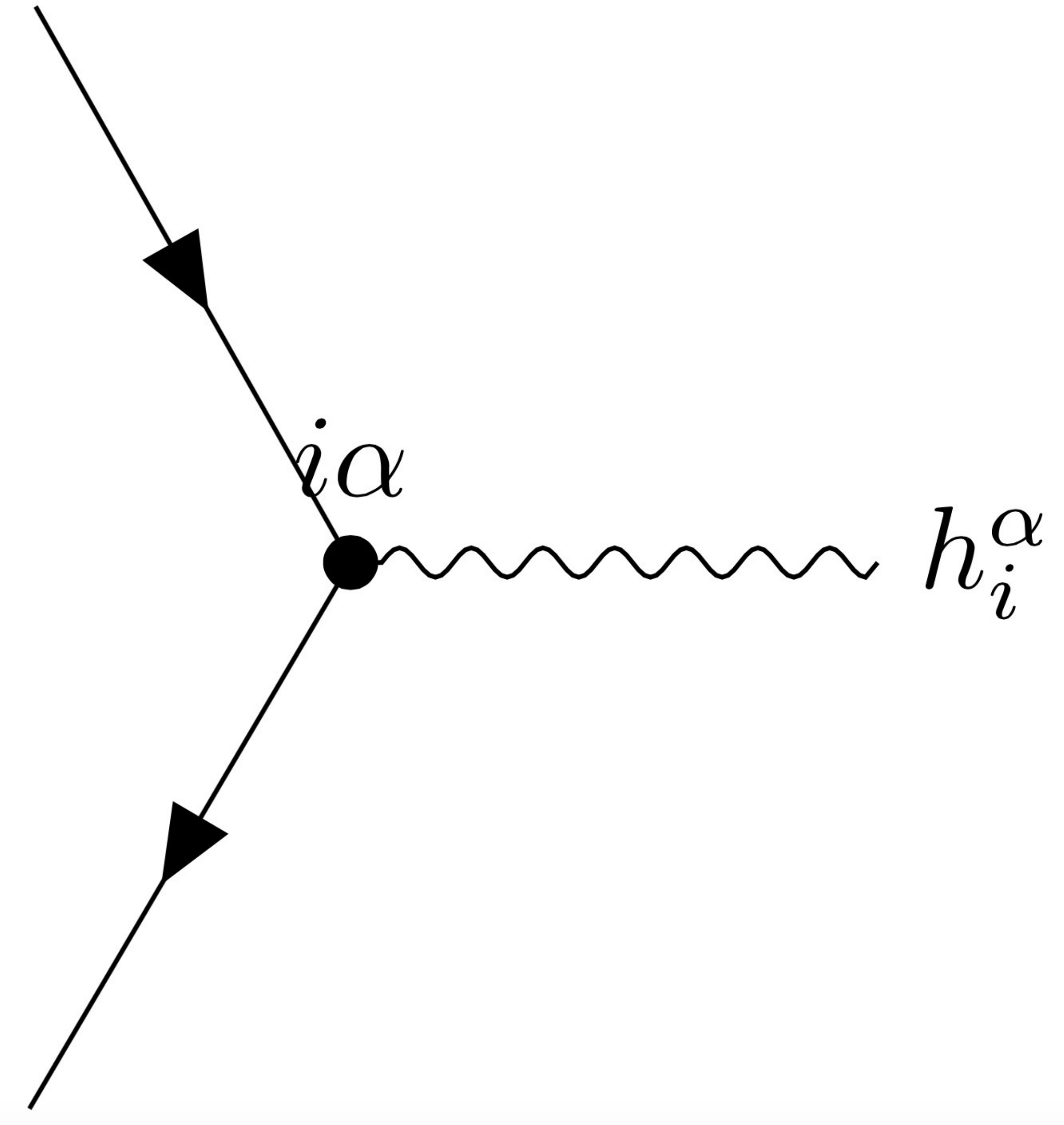}
\caption{Vertex for the scaling fields.}
\label{fig:field}
\end{figure}
%--------------------------------------------

Now we proceed to a momentum shell RG treatment.
Let's first write the terms within the action in the frequency-momentum space. 
The $\Delta$-term is represented as the diagram in Fig. \ref{fig:delta},
in which $i,j=1,2,3$, and $<ij>=x,z,y$ for the bonds $<12>,<23>,<31>$.
In the frequency-momentum space, the expression corresponding to Fig. \ref{fig:delta} is (denoting $<ij>$ as $\gamma$ for short)
\begin{flalign}
&\Delta \int d\tau \sum_n S_i^\gamma(\tau,n) S_j^\gamma(\tau,n)\nn\\
=&\Delta \int d\tau \sum_n S^\gamma(\tau,i+3n) S^\gamma(\tau,j+3n)\nn\\
=&\frac{\Delta}{N\beta}\int d\tau \sum_n \sum_{k_1,k_2,k_3,k_4}  c^\dagger(k_1) \frac{1}{2}\sigma^\gamma c(k_2) \cdot c^\dagger(k_3) \frac{1}{2}\sigma^\gamma c(k_4)
 e^{i(\omega_1-\omega_2+\omega_3-\omega_4) \tau}  e^{i(\vec{k}_1-\vec{k}_2)(i+3n)a\hat{x}} e^{i(\vec{k}_3-\vec{k}_4)(j+3n)a\hat{x}},
\label{eq:Delta_k}
\end{flalign}
in which $k=(i\omega,\vec{k})$, $a$ is the lattice spacing,
$N$ is the total number of sites, $\beta$ is the inverse of temperature,
$n$ is summed over the unit cells, and $\hat{x}$ is the unit vector in the spatial direction.
By integrating over $\tau$ and summing over $n$, and using
\bea
\frac{1}{N}\sum_n e^{i(\vec{k}_1-\vec{k}_2+\vec{k}_3-\vec{k}_4)\cdot 3na\hat{x}}=\frac{1}{3}\sum_{m=1}^3 \delta_{\vec{k}_1-\vec{k}_2+\vec{k}_3-\vec{k}_4,\frac{2\pi}{3a}m\hat{x}},\nn\\
\eea
Eq. (\ref{eq:Delta_k}) is equal to
\begin{flalign}
%&\Delta \int d\tau \sum_n S_i^\gamma(\tau,n) S_j^\gamma(\tau,n)\nn\\
&\frac{\Delta}{N\beta}\frac{1}{3}\sum_m\sum_{k_1,k_2,k_3,k_4}  e^{i(\vec{k}_1-\vec{k}_2)(i-j) a\hat{x}}
e^{-i \frac{2\pi }{3}m j} \delta_{k_1-k_2+k_3-k_4+\frac{2\pi}{3a}m\hat{x},0}
 c^\dagger(k_1) \frac{1}{2}\sigma^\gamma c(k_2) \cdot c^\dagger(k_3) \frac{1}{2}\sigma^\gamma c(k_4).
\end{flalign}
Hence we obtain
\begin{flalign}
&\Delta \int d\tau \sum_n S_i^\gamma(\tau,n) S_j^\gamma(\tau,n)
=\Delta\frac{1}{3N\beta} \sum_m e^{-i\frac{2\pi}{3}mj}\sum_{k,p,q} e^{i\vec{q}\cdot (i-j)a\hat{x}}    c^\dagger(k+q) \frac{1}{2}\sigma^\gamma c(k) \cdot c^\dagger(p-q) \frac{1}{2}\sigma^\gamma c(p+\frac{2\pi}{3a}m\hat{x}).
\end{flalign}

The magnetic field term is represented as the diagram in Fig. \ref{fig:field}.
The expression in the frequency-momentum space is
\begin{flalign}
&\int d\tau \sum_n h_l^\alpha (\tau,n) S_l^\alpha(\tau,n)\nn\\
=& \int d\tau \sum_n h_l^\alpha(\tau,n) S^\alpha (\tau,l+3n)\nn\\
=& \int d\tau \sum_n \frac{1}{N\beta} \sum_{q} h_l^\alpha (-q) e^{-i\omega\tau} e^{-i\vec{q}\cdot 3na\hat{x}} 
\sum_{q^\prime} S^\alpha (q^\prime) e^{i\omega \tau} e^{i\vec{q}^\prime\cdot (l+3n)a\hat{x}}\nn\\
=&\frac{1}{3}\sum_q\sum_m e^{i\vec{q}\cdot la \hat{x} } e^{i\frac{2\pi}{3}ml} h_l^\alpha(-q) S^\alpha (q+\frac{2\pi}{3a}m\hat{x}).
\label{eq:field_expr1}
\end{flalign}
In terms of the fermion operators, we have
\begin{flalign}
&\int d\tau \sum_n h_l^\alpha (\tau,n) S_l^\alpha(\tau,n)
=\frac{1}{3}\sum_{kq}\sum_m e^{i\vec{q}\cdot la \hat{x} } e^{i\frac{2\pi}{3}ml} h_l^\alpha(-q) c^\dagger(k+\frac{2\pi}{3a}m\hat{x})\frac{1}{2}\sigma^\alpha c(k-q),
\label{eq:field_expr2}
\end{flalign}
in which $c^\dagger(k) = (c_\uparrow^\dagger(k),c_\downarrow^\dagger(k))$.
In what follows, we focus on the uniform part $h_{u,i}^\alpha(\tau,n)$.
Then the momentum transfer $\vec{q}$ is $|\vec{q}|\sim 0$.
If we want to consider the staggered part $h_{s,i}^\alpha$,
we should write $\vec{q}=\vec{q}^\prime+\pi/3$, with $|\vec{q}^\prime|\sim 0$.

%-------------------------------------------- 
\begin{figure}[h]
\includegraphics[width=8cm]{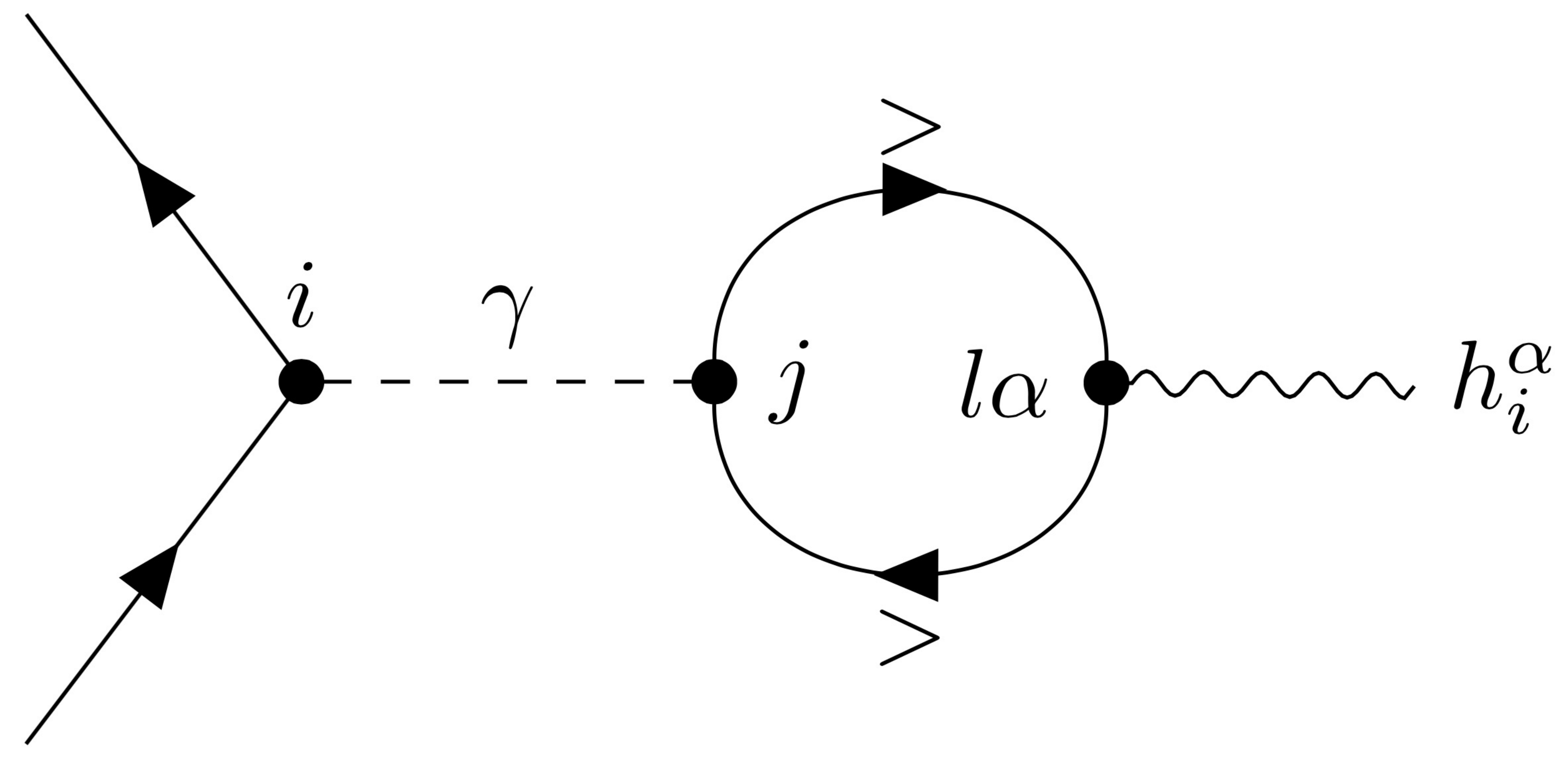}
\caption{Diagram of the renormalization of the scaling fields due to the SU(2) breaking $\Delta$-term.
} \label{fig:renormalization}
\end{figure}
%--------------------------------------------

Next we consider the renormalization of $h_{u,i}^\alpha$  due to the effect of the $\Delta$-term. 
In what follows, we will drop the subscript ``u" for simplicity. 
The diagram is shown in Fig. \ref{fig:renormalization} in which the momentum integrated within the loop corresponds to the fast mode (represented as $>$ in the figure) in the treatment of a momentum shell RG.
Take the renormalization of $h_1^x$ as an example. 
Notice that the term $\Delta \int d\tau \sum_n S_{1+3n}^x(\tau)S_{2+3n}^x(\tau)$ contributes to
 the renormalization of $h_1^x$ by contracting $S^x_{2+3n}(\tau)$ with  $S^x_{l+3n^\prime}(\tau^\prime)$ ($l=1,2,3$).
Thus in Fig. \ref{fig:renormalization}, we should make the substitution $i\rightarrow 1$, $j\rightarrow 2$, $l\rightarrow l$. 

The analytic expression corresponding to Fig. \ref{fig:renormalization} is
\begin{flalign}
&\Delta \frac{1}{3N\beta} \sum_m e^{-i\frac{2\pi}{3a} mj} \sum_{kpq} e^{i\vec{q}\cdot(i-j)a\hat{x} } c^\dagger (k+q) \frac{1}{2}\sigma^\gamma c(k) \cdot \frac{1}{3} \sum_{k^\prime q^\prime m^\prime} e^{i\vec{q}^\prime \cdot la\hat{x}} e^{i\frac{2\pi}{3} m^\prime l} h_l^\alpha(-\vec{q}^\prime)\nn\\
&\times
\langle c^\dagger(p-q) \frac{1}{2} \sigma^\gamma c(p+\frac{2\pi}{3a}m  \hat{x}) c^\dagger(k^\prime+\frac{2\pi}{3a}  m^\prime\hat{x})\frac{1}{2}\sigma^\alpha c(\vec{k}^\prime-q^\prime)\rangle_{\text{f}},
\label{eq:expression_renorm}
\end{flalign}
in which $\langle\rangle_{f}$ represents averaging over fast modes.
The averaging leads to the following momentum constraints,
\bea
p+\frac{2\pi}{3a}m &=& k^\prime+\frac{2\pi}{3a} m^\prime,\nn\\
p-q &=&k^\prime-q^\prime,
\eea
which gives
\bea
m^\prime &=& m+\bar{m}\nn\\
p&=& k^\prime +\frac{2\pi}{3a} \bar{m}\nn\\
q&=& q^\prime +\frac{2\pi}{3a} \bar{m},
\label{eq:momentum_conserv}
\eea
in which $\bar{m}=1,2,3$.
Then Eq. (\ref{eq:expression_renorm}) becomes
\begin{flalign}
&\Delta \frac{1}{9N\beta} \sum_{m,\bar{m}} \sum_{kk^\prime q^\prime} e^{-i\frac{2\pi}{3a} mj}  e^{i\vec{q}^\prime\cdot(i-j)a\hat{x} } e^{i\frac{2\pi}{3}\bar{m}(i-j) } e^{i\vec{q}^\prime \cdot la\hat{x}} e^{i\frac{2\pi}{3} (m+\bar{m}) l} h_l^\alpha(-q^\prime) c^\dagger (k+q^\prime+\frac{2\pi}{3a}\bar{m}\hat{x}) \frac{1}{2}\sigma^\gamma c(k)\nn\\
&\times \langle c^\dagger(k^\prime-q^\prime) \frac{1}{2} \sigma^\gamma c(k^\prime+\frac{2\pi}{3a}(m+\bar{m})  \hat{x}) c^\dagger(k^\prime+\frac{2\pi }{3a} (m+\bar{m}) \hat{x})\frac{1}{2}\sigma^\alpha c(\vec{k}^\prime-q^\prime)\rangle_{\text{f}}.
\label{eq:expression_renorm2}
\end{flalign}
To further simplify the expression,
$e^{-i\frac{2\pi}{3}mj}$ and $e^{-i\frac{2\pi}{3}\bar{m}j}$ are first collected together,
 then combined with $e^{i\frac{2\pi}{3}(m+\bar{m}) l}$.
The combined factor is then put together with the $\langle\rangle_f$ term and the result depends on $m+\bar{m}$ only.
The remaining terms only depend on $\bar{m}$.
Doing these, we obtain
\begin{flalign}
& \frac{1}{3} \sum_{\bar{m}}e^{i\frac{2\pi}{3a} \bar{m}i} \sum_{k q^\prime}  e^{i\vec{q}^\prime\cdot ia\hat{x} } h_l^\alpha(-q^\prime) c^\dagger (k+q^\prime+\frac{2\pi}{3a}\bar{m}\hat{x}) \frac{1}{2}\sigma^\gamma c(k) \nn\\
&\times \frac{\Delta}{3N\beta} e^{-i\vec{q}^\prime \cdot (j-l) a \hat{x}} \sum_m\sum_{k^\prime} e^{-i\frac{2\pi}{3}(m+\bar{m})(j-l)}
 \langle c^\dagger(k^\prime-q^\prime)\frac{1}{2} \sigma^\gamma c(k+\frac{2\pi}{3a}(m+\bar{m})  \hat{x}) c^\dagger(k^\prime+\frac{2\pi }{3a} (m+\bar{m}) \hat{x})\frac{1}{2}\sigma^\alpha c(\vec{k}^\prime-q^\prime)\rangle_{\text{f}}.
\label{eq:expr2}
\end{flalign}

Notice that in Eq. (\ref{eq:expr2}), the average $\langle\rangle_f$ is non-vanishing only when $\alpha=\gamma$.
Hence the first line in Eq. (\ref{eq:expr2}) is simply $\int d\tau \sum_n h_l^\alpha (\tau,n) S_l^\alpha(\tau,n)$
as can be seen from Eq. (\ref{eq:field_expr2}).
This confirms that the diagram in Fig. \ref{fig:renormalization} indeed renormalizes $h_i^\alpha$.
Since $q^\prime$ is a slow wavevector, we can ignore $q^\prime$ in the remaining part of Eq. (\ref{eq:expr2}) other than the field term.
In summary, we conclude that Eq. (\ref{eq:expr2}) is equal to
\bea
\Delta\lambda_{jl}\delta_{\alpha\gamma} \ln b\cdot \int d\tau \sum_n h_l^\alpha (\tau,n) S_l^\alpha(\tau,n).
\eea
The coefficient is 
\begin{flalign}
\lambda_{jl} \ln b =  -\frac{a}{6}  \sum_m e^{-i\frac{2\pi}{3}m(j-l)}\int_{\Lambda/b}^\Lambda d^2k  \mathcal{G}(k) \mathcal{G}(k+\frac{2\pi }{3a} m \hat{x}), \nn\\
\label{eq:lambda_coefficient}
\end{flalign}
in which at zero temperature the sum over $k^\prime$ is  turned into an integral restricted within the momentum shell $[\Lambda/b,\Lambda]$, 
``$a$" is the lattice constant,
the minus sign comes from the fermion loop, and
\bea
\mathcal{G}(k)=\frac{1}{i\omega-\epsilon(k)}
\eea
is the free fermion Green's function where $\epsilon(k)$ is the dispersion.
In conclusion, the RG flow equation for the scaling field is 
\bea
\frac{dh_{u,i}^\alpha}{d\ln b}=h_{u,i}^\alpha-\lambda_{jl} \Delta h_{u,l}^\alpha,
\eea
in which $\alpha=<ij>$.

%-------------------------------------------- 
\begin{figure}[h]
\includegraphics[width=14cm]{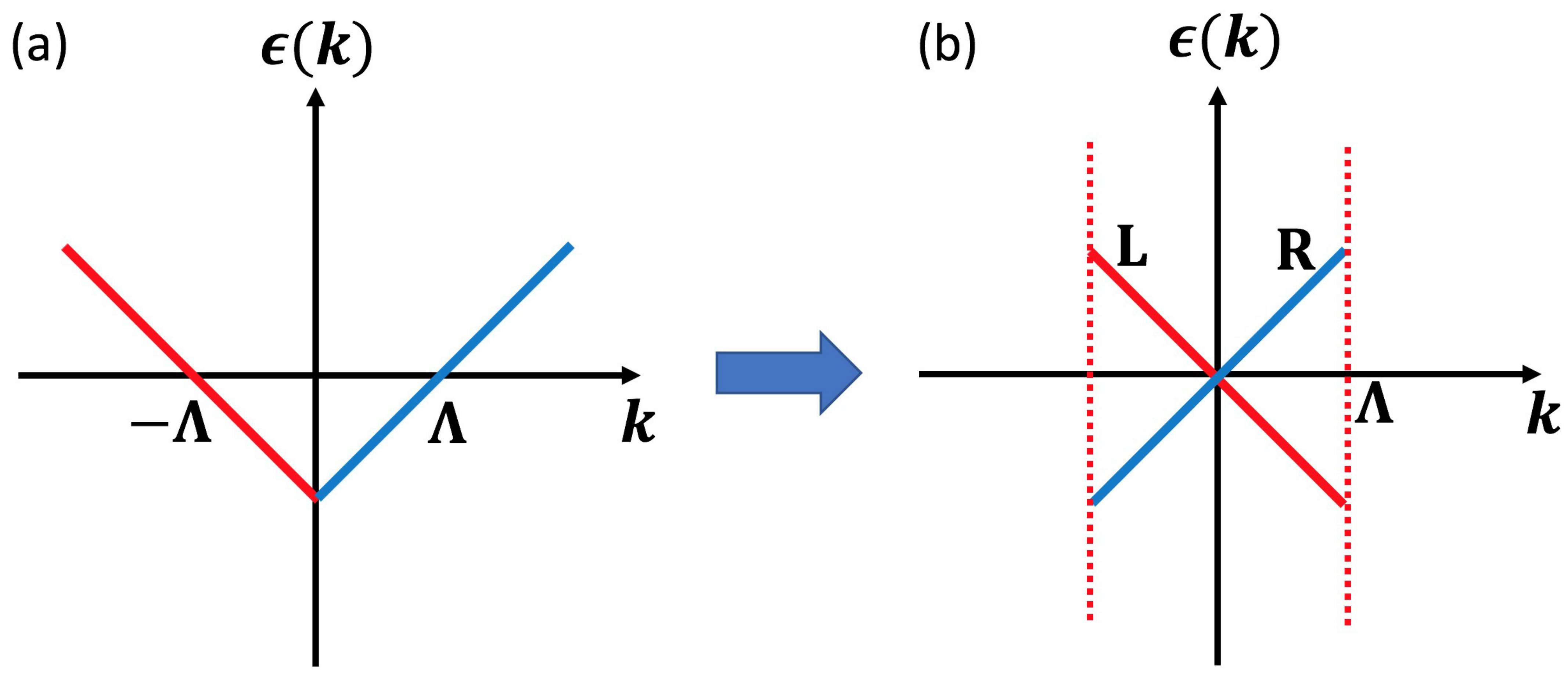}
\caption{Free fermion fixed point dispersion.
} \label{fig:dispersion}
\end{figure}
%--------------------------------------------

Next we proceed to calculating the coefficients $\lambda_{jl}$ in the flow equations.
At the free fermion fixed point, the dispersion is linear as shown in Fig. \ref{fig:dispersion} (a).
We ignore the nonlinear terms in the band structure since they are of higher dimensions hence irrelevant in the vicinity of the free fermion fixed point.
The dispersion can be folded into a Dirac fermion with left and right movers as shown in Fig. \ref{fig:dispersion} (a),
in which the cutoff $\Lambda=\frac{\pi}{2a}$.

In momentum shell RG, the modes $(\omega,\vec{k})$ satisfying $\sqrt{\omega^2+(v\vec{k})^2}\in [\Lambda/b,\Lambda]$
are integrated over, where $v=ta$ is the free fermion velocity and $t$ is the hopping strength. 
By rescaling $(\omega,\vec{k})$ to $(\omega/(v\Lambda),\vec{k}/\Lambda)$,
Eq. (\ref{eq:lambda_coefficient}) becomes
\begin{flalign}
\lambda_{jl}\ln b&=-\frac{1}{6t} \sum_m e^{-i\frac{2\pi}{3}m(j-l)}\ln b \sum_{\nu=\pm 1} \int \frac{d\theta}{4\pi^2}\frac{1}{i\cos\theta-\bar{\epsilon}(\sin\theta+\nu)} \frac{1}{i\cos\theta-\bar{\epsilon}(\sin\theta+\nu+\frac{4}{3}m)},
\end{flalign}
in which $\nu=\pm1$ corresponds to left and right movers, and 
$\bar{\epsilon}(x)= \text{sgn}(x)-1$, $-2\leq x\leq 2$ ($x$ is mod $4$). 
Thus 
\begin{flalign}
\lambda_{jl}&=\frac{1}{24\pi^2t} \sum_{m=0,\pm 1} e^{-i\frac{2\pi}{3}m(j-l)} \sum_{\nu=\pm 1} \int_0^{2\pi} d\theta
\frac{1}{\cos^2\theta+\bar{\epsilon}^2(\sin\theta+\nu)} \frac{1}{\cos^2\theta+\bar{\epsilon}^2(\sin\theta+\nu+\frac{4}{3}m)}\nn\\
&\times \big[ \cos^2\theta -\bar{\epsilon} (\sin \theta+\nu)\bar{\epsilon} (\sin \theta+\nu+\frac{4}{3}m) -i\cos\theta \big(\bar{\epsilon} (\sin \theta+\nu)+\bar{\epsilon} (\sin \theta+\nu+\frac{4}{3}m)\big)
\big].
\end{flalign}
The imaginary part vanishes, as can be seen by performing a change of variable $\theta \rightarrow \pi-\theta$
under which $\sin \theta$ is invariant but $\cos \theta$ changes a sign.
The $m=0$ term also vanishes. 
This is because the $\theta$-integral for both the left and right movers is equal to $\int d\theta (\cos^2\theta-\sin^2\theta)=0$.
Thus we obtain
\bea
\lambda_{jl}&=\frac{1}{24\pi^2t} \sum_{m=\pm1} e^{-i\frac{2\pi}{3}m(j-l)} \sum_{\nu=\pm 1} \int_0^{2\pi} d\theta
\frac{\cos^2\theta -\bar{\epsilon} (\sin \theta+\nu)\bar{\epsilon} (\sin \theta+\nu+\frac{4}{3}m)}{\big[\cos^2\theta+\bar{\epsilon}^2(\sin\theta+\nu)\big]\cdot \big[\cos^2\theta+\bar{\epsilon}^2(\sin\theta+\nu+\frac{4}{3}m)\big]}.
\label{eq:lambda_int}
\eea
Using $\bar{\epsilon}(x)=\bar{\epsilon}(-x)$, it can be further shown that in Eq. (\ref{eq:lambda_int}), 
$(m=1,\nu=1)$ is equal to $(m=-1,\nu=-1)$,
and $(m=1,\nu=-1)$ is equal to $(m=-1,\nu=1)$.
Hence we get
\bea
\lambda_{jl}&=\frac{1}{t} E  \cos(\frac{2\pi}{3}(j-l)), 
\eea
in which
\begin{flalign}
E=&\frac{1}{12\pi^2}\sum_{\nu=\pm 1} \int_0^{2\pi} d\theta
\frac{\cos^2\theta -\bar{\epsilon} (\sin \theta+\nu)\bar{\epsilon} (\sin \theta+\nu+\frac{4}{3})}{\big[\cos^2\theta+\bar{\epsilon}^2(\sin\theta+\nu)\big]\cdot \big[\cos^2\theta+\bar{\epsilon}^2(\sin\theta+\nu+\frac{4}{3})\big]}.
\end{flalign}
The numerical evaluation of $E$ gives $E=0.14$.

The flow equation of $h_1^x$ now becomes
\bea
\frac{dh_{u,1}^x}{d\ln b}&=&h_{u,1}^x-\lambda_{22} \Delta h_{u,2}^x-\lambda_{23} \Delta h_{u,3}^x\nn\\
&=&h_{u,1}^x-E\frac{\Delta}{t} h_{u,2}^x-E\cos(\frac{2\pi}{3})\frac{\Delta}{t}  h_{u,3}^x.
\eea
Comparing with the flow equation in the main text, we see that
\bea
\lambda&=&\frac{1}{t}E>0,\nn\\
\nu&=&\frac{1}{t}E\cos(\frac{2\pi}{3})<0.
\eea

Finally we discuss the flow of the staggered part $h_{s,i}^\alpha$.
We give a quick derivation for the flow equation of $h_{s,i}^\alpha$, 
only highlighting the difference from the derivation for the flow of the uniform part.

By replacing $h_l^\alpha(\tau,n)$ by $(-)^n h_{s,l}^\alpha(\tau,n)$ in Eq. (\ref{eq:field_expr2}),
we obtain
\begin{flalign}
&\int d\tau \sum_n h_l^\alpha (\tau,n) S_l^\alpha(\tau,n)
=\frac{1}{3}\sum_{kq}\sum_m e^{i\vec{q}\cdot la \hat{x} } e^{i\frac{2\pi}{3}(m+\frac{1}{2})l} (-)^l h_{s,l}^\alpha(-q) c^\dagger(k+\frac{2\pi}{3a}m\hat{x} +\frac{\pi}{3a}\hat{x})\frac{1}{2}\sigma^\alpha c(k-q).
\label{eq:field_expr3}
\end{flalign}
In what follows, we will drop the subscript ``s" for simplicity.
The renormalization expression in Eq. (\ref{eq:expression_renorm}) becomes
\begin{flalign}
&\Delta \frac{1}{3N\beta} \sum_m e^{-i\frac{2\pi}{3a} mj} \sum_{kpq} e^{i\vec{q}\cdot(i-j)a\hat{x} } c^\dagger (k+q) \frac{1}{2}\sigma^\gamma c(k) \cdot \frac{1}{3} \sum_{k^\prime q^\prime m^\prime} e^{i\vec{q}^\prime \cdot la\hat{x}} e^{i\frac{2\pi}{3} (m^\prime +\frac{1}{2}) l} (-)^l h_l^\alpha(-\vec{q}^\prime)\nn\\
&\times
\langle c^\dagger(p-q) \frac{1}{2} \sigma^\gamma c(p+\frac{2\pi}{3a}m  \hat{x}) c^\dagger(k^\prime+\frac{2\pi}{3a}  m^\prime\hat{x}+\frac{\pi}{3a} \hat{x})\frac{1}{2}\sigma^\alpha c(\vec{k}^\prime-q^\prime)\rangle_{\text{f}},
\label{eq:expression_renorm2}
\end{flalign}
and the momentum conservations in Eq. (\ref{eq:momentum_conserv}) are now 
\bea
m^\prime &=& m+\bar{m}\nn\\
p&=& k^\prime +\frac{2\pi}{3a} \bar{m}+\frac{\pi}{3a}\nn\\
q&=& q^\prime +\frac{2\pi}{3a} \bar{m}+\frac{\pi}{3a}.
\label{eq:momentum_conserv2}
\eea
Then Eq. (\ref{eq:expr3}) becomes
\begin{flalign}
& \frac{1}{3} \sum_{\bar{m}}e^{i\frac{2\pi}{3a} (\bar{m}+\frac{1}{2})i} (-)^i \sum_{k q^\prime}  e^{i\vec{q}^\prime\cdot ia\hat{x} } h_l^\alpha(-q^\prime) c^\dagger (k+q^\prime+\frac{2\pi}{3a}\bar{m}\hat{x}) \frac{1}{2}\sigma^\gamma c(k) \times \frac{\Delta}{3N\beta}  e^{-i\vec{q}^\prime \cdot (j-l) a \hat{x}} (-)^{i-l} \nn\\
&\cdot \sum_m\sum_{k^\prime} e^{-i\frac{2\pi}{3}(m+\bar{m}+\frac{1}{2})(j-l)}
 \langle c^\dagger(k^\prime-q^\prime)\frac{1}{2} \sigma^\gamma c(k+\frac{2\pi}{3a}(m+\bar{m}+\frac{1}{2})  \hat{x}) c^\dagger(k^\prime+\frac{2\pi }{3a} (m+\bar{m}+\frac{1}{2}) \hat{x})\frac{1}{2}\sigma^\alpha c(\vec{k}^\prime-q^\prime)\rangle_{\text{f}}.
\label{eq:expr3}
\end{flalign}
Instead of Eq. (\ref{eq:lambda_coefficient}), Eq. (\ref{eq:expr3}) leads to an RG coefficient
\begin{flalign}
\lambda_{jl} \ln b =  -\frac{a}{6} (-)^{i-l} \sum_m e^{-i\frac{2\pi}{3}(m+\frac{1}{2})(j-l)}\int_{\Lambda/b}^\Lambda d^2k  \mathcal{G}(k) \mathcal{G}(k+\frac{2\pi }{3a} (m+\frac{1}{2}) \hat{x}). \nn\\
\label{eq:lambda_coefficient2}
\end{flalign}
Correspondingly, Eq. (\ref{eq:lambda_int}) is changed to
\bea
\lambda_{jl}&=\frac{1}{24\pi^2t} (-)^{i-l} \sum_{m=0,\pm1} e^{-i\frac{2\pi}{3}(m+\frac{1}{2})(j-l)} \sum_{\nu=\pm 1} \int_0^{2\pi} d\theta
\frac{\cos^2\theta -\bar{\epsilon} (\sin \theta+\nu)\bar{\epsilon} (\sin \theta+\nu+\frac{4}{3}(m+\frac{1}{2}))}{\big[\cos^2\theta+\bar{\epsilon}^2(\sin\theta+\nu)\big]\cdot \big[\cos^2\theta+\bar{\epsilon}^2(\sin\theta+\nu+\frac{4}{3}(m+\frac{1}{2}))\big]},
\label{eq:lambda_int2}
\eea
where in particular, the $m=0$ term does not vanish in the current situation. 
Define $E_m$ as
\bea
E_m=\frac{1}{24\pi^2}\sum_{\nu=\pm 1} \int_0^{2\pi} d\theta
\frac{\cos^2\theta -\bar{\epsilon} (\sin \theta+\nu)\bar{\epsilon} (\sin \theta+\nu+\frac{4}{3}(m+\frac{1}{2}))}{\big[\cos^2\theta+\bar{\epsilon}^2(\sin\theta+\nu)\big]\cdot \big[\cos^2\theta+\bar{\epsilon}^2(\sin\theta+\nu+\frac{4}{3}(m+\frac{1}{2}))\big]},
\label{eq:Em}
\eea
we have 
\bea
\lambda_{jl}=\frac{1}{t} (-)^{i-l} \sum_{m=0,\pm1} e^{-i\frac{2\pi}{3}(m+\frac{1}{2})(j-l)} E_m.
\eea
The values of $E_m$ can be evaluated numerically as
\bea
E_0=-0.069, E_{+1}=0.053,E_{-1}=-0.069.
\eea
Notice that $\lambda_{jl}$ only depends on $j-l$, where $j-l =0,\pm1 \mod 3$.
It is straightforward to obtain $\lambda_{ii}=-0.039/t$, $\lambda_{i,i\pm 1}=0.060/t$.

%The derivation of the flow equation is exactly similar to $h_{u,i}^\alpha$
%except that now $\frac{2\pi}{3m}$ should be replaced by $\frac{2\pi}{3m}+\pi$.
%The flow equation now is
%\bea
%\frac{dh_{s,i}^\alpha}{d\ln b}=h_{s,i}^\alpha-\lambda^\prime_{jl} \Delta h_{s,l}^\alpha,
%\eea
%in which 
%\bea
%\lambda^\prime_{jl}=\frac{1}{t} E^\prime  \cos(\frac{2\pi}{3}(j-l)), 
%\eea
%where
%\begin{flalign}
%E^\prime=&\frac{1}{12\pi^2}\sum_{\nu=\pm 1} \int_0^{2\pi} d\theta
%\frac{\cos^2\theta -\bar{\epsilon} (\sin \theta+\nu)\bar{\epsilon} (\sin \theta+\nu+\frac{10}{3})}{\big[\cos^2\theta+\bar{\epsilon}^2(\sin\theta+\nu)\big]\cdot \big[\cos^2\theta+\bar{\epsilon}^2(\sin\theta+\nu+\frac{10}{3})\big]}.
%\end{flalign}
%The numerical evaluation of $E^\prime$ gives $E^\prime=-0.014$.
%Thus we see that the slope of $C_2/C_1$ around $\Delta=0$
%is opposite to that of $D_2/D_1$, 
%which is consistent with the DMRG numerics.
%Furthermore, since the magnitude of $E^\prime$ is one order smaller than $E$,
%the magnitude of the slope of $C_2/C_1$ should be much less than that of $D_2/D_1$,
%which is again consistent with the numerics.

%%%%%%%%%%%%%%%%%%%
\subsection{Solving the flow equations}
\label{app:flow_solve}

In this section, we solve the flow equations
\bea
\frac{d h_{1}^x}{dt} &=& (1-\nu \Delta)h_{1}^x - \lambda \Delta h_{2}^x- \nu \Delta h_{3}^x,\nn\\
\frac{d h_{2}^x}{dt} &=& (1-\nu \Delta)h_{2}^x - \lambda \Delta h_{1}^x- \nu \Delta h_{3}^x,\nn\\
\frac{d h_{3}^x}{dt} &=& h_{3}^x,
\label{eq:flow}
\eea
in which $t=\ln b$, and the subscripts $u,s$ are dropped in $h$ for simplicity.
The initial conditions are $h_j^x(t=0)=h_j^x(0)$, $j=1,2,3$.

The equation for $h_3^x(t)$ is easily solved as
\bea
h_3^x(t)=e^th_3^x(0).
\eea
The sum and difference of the equations for $h_1^x,h_2^x$ are
\bea
\frac{d(h_1^x+h_2^x)}{dt} &=& (1-(\nu+\lambda)\Delta) (h_x^x+h_2^x)-2\nu\Delta h_3^x,\nn\\
\frac{d(h_1^x-h_2^x)}{dt} &=& (1-(\nu-\lambda)\Delta) (h_x^x-h_2^x).
\label{eq:h1h2}
\eea
The second equation in Eq. (\ref{eq:h1h2}) can be readily solved as
\bea
h_1^x(t)-h_2^x(t)= e^{(1-(\nu-\lambda)\Delta )t} (h_1^x(0)-h_2^x(0)).
\eea
To solve the first equation in Eq. (\ref{eq:h1h2}), let 
\bea
h_1^x(t)+h_2^x(t)=u(t) e^{(1-(\nu+\lambda)\Delta )t}.
\eea
The equation for $u(t)$ is
\bea
\frac{du}{dt}=-2\nu\Delta h_3^x(0) e^{(\lambda+\nu)\Delta t},
\eea
which can be solved as
\bea
u(t)& = h_1^x(0)+h_2^x(0) +\frac{2\nu}{\nu+\lambda} h_3^x(0) (1-e^{(\nu+\lambda)\Delta t}).
\eea

Keeping terms only up to first order in $\Delta$, we get
\bea
h_1^x(b)&=& b\big[(1-\nu\Delta \ln b)h_1^x(0)-\lambda\Delta \ln b h_2^x(0)-\nu\Delta\ln b h_3^x(0) \big],\nn\\
h_2^x(b)&=& b\big[(1-\nu\Delta \ln b)h_2^x(0)-\lambda\Delta \ln b h_1^x(0) -\nu\Delta\ln b h_3^x(0) \big],\nn\\
h_3^x(b)& =& bh_3(0).
\eea

%%%%%%%%%%%%%%%%%%%
\section{More on AFM phase}

%%%%%%%%%%%%%%%%%%%
\subsection{DMRG numerics}

Our numerics show that periodic boundary conditions (PBC) are more efficient than open boundary conditions (OBC) in demonstrating 
the $1/r^2$ behavior and the logarithmic corrections.
We stress that although 
DMRG simulations are more challenging with PBC,
in the AFM phase a choice of $L=144$ sites with PBC is sufficient for the purpose of demonstrating the 
modified nonabelian bosonization formula.
To reach numerical convergence,
up to $m=1000$ DMRG states were kept and tens of finite size sweeps were performed with a  truncation error of $10^{-6}$.

%%%%%%%%%%%%%%%%%%%
\subsection{Numerical study on the $\pi/3$-oscillating components of the correlation functions}

%-------------------------------------
\begin{figure}[h]
\includegraphics[width=14cm]{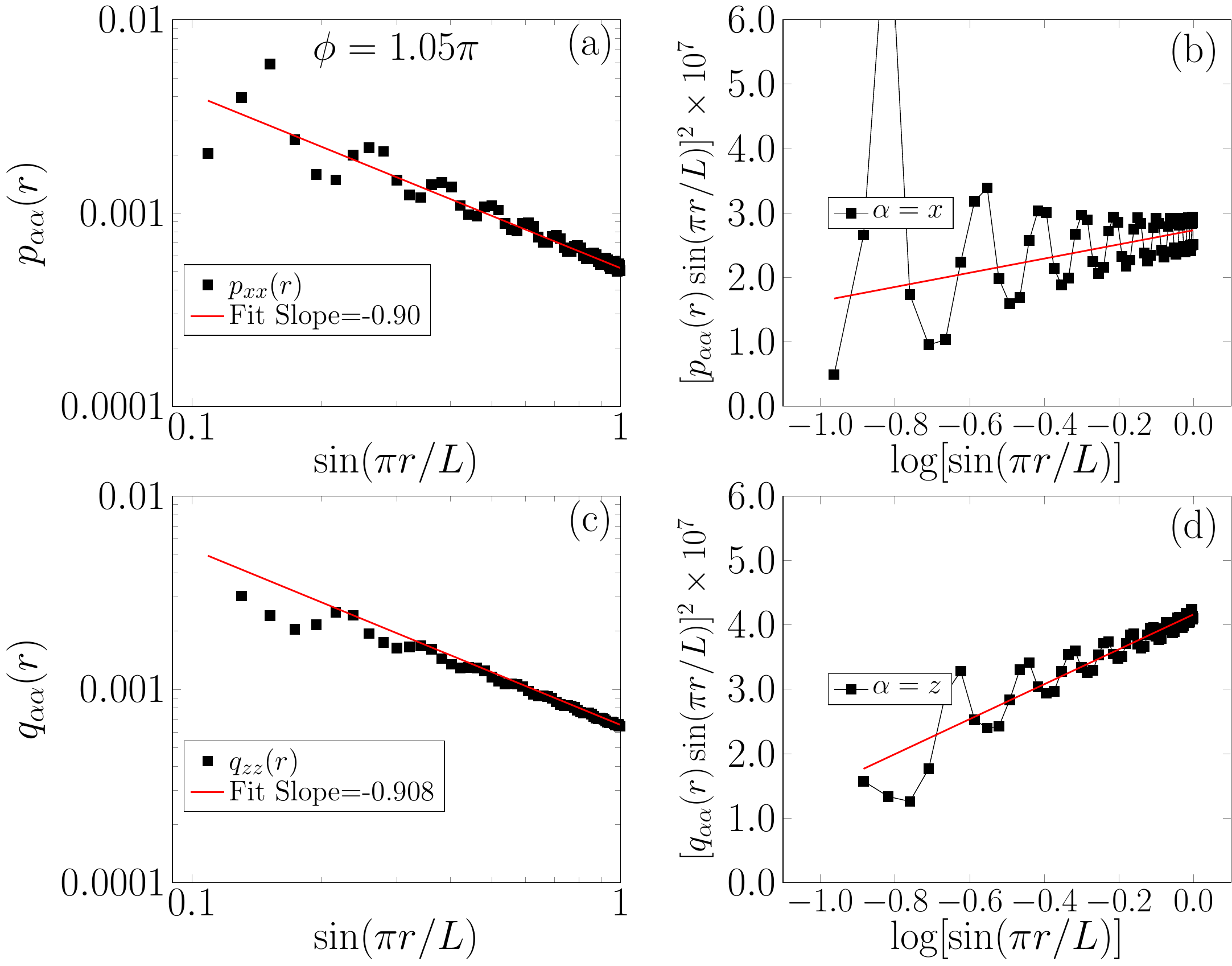}
\caption{(a) $p_{xx}(r)$ and (c) $q_{zz}(r)$ vs. $\sin(\pi r/L)$ on a log-log scale;
(b) $\big(p_{xx}(r) \sin(\pi r/L)\big)^2$ and (d) $\big(q_{zz}(r) \sin(\pi r/L)\big)^2$ plotted against $\ln\big(\sin(\pi r/L)\big)$.
DMRG numerics are carried out on a system of $L=144$ sites with periodic boundary conditions at $\phi=1.05\pi$.
$p_{\alpha\alpha}(r)$ and $q_{\alpha\alpha}(r)$ are then extracted from $\langle S_1^\alpha S_{1+r}^\alpha\rangle$ using the nine-point formula.
} 
\label{fig:2suppl}
\end{figure}
%-------------------------------------

In this section, we study the momentum $\pm\pi/3$ oscillating components of the spin-spin correlation functions.
An angle $\phi=1.05\pi$ is chosen as a representative example.
The correlation functions $\langle S_1^\alpha S_{1+r}^\alpha\rangle$ are calculated from DMRG numerics on a system of $L=144$ sites with a periodic boundary condition. As throughout the manuscript, to reach numerical convergence, up to $m=1000$ DMRG states were kept and tens of finite size sweeps were performed with a final truncation error of $10^{-6}$.
We will neglect the uniform and momentum $\pm 2\pi/3$ oscillating components of the correlation functions, 
since they decay faster than the staggered and momentum $\pm \pi/3$ oscillating components at long distances.
%Using the notation introduced in Sec. \ref{sec:ninepoint},
We denote the staggered component as $s_{\alpha\alpha}(r)$ and the two momentum $\pm \pi/3$ oscillating components as $p_{\alpha\alpha}(r)$ and $q_{\alpha\alpha}(r)$. 
We expect that all of these nine correlation functions $s_{\alpha\alpha},p_{\alpha\alpha},q_{\alpha\alpha}$ ($\alpha=x,y,z$) behave as $\sim \ln^{1/2}\big(\sin(\pi r/L)\big)/\sin(\pi r/L)$
at long distances.
Since $s_{\alpha\alpha}$ has already been studied in the maintext, here we focus on $p_{\alpha\alpha}$ and $q_{\alpha\alpha}$.
A representative direction $\alpha=x$ is chosen for $p_{\alpha\alpha}$ and  $\alpha=z$ is chosen for $q_{\alpha\alpha}$.

In Fig. \ref{fig:2suppl} (a) and (c), $p_{xx}(r)$ and $q_{zz}(r)$ are plotted against $\sin(\pi r/L)$ on a log-log scale,
both exhibiting a good linear relation with a slope $\sim -0.9$ which is close to $-1$ within $10\%$ error.
Due to the logarithmic correction, it is expected that the observed exponent is slightly smaller than the predicted value $1$. 
To study the logarithmic factor, in Fig. \ref{fig:2suppl} (b) and (d),
$\big(p_{xx}(r) \sin(\pi r/L)\big)^2$ and $\big(q_{zz}(r) \sin(\pi r/L)\big)^2$ are plotted against $\log\big(\sin(\pi r/L)\big)$.
If the logarithmic factor is $\ln^{1/2}\big(\sin(\pi r/L)\big)$, then a linear relation will be observed.
We see from Fig. \ref{fig:2suppl} (b) and (d) that the linearity is not good due to an oscillation with a six-site periodicity.
In fact, such six-site oscillation is not unexpected.
When applying the nine-point formula, the uniform component and the momentum $\pm 2\pi/3$ components are neglected.
These naturally introduce oscillations into the extracted values of $s_{\alpha\alpha},p_{\alpha\alpha},q_{\alpha\alpha}$ with a six-site periodicity. 
On the other hand, since $C_2/C_1$ is still close to $1$ even very far away from the SU(2) symmetric point $\phi=5\pi/4$,
$s_{\alpha\alpha}$ dominates over $p_{\alpha\alpha},q_{\alpha\alpha}$. 
The smallness of $p_{\alpha\alpha},q_{\alpha\alpha}$ means that they are more sensitive to the influence of the uniform and the momentum $\pm 2\pi/3$ components.
Indeed, Fig. 3 (d) in the maintext also contains oscillations, but much less prominent than those in Fig. \ref{fig:2suppl} (b) and (d).
We expect that the oscillations in Fig. \ref{fig:2suppl} (b) and (d) can be reduced by going to larger system sizes.

%%%%%%%%%%%%%%%%%%%
\subsection{Finite size scaling of the exponents for the staggered parts of the correlation functions}

\begin{figure}[h]
\includegraphics[width=18cm]{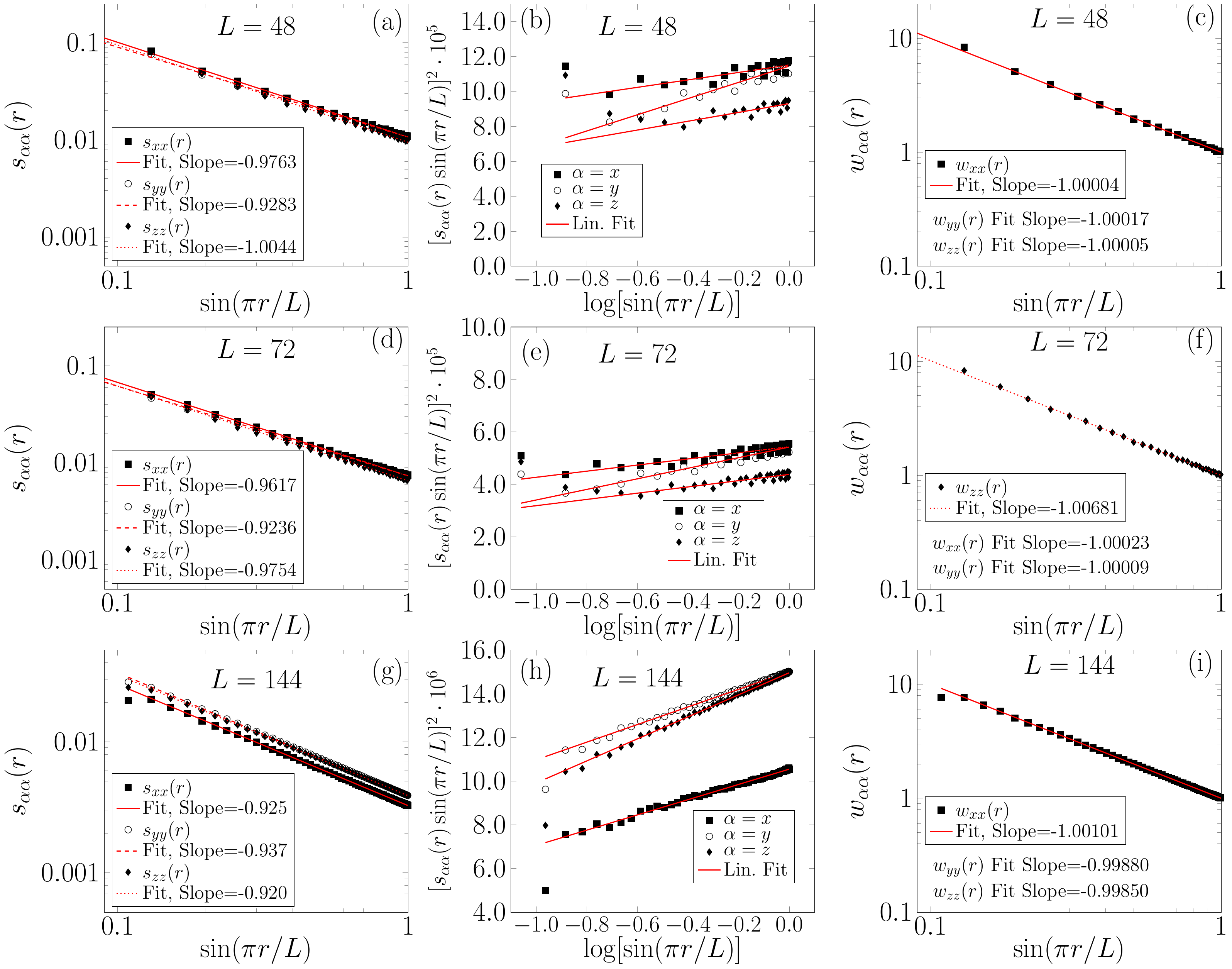}
\caption{$s_{\alpha\alpha}(r)$ ($\alpha=x,y,z$) vs. $\sin(\pi r/L)$ on a log-log scale for a finite size system of (a) 48, (d) 72 and (g) 144 sites; 
$[s_{\alpha\alpha}(r) \sin(\pi r/L)]^2$ vs. $\log (\sin(\pi r/L))$ for a finite size system of (b) 48, (e) 72 and (h) 144 sites; 
$w_{\alpha\alpha}(r)$ vs. $\sin(\pi r/L)$ on a log-log scale for a finite size system of (c) 48, (f) 72 and (i) 144 sites, where $w_{\alpha\alpha}(r)$ is defined as $s_{\alpha\alpha}(r)$ divided by the logarithmic factor as explained in the text.
In (a,b,d,e,g,h), $s_{\alpha\alpha}(r)$ is the staggered part of the correlation function $\mathopen{<}S_{1}^\alpha S_{r+1}^\alpha\mathclose{>}$ extracted from a three-point formula as explained in the main text. 
In (c,f,i), all three curves of $w_{\alpha\alpha}$ for $\alpha=x,y,z$ collapse, hence the data of only one spin direction are shown for each system size. 
Periodic boundary conditions are used in DMRG numerical computations.
}
\label{fig:fig3_fin}
\end{figure}

In this section, we study the finite size scaling behaviors of the exponents for the staggered components $s_{\alpha\alpha}(r)$ ($\alpha=x,y,z$) of the spin-spin correlation functions $\mathopen{<}S_1^\alpha S_{r+1}^\alpha\mathclose{>}$ varying system size. 
The DMRG numerical results are shown in Fig. \ref{fig:fig3_fin},
where $s_{\alpha\alpha}$ is extracted from the numerically calculated $\mathopen{<}S_1^\alpha S_{r+1}^\alpha\mathclose{>}$ using a three-point formula as explained in the main text. 
We show that by properly dividing $s_{\alpha\alpha}$ by the factor of the logarithmic correction, the exponents are close to $-1$ with remarkable precisions (error less than $0.2\%$) for all three system sizes $L=48,72,144$.
Thus the exponent nearly has no dependence on the system size at least when $L$ is greater than $48$.

In Fig. \ref{fig:fig3_fin} (a, d, g), $s_{\alpha\alpha}$ vs. $\sin(\pi r/L)$ are plotted on a log-log scale for a system of $L=48,72,144$ sites, respectively. 
It can seen that the slopes are already very close to $-1$ (with an error within $\sim 8\%$) for all three system sizes. 
The small deviations from $-1$ is due to the logarithmic factor arising from the marginally irrelevant operator $-\lambda \vec{J}_L\cdot \vec{J}_R$ as explained in the main text, where $\lambda>0$ is the coupling constant.

To get a better exponent, we will extract the logarithmic factor first, and then divide $s_{\alpha\alpha}$ by the extracted logarithmic factor.
In Fig. (b, e, h), $[s_{\alpha\alpha}(r) \sin(\pi r/L)]^2$ vs. $\log (\sin(\pi r/L))$ are plotted for a system of $L=48,72,144$ sites, respectively. 
The data can be fitted with a linear relation $[s_{\alpha\alpha}(r) \sin(\pi r/L)]^2=A_{\alpha} \log (\frac{L}{\pi} \sin(\pi r/L))+B_{\alpha}$,
and the good linear fits  indicate logarithmic factors with a power of $1/2$ consistent with Eq. (4) in the main text.
Next define $w_{\alpha\alpha}$ as 
\bea
w_{\alpha\alpha}(r)=\frac{s_{\alpha\alpha}(r)}{\sqrt{A_{\alpha} \log (\frac{L}{\pi}\sin(\frac{\pi r}{L}))+B_{\alpha}}},
\eea
so that the logarithmic factor is removed from $s_{\alpha\alpha}$.
In Fig. 3 (c, f, i), $w_{\alpha\alpha}(r)$ vs. $\sin(\pi r/L)$ are plotted on a log-log scale for a system of $L=48,72,144$ sites, respectively, all exhibiting slopes close to $-1$ with remarkable precisions (less than $0.2\%$). 
Therefore we see that essentially there is no finite size dependence of the exponent.

%%%%%%%%%%%%%%%%%%%
\subsection{Finite size scaling of $C_1/C_2$}
\label{app:scaling}

%--------------------------------------------------------
\begin{figure*}[h]
\includegraphics[width=\textwidth]{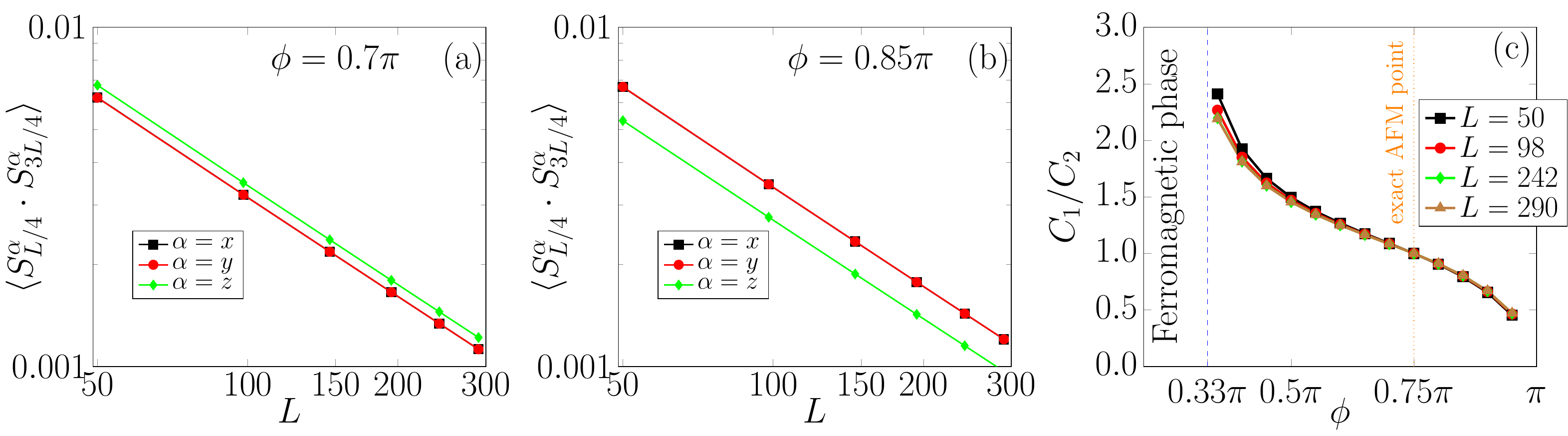}
\caption{$\langle S_{L/4}^\alpha S_{3L/4}^\alpha \rangle$ ($\alpha=x,y,z$) as functions of system size $L$
in a log-log plot at (a) $\phi=0.7\pi$ and (b) $\phi=0.85\pi$,
 and (c) $C_1/C_2$ extracted from the spin-spin correlations
as a function of $\phi$ within the AFM phase.}
\label{fig:4}
\end{figure*}
%--------------------------------------------------------

In this section, we study the dependence of the ratio $C_1/C_2$ on the system size for different angles $\phi$. 
Here we use a method independent from the one used in the main text. 
An open boundary condition is adopted here,
and spin-spin correlations are evaluated between the sites at $r_1=L/4$ and $r_2=3L/4$.
To reach numerical convergence, 
$m=800$ DMRG states are used and up to 10 finite size sweeps are performed with a final truncation error of $10^{-9}$.

Given the above setup, $\langle S_{r_1}^\alpha S_{r_2}^\alpha \rangle$
are computed for different $L$.
As shown in Fig. \ref{fig:4}, when displayed in a log-log scale, a perfect linear behavior is observed as in the main text for the staggered part of the correlation functions.
Similarly to Fig. 3 in the maintext, 
$x$ and $y$ correlations numerically coincide at large distances, 
while the $z$ correlation appears as a parallel straight line with approximately 
the same slope but different intercept compared with the other two correlations. 

By extrapolating the intercepts with the y-axis, the ratio $ C_1/C_2$ can be extracted. 
Fig. \ref{fig:4} (c) shows a very weak dependence of $C_1/C_2$ on 
system sizes for all the values of $\phi$ within the AFM phase of the model. 
We have verified that the results are consistent with those obtained using chains with 
periodic boundary conditions, and therefore provide further evidence that the numerical results 
do not depend on the choice of boundary conditions.

%%%%%%%%%%%%%%%%%%%
\section{The FM phase}

In this section, we discuss the ordered phase. 
Due to the equivalence established by $\mathcal{T}_3$ as mentioned in the main text, 
we will use interchangeably the angles $\phi$ and $2\pi-\phi$ and do not distinguish between the two.

%%%%%%%%%%%%%%%%%%%%%%%%%%%%%
\subsection{Spin orientations with $O_h\rightarrow D_4$ symmetry breaking}

We discuss the $O_h\rightarrow D_4$ symmetry breaking.
We show that the spin orientations 
\bea
\langle\vec{S}_1\rangle=S^\prime \hat{z}, \langle\vec{S}_2\rangle=S^{\prime\prime} \hat{z}, \langle\vec{S}_3\rangle=S^{\prime\prime} \hat{z},  
\label{eq:spin_orient2}
\eea
are invariant under $\mathopen{<}R_aT_a R(\hat{z},\pi)R_I I R(\hat{z},\pi), T(R_aT_a)^{-1}R_I I R_aT_a\mathclose{>}\cong D_4$,
where $D_4$ is the symmetry group of a square.

%-------------------------------------------- 
\begin{figure}[h]
\includegraphics[width=7.5cm]{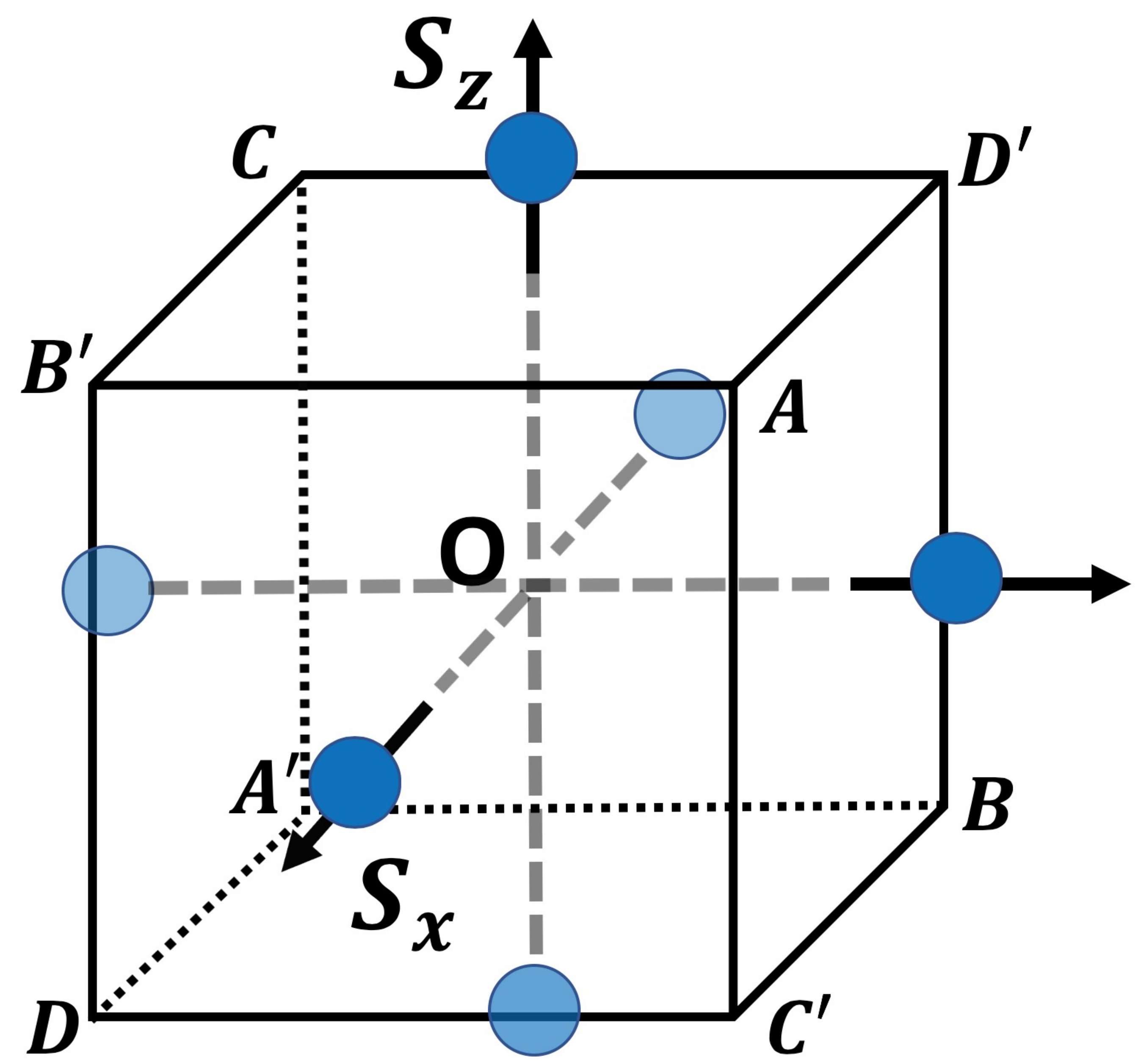}
\caption{"Center of mass" directions represented by the six blue solid circles for the $O_h\rightarrow D_4$ symmetry breaking.
} \label{fig:order}
\end{figure}
%--------------------------------------------

Consider a general spin configuration within a unit cell,
\bea
\vec{S}_1=\left(\begin{array}{c}
x_1\\
y_1\\
z_1
\end{array}
\right),
\vec{S}_2=\left(\begin{array}{c}
x_2\\
y_2\\
z_2
\end{array}
\right),
\vec{S}_3=\left(\begin{array}{c}
x_3\\
y_3\\
z_3
\end{array}
\right). 
\label{eq:Spin_general}
\eea
Under $ R_aT_a \cdot R(\hat{z},\pi)\cdot R_I I \cdot R(\hat{z},\pi)$, Eq. (\ref{eq:Spin_general}) is mapped to 
\bea
R_aT_a\cdot  R(\hat{z},\pi) \cdot  R_I I  \cdot R(\hat{z},\pi) : &\vec{S}_1\rightarrow \left(\begin{array}{c}
-x_1\\
-y_1\\
z_1
\end{array}
\right) 
\rightarrow \left(\begin{array}{c}
z_3\\
y_3\\
-x_3
\end{array}
\right)
\rightarrow \left(\begin{array}{c}
 z_3\\
 -y_3\\
 x_3
\end{array}
\right)
\rightarrow \left(\begin{array}{c}
 y_1\\
 -x_1\\
 z_1
\end{array}
\right)\nn\\
&\vec{S}_2\rightarrow \left(\begin{array}{c}
-x_2\\
-y_2\\
z_2
\end{array}
\right) 
\rightarrow \left(\begin{array}{c}
z_2\\
y_2\\
-x_2
\end{array}
\right)
\rightarrow \left(\begin{array}{c}
 z_2\\
 -y_2\\
 x_2
\end{array}
\right)
\rightarrow \left(\begin{array}{c}
 y_3\\
 -x_3\\
 z_3
\end{array}
\right)\nn\\
&\vec{S}_3\rightarrow \left(\begin{array}{c}
-x_3\\
-y_3\\
z_3
\end{array}
\right) 
\rightarrow \left(\begin{array}{c}
z_1\\
y_1\\
-x_1
\end{array}
\right)
\rightarrow \left(\begin{array}{c}
 z_1\\
 -y_1\\
 x_1
\end{array}
\right)
\rightarrow \left(\begin{array}{c}
 y_2\\
 -x_2\\
 z_2
\end{array}
\right),
\eea
in which the arrows indicate subsequent applications of the operators in $R_aT_a\cdot  R(\hat{z},\pi) \cdot  R_I I  \cdot R(\hat{z},\pi)$ separated by dot.
Under $T\cdot (R_aT_a)^{-1}\cdot R_I I \cdot R_aT_a$, Eq. (\ref{eq:Spin_general}) is mapped to 
\bea
T\cdot (R_aT_a)^{-1}\cdot R_I I \cdot R_aT_a: &\vec{S}_1\rightarrow \left(\begin{array}{c}
z_2\\
x_2\\
y_2
\end{array}
\right)
\rightarrow \left(\begin{array}{c}
-x_2\\
-z_2\\
-y_2
\end{array}
\right) 
\rightarrow \left(\begin{array}{c}
-y_1\\
-x_1\\
-z_1
\end{array}
\right)
\rightarrow \left(\begin{array}{c}
y_1\\
x_1\\
z_1
\end{array}
\right),\nn\\
&\vec{S}_2\rightarrow \left(\begin{array}{c}
z_3\\
x_3\\
y_3
\end{array}
\right)
\rightarrow \left(\begin{array}{c}
-x_1\\
-z_1\\
-y_1
\end{array}
\right)
\rightarrow \left(\begin{array}{c}
-y_3\\
-x_3\\
-z_3
\end{array}
\right)
\rightarrow \left(\begin{array}{c}
y_3\\
x_3\\
z_3
\end{array}
\right),\nn\\
&\vec{S}_3\rightarrow \left(\begin{array}{c}
z_1\\
x_1\\
y_1
\end{array}
\right)
\rightarrow \left(\begin{array}{c}
-x_3\\
-z_3\\
-y_3
\end{array}
\right)
\rightarrow \left(\begin{array}{c}
-y_2\\
-x_2\\
-z_2
\end{array}
\right)
\rightarrow \left(\begin{array}{c}
y_2\\
x_2\\
z_2
\end{array}
\right).\nn\\
\eea
Clearly, Eq. (\ref{eq:spin_orient2}) is invariant under $R_aT_a\cdot R(\hat{z},\pi)\cdot R_I I\cdot R(\hat{z},\pi)$ and $T\cdot (R_aT_a)^{-1}\cdot R_I I\cdot R_aT_a$.
In addition, the invariant spin configurations under both operations can only be of the form given in Eq. (\ref{eq:spin_orient2}).

Next we  prove that $\mathopen{<}R_aT_a R(\hat{z},\pi)R_I I R(\hat{z},\pi), T(R_aT_a)^{-1}R_I I R_aT_a\mathclose{>}$ is isomorphic to $D_4$.
The generator-relation representation of $D_4$ is 
\bea
D_4=\mathopen{<}a,b|a^4=b^2=(ab)^2=e\mathclose{>}.
\label{eq:D8}
\eea
We make the following identification:
$a= R_aT_a \cdot R(\hat{z},\pi)\cdot R_I I \cdot R(\hat{z},\pi)$, and
$b= T\cdot (R_aT_a)^{-1}\cdot R_I I \cdot R_aT_a$,
We show that $a$ and $b$ satisfy the relations in Eq. (\ref{eq:D8}).
Since the actions of $a$ and $b$ in the spin space are $R(\hat{z},\pi/2)$ and the reflection to the plane $ACA^\prime C^\prime$ shown in Fig. \ref{fig:order}, respectively, 
it is straightforward to verify that the relations in Eq. (\ref{eq:D8})
are satisfied by restricting the actions to the spin space.
Then it is enough to verify the relations for the spatial components.
Firstly, for $a^4$ we have
\bea
(T_aI)^4=T_a(IT_aI)T_a(IT_aI)=T_aT_{-a}T_aT_{-a}=e.
\eea
Secondly, for $b^2$, we have
\bea
(T_a^{-1}IT_a)^2=T_{-a}I(T_aT_{-a})IT_a=T_{-a}I^2T_a=T_{-a}T_a=e.
\eea
Thirdly, for $(ab)^2$, we have
\bea
(T_aIT_a^{-1}IT_a)^2=T_{3a}^2,
\eea
which is $e$ modulo $T_{3a}$.
This shows that $\mathopen{<}R_aT_a R(\hat{z},\pi)R_I I R(\hat{z},\pi), T(R_aT_a)^{-1}R_I I R_aT_a\mathclose{>}$ is isomorphic to a subgroup of $D_4$.
On the other hand, by only considering the actions within the spin space, 
one can show that there are at least eight elements within the group $\mathopen{<}R_aT_a R(\hat{z},\pi)R_I I R(\hat{z},\pi), T(R_aT_a)^{-1}R_I I R_aT_a\mathclose{>}$.
Thus we conclude that it is isomorphic to $D_4$.

%%%%%%%%%%%%%%%%%%%
\subsection{ED results on the ground state degeneracies}
\label{app:degeneracy}

The model is equivalent to a ferromagnetic Heisenberg model at $\phi=0.25\pi$. 
We have verified numerically that the ferromagnetic phase extends 
in the region $0.12\pi \lesssim\phi< \phi_c\approx 0.33\pi$. 
Therefore, without loss of generality, 
in this section we investigate the low energy
properties of the model for $\phi=0.2\pi$.

%In the entire region $0<\phi\lessapprox0.33\pi$, classical mean field calculations 
%predict a gapped ferromagnetic phase, with a characteristic symmetry 
%breaking pattern from $O_h$ to $D_6$, corresponding to a degenerate 
%ground state subspace with dimension 8. Geometrically, this is equivalent 
%to the number of corners of a cube, pointing
%to the directions $\hat{n}_{\alpha}=\sqrt{1/3}(\pm 1,\pm 1, \pm 1)$.

%In our numerical analysis, we foused our attention on $\phi=0.2\pi$ but verified
%that the properties elucidated below are valid in the entire region $0<\phi\lessapprox 0.33\pi$.

\begin{table*}[h]
                \begin{tabular}[t]{|c|c|c|c|c|}
                \hline
		 & h=0 &  $h_{\hat{y}}=5\times10^{-6}$ & $h_{\hat{n}_a}=5\times10^{-6}$ & $h_{\hat{n}_I}=5\times10^{-6}$\\ \hline 
		1 &\tikzmark{top left 1}-3.54113& \tikzmark{top left 3}-3.54118\tikzmark{bottom 
                right 3} & \tikzmark{top left 5}-3.54118&\tikzmark{top left 7}-3.54118\\ 
		2 &-3.54113& -3.54113 &-3.54118 &-3.54118\tikzmark{bottom right 7}\\
		3 &-3.54113& -3.54113 &-3.54118\tikzmark{bottom right 5}& -3.54113 \\ 
		4 &-3.54113&-3.54113 &-3.54108 &-3.54113 \\ 
		5 &-3.54113& -3.54113 &-3.54108 & -3.54108\\
		6 &-3.54113\tikzmark{bottom right 1}&-3.54108 & -3.54108 & -3.54108\\ 
		7 & -3.53799 & -3.53801& -3.53802 & -3.53802\\ 
		8 & -3.53799 &-3.53798 & -3.53799 & -3.53798\\  \hline
                \end{tabular}
                \DrawBox[ultra thick, blue]{top left 1}{bottom right 1}
                \DrawBox[ultra thick, blue]{top left 3}{bottom right 3}
                \DrawBox[ultra thick, blue]{top left 5}{bottom right 5}
                \DrawBox[ultra thick, blue]{top left 7}{bottom right 7}
                \hfill		
\caption{Energies of several lowest lying states computed with 
Lanczos Exact Diagonalization. 
Numerics are performed on a system containing $L=21$ sites with a periodic boundary condition. }
\label{Table:energy}
\end{table*}

ED on spin $S=1/2$ chain with periodic 
boundary conditions finds a degenerate ground state subspace with 
dimension 6 at zero field as shown by the blue box under the $h=0$ column in Table \ref{Table:energy}. 
This subspace is separated from the 
excited states by a relatively small gap $\sim 10^{-3}$ for a chain with length
$L=21$ sites.
These results are compatible with a symmetry breaking pattern from $O_h$ to $D_4$.
The 6-fold degeneracy of $O_h\rightarrow D_4$ symmetry breaking is equivalent to the number of faces of a cube shown in Fig. \ref{fig:order}, 
with normal directions pointing along the cartesian axes directions
$\hat{\alpha}=\pm\hat{x},\pm\hat{y},\pm\hat{z}$. 

To further test the symmetry breaking patterns, we apply a small uniform magnetic field
in such a way that the low lying sextet do not hybridize with excited states above 
the gap in the spectrum. In particular, 
we have applied a magnetic field of strength 
$h_{\hat{y}}=5\times 10^{-6}$ along the $y$-direction such that the constraint 
$10^{-6}\ll h_{\hat{y}} L\simeq 10^{-4}\ll 10^{-3}$ is fulfilled.
The column of $h_{\hat{y}}$ in Table \ref{Table:energy} shows that there is a unique ground state separated by a gap 
$E_2-E_1\simeq 5\times 10^{-5}$ from a quartet of excited states. 
Above the first six states, there is a gap $E_7-E_6\simeq 10^{-3}$ 
as without field, showing that the field is just acting within
the low energy sextet. 

We also apply a small magnetic field along the direction of one of the eight corners of the cube,
$h_{\hat{n}}=\frac{1}{\sqrt{3}}(1,1,1)$. 
As shown in the column $h_{\hat{n}_a}$ in  Table \ref{Table:energy}, the 6-fold degenerate manifold splits 
into two triplet of states: Three states
pointing to the directions $+\hat{x},+\hat{y},+\hat{z}$ and the other three to
opposite direction $-\hat{x},-\hat{y},-\hat{z}$. 
This is again consistent with the symmetry breaking pattern shown in Fig. \ref{fig:order}.

We finally apply a small magnetic field point to the middle point of one of the twelve edges of the cube,
$h_{\hat{n}_I}=\frac{1}{\sqrt{2}}(1,0,-1)$. 
As shown in the column $h_{\hat{n}_I}$ in  Table \ref{Table:energy}, the 6-fold degenerate manifold splits 
into three doublets of states: Two states
pointing to the directions $+\hat{x},-\hat{z}$, 
two states pointing to $+\hat{y},-\hat{y}$,
and two states pointing to $-\hat{x},\hat{z}$.
This is again consistent with the symmetry breaking pattern shown in Fig. \ref{fig:order}.

%%%%%%%%%%%%%%%%%%%%%%%%%%%%%%%%%%%
\subsection{DMRG results on the correlation functions in the FM phase}

%-------------------------------------
\begin{figure}[h]
\includegraphics[width=8cm]{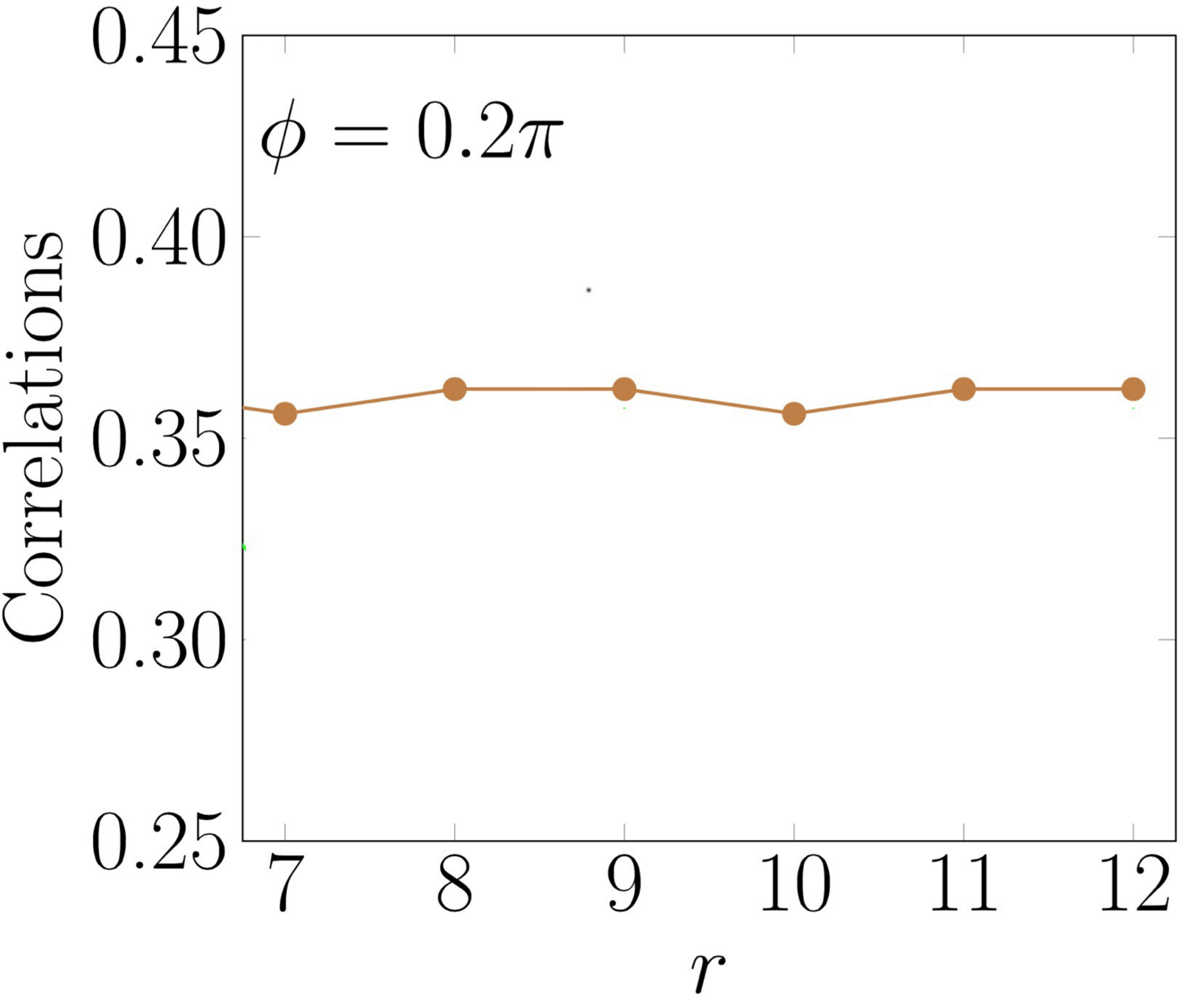}
\caption{$\langle S_1^z S_{r+1}^z\rangle$ at $h_z=10^{-4}$ along $\hat{z}$.
All other correlation functions vanish including $\langle S_1^\alpha S_{r+1}^\alpha\rangle$ ($\alpha=x,y$) and all cross correlations.
Hence, they are not displayed.
} 
\label{fig:corr_FM}
\end{figure}
%-------------------------------------

We have numerically computed the correlation functions under different fields using DMRG numerics on a system with $L=24$ sites with a periodic boundary condition.
Fig. \ref{fig:corr_FM} shows $\langle S_1^z S_{r+1}^z\rangle$ with $h_z=10^{-4}$ along $\hat{z}$,
and the pattern is consistent with Eq. (\ref{eq:spin_orient2}).
All other correlation functions vanish, including $\langle S_1^x S_{r+1}^x\rangle$, $\langle S_1^y S_{r+1}^y \rangle$,
and all cross correlations $\langle S_1^\alpha S_{r+1}^\beta\rangle$ ($\alpha\neq \beta$). 
The results are again consistent with the $O_h\rightarrow D_4$ symmetry breaking.

%%%%%%%%%%%%%%%%%%%%%%%%%%%%%%%%%%%%%%%%%%%%%%%%%%%%%%%%%%%%%%%%%

\end{widetext}


\begin{thebibliography}{10}
\bibitem{Logarithmic}
The logarithmic correction in Eq. (\ref{eq:modified_correlation}) is neglected in this treatment which does not affect the ratio $C_1/C_2$ displayed in Fig. \ref{fig:CnD} (f).
\bibitem{LargeU}
Since a charge gap opens at infinitesimally small $U$, 
the low energy physics in the gapless spin sector of $H_F$ when $\Delta=0$
is the same for both the weak and the large $U$, where the large $U$ limit reproduces the AFM Heisenberg model.
\end{thebibliography}
\end{document}